\documentclass[%
  reprint,           %
  aps,
  rmp,
  superscriptaddress,
  floatfix,
  longbibliography,
]{revtex4-2}

\usepackage[utf8]{inputenc}
\usepackage[T1]{fontenc}
\usepackage{amsmath,amssymb,amsfonts}
\usepackage{graphicx}
\usepackage{hyperref}
\usepackage{microtype}
\usepackage{bm}
\usepackage{xcolor}
\usepackage{slashed}

\hypersetup{
  colorlinks=true,
  citecolor=blue,
  linkcolor=blue,
  urlcolor=blue
}

\newcommand \widebar [1] {\overline{#1}}

\newcommand{\Lbra}[1]{\left<#1\left|}
\newcommand{\Rket}[1]{\right|#1\right>}

\begin{document}

\title{Generalized parton distributions: Theory meets experiment}
\author{Yuxun Guo}
\email{yuxunguo@lbl.gov}
\affiliation{Nuclear Science Division, Lawrence Berkeley National
Laboratory, Berkeley, CA 94720, USA}
\affiliation{Physics Department, University of California, Berkeley, CA 94720, USA}
\affiliation{RIKEN BNL Research Center and Physics Department, Brookhaven National Laboratory, Upton, NY 11973, USA}
\author{Xiangdong Ji}
\email{xdji@sjtu.edu.cn, corresponding author}
\affiliation{Tsung-Dao Lee Institute and School of Physics and Astronomy,
Shanghai Jiao Tong Unviersity, Shanghai, China}

\author{Yao Ji}
\email{yaoji@cuhk.edu.cn, corresponding author}
\affiliation{School of Science and Engineering, The Chinese University of Hong Kong, Shenzhen 518172, China}

\author{Jialu Zhang}
\email{elpsycongr00@sjtu.edu.cn,corresponding author}
\affiliation{Tsung-Dao Lee Institute and School of Physics and Astronomy,
Shanghai Jiao Tong Unviersity, Shanghai, China}

\begin{abstract}

Over the past three decades, generalized parton distributions (GPDs) have emerged as one of the most active and important areas of research in nucleon structure and quantum chromodynamics (QCD). Since the last comprehensive review two decades ago, substantial progress has been made in experimental measurements of hard exclusive processes, such as deeply virtual Compton scattering and near-threshold $J/\psi$ production, as well as in increasingly sophisticated phenomenological analyses of GPDs that enable three-dimensional nucleon tomography. Theoretical advances in perturbative coefficient functions, scale evolutions, and kinematic and power corrections have considerably improved the precision of GPD phenomenology, while new hard exclusive processes for probing GPDs have been explored. More interestingly, lattice QCD can now directly access GPDs at fixed parton momentum fractions $x$ and skewness $\xi$ through large-momentum expansions, in addition to the traditional calculations of their moments, or generalized form factors. Significant progress has also been made in exploring the QCD energy-momentum tensor that encodes fundamental information on the nucleon's mass distribution, complete spin structure, and spatial distributions of momentum current and color-Lorentz forces acting on quarks and gluons.

\end{abstract}

\maketitle

\tableofcontents
\vspace{2em}

\section{Introduction}
\label{sec:1_intro}

Generalized parton distributions (GPDs) have been studied for more than three decades, beginning with the pioneering work of \textcite{Muller:1994ses}. They are defined through matrix elements of light-ray (or light-cone) correlators between off-forward hadron states, with perturbative evolution that interpolates between those of parton distribution functions (PDFs) and meson light-cone distribution amplitudes (DAs). This early work, however, was not motivated by the role of GPDs in understanding nucleon structure. Shortly thereafter, GPDs were independently introduced as off-forward parton distributions by one of the present authors~\cite{Ji:1996ek} in the context of studies of the nucleon spin structure. These studies require knowledge of the energy-momentum tensor (EMT) form factors in quantum chromodynamics (QCD). Although the nucleon EMT form factors cannot be measured directly, partonic sum rules derived from GPDs provide an indirect means of accessing them experimentally. This remarkable connection between GPDs and fundamental aspects of nucleon structure subsequently stimulated rapid growth of the field. Approximately 300 papers on GPDs appeared between 1996 and 2005, with several review articles documenting these early and fruitful developments~\cite{Ji:1998pc,Goeke:2001tz,Diehl:2003ny,Ji:2004gf,Belitsky:2005qn}. An early history of GPDs can be found in the APS 2016 Feshbach prize talk by one of the authors~\cite{Ji:2016djn}.

While the theoretical foundations of GPDs were largely established by 2005, substantial progress has followed. A major advancement is the direct calculation of $x$-dependent GPDs in lattice QCD using large-momentum effective theory (LaMET)~\cite{Ji:2013dva,Ji:2014gla,Ji:2020ect,Alexandrou:2020zbe, Lin:2020rxa}. In parallel, EMT form factors have become central to studies of the nucleon mass decomposition and mechanical structure~\cite{Polyakov:2018zvc,Burkert:2018bqq,Shanahan:2018nnv, Ji:2021mtz,Ji:2025qax,Ji:2026lyj}, while the range of processes considered as probes of GPDs and their moments has broadened, notably through near-threshold heavy-quarkonium production~\cite{Hatta:2018ina,Guo:2021ibg,GlueX:2019mkq,Duran:2022xag,GlueX:2023pev,007:2026dow,Tyson:2026gnd}. At the same time, the steadily expanding experimental and lattice-QCD constraints have enabled increasingly comprehensive global analyses of GPDs~\cite{Kumericki:2016ehc,Moutarde:2019tqa,Guo:2025muf}. This review aims to bridge the roughly two-decade gap since the last major reviews by providing a unified account of these theoretical, computational, phenomenological, and experimental advances.

The nucleon---the proton and neutron---is the principal building block of visible matter in the Universe. The study of nucleon structure spans more than a century and has been recognized by at least five Nobel Prizes. These include Stern’s measurement of the proton magnetic moment~\cite{Frisch:1933i,Estermann:1933i}, Hofstadter’s determination of the proton’s finite size~\cite{Hofstadter:1956qs}, Gell-Mann’s quark model~\cite{Gell-Mann:1964ewy}, the discovery of quarks through deep-inelastic scattering (DIS)~\cite{Bloom:1969kc,Breidenbach:1969kd}, and the establishment of QCD as the fundamental theory of strong interactions through the discovery of asymptotic freedom~\cite{Gross:1973id,Politzer:1973fx}. The last two, however, marked only the beginning of a quantitative understanding of nucleon structure in terms of its fundamental constituents: quarks and gluons.

Historically, two broad classes of experiments have played central roles in the study of nucleon structure: elastic electron scattering and related electromagnetic processes, pioneered by Hofstadter~\cite{Hofstadter:1956qs}, and DIS and related high-energy reactions, pioneered by Friedman, Kendall, and Taylor~\cite{Bloom:1969kc,Breidenbach:1969kd}. Elastic electron scattering measures the nucleon electromagnetic form factors and thereby probes the spatial distributions of charge and magnetization. In contrast, DIS reveals the longitudinal momentum distributions of quarks and gluons within the nucleon. Together, these approaches have provided more than five decades of experimental data and established much of our present understanding of nucleon structure. Meanwhile, it has become increasingly clear that the nucleon is far more complex than suggested by a simple quark model: it is a relativistic many-body system governed by rich nonperturbative QCD dynamics.

A more complete description of nucleon structure therefore requires knowledge of correlations between parton momentum and spatial distributions that are inaccessible through either elastic scattering or inclusive DIS alone. Such information is essential for understanding how the proton spin arises from quark and gluon helicities and orbital angular momenta (OAM), as well as for resolving the spatial distributions of energy, momentum current, and internal forces among its constituents. 
GPDs provide a natural framework for addressing these questions by unifying the information contained in collinear parton distributions and elastic form factors while correlating longitudinal momentum with transverse spatial structure. Moreover, their Mellin moments yield generalized form factors of local ``twist-two'' operators, including the quark and gluon components of the QCD EMT, thereby connecting the partonic structure of the nucleon to its spin, mass, and mechanical properties.

Another crucial aspect of GPDs lies in the multidimensional imaging of the nucleon. A natural framework for visualizing the multidimensional structure of the nucleon is provided by Wigner distributions, which describe correlations between the position and momentum of quarks and gluons~\cite{Belitsky:2003nz}. In quantum mechanics, a Wigner distribution is a phase-space quasi-probability distribution and need not be positive definite; in QCD, its partonic generalization depends on the longitudinal momentum fraction $x$, transverse position $\boldsymbol{b}_\perp$, and transverse momentum $\boldsymbol{k}_\perp$. GPDs arise as projections of these more general distributions after integration over transverse momentum. A particularly transparent interpretation emerges at zero skewness, $\xi=0$, where \textcite{Burkardt:2000za,Burkardt:2002hr} showed that the Fourier transform of a GPD with respect to the transverse momentum transfer $\boldsymbol{\Delta}_\perp$ admits a probability-density interpretation in impact-parameter space. In particular, the Fourier transform of, e.g, the quark GPD $H^q(x,0,t)$ with $t=-\boldsymbol\Delta_\perp^2$, describes the transverse spatial distribution of quarks carrying longitudinal momentum fraction $x$ at an impact parameter $\boldsymbol{b}_\perp$ relative to the nucleon's transverse center of momentum. This interpretation provides the basis for nucleon tomography, in which longitudinal momentum and transverse spatial information are correlated within a unified partonic description.

The need to access the full spin and spatial structure of the nucleon motivates the exploration of new experimental processes and observables. Deeply virtual Compton scattering (DVCS), $\gamma^* N\to\gamma N$, first proposed and named in~\textcite{Ji:1996ek,Ji:1996nm}, provides a particularly clean realization of this idea. In the Bjorken limit, the highly virtual photon acts as a hard probe that couples directly to quarks, while the exclusivity of the process preserves the spatial information encoded in the momentum transferred to the nucleon. The resulting amplitudes are then described in terms of GPDs, which correlate the longitudinal momentum of partons with their transverse spatial distribution. Closely related deeply virtual meson-production (DVMP) processes, $\gamma^*N\to MN$, provide complementary sensitivity to the flavor, spin, and gluon structure of the nucleon~\cite{Radyushkin:1996ru,Collins:1996fb}. 

Following these theoretical achievements, a decisive experimental milestone was reached in 2001, when the CLAS at Jefferson Lab (JLab) and HERMES collaborations at Hadron--Electron Ring Accelerator (HERA) observed beam-spin asymmetries in exclusive photon electroproduction, while H1 at HERA measured the DVCS cross section at high energy~\cite{CLAS:2001wjj,HERMES:2001bob,H1:2001nez}. The observed beam-spin asymmetries exhibited the characteristic interference between the DVCS amplitude and the Bethe--Heitler (BH) process, in which the final-state photon is radiated from the lepton. Since the BH amplitude is calculable in quantum electrodynamics in terms of the nucleon elastic form factors, the interference provides direct sensitivity, at leading twist, to the imaginary part of the DVCS amplitude. These measurements demonstrated that spin asymmetries can reveal GPD-sensitive information even in kinematics where the BH contribution dominates the cross section. Complementarily, the H1 measurements of the DVCS cross section over a broad range of $Q^2$ and photon--proton center-of-mass energy $W$ established the high-energy, small-$x$ regime as an important testing ground for QCD-based descriptions of DVCS.

The development of GPDs was accompanied by several influential reviews that established the conceptual and phenomenological foundations of the field~\cite{Ji:1998pc,Goeke:2001tz,Diehl:2003ny,Ji:2004gf,Belitsky:2005qn}, which differ substantially in scope, technical depth, and intended audience.

\textcite{Ji:1998pc} appeared at an early stage of the field. It identifie GPDs 
as the relevant nonperturbative quantities in DVCS and hard diffractive vector-meson production and established the connection between their Mellin moments and generalized form factors through polynomiality. Of particular importance is the relation to the energy-momentum tensor and, consequently, to the quark and gluon angular momenta in the nucleon. The review is concise and historically important, although it naturally predates later advances.

\textcite*{Goeke:2001tz} provided one of the first broad reviews of GPDs, covering their defining properties, polynomiality, the $D$-term, the spin sum rule, chiral dynamics, and transition GPDs. The review also examined DVCS and hard meson electroproduction in detail and identified experimentally useful observables. A distinctive feature is its emphasis on the large-$N_c$ limit and the chiral quark-soliton model as guides for phenomenological parametrizations, making it particularly valuable for understanding early model-based approaches and connections to low-energy hadronic dynamics.

\textcite{Diehl:2003ny} became a standard theoretical reference through its systematic and comprehensive treatment of the subject. It covers operator definitions, helicity structure, polynomiality, evolution, positivity bounds, double distributions, light-cone wave-function overlaps, and the impact-parameter space representation. It also discusses generalized distribution amplitudes and their relation to GPDs through crossing symmetry. On the phenomenological side, the review addresses factorization in exclusive processes, power corrections, small-$x$ dynamics, and strategies for extracting GPD information from observables. Compared with \textcite*{Goeke:2001tz}, it places less emphasis on specific hadronic models and more on the general field-theoretical framework.

\textcite{Ji:2004gf} adopted a more pedagogical and conceptually oriented perspective. Rather than providing an exhaustive technical survey, it introduced GPDs through their connections to form factors, parton distributions, and quantum phase-space, or Wigner, distributions. Particular emphasis was placed on transverse spatial imaging in impact-parameter space and on the physical content of GPD moments. The review also briefly discussed hard exclusive processes, lattice-QCD calculations, and phenomenological parametrizations. Its principal strength lies in elucidating the role of GPDs in constructing multidimensional images of partonic structure.

Finally, \textcite{Belitsky:2005qn} provided the most encyclopedic account among these early reviews. They combined a detailed discussion of GPD properties, modeling, evolution, double distributions, positivity constraints, and phase-space interpretations with a comprehensive treatment of exclusive reactions. Particular attention was devoted to higher-order and higher-twist effects in photon, lepton-pair, and meson production, as well as to methods for deriving the corresponding coefficient functions and observables. The review remains an especially valuable technical reference for the perturbative-QCD foundations of GPD phenomenology and the broad range of processes sensitive to GPDs.

Since 2005, GPD research has progressed from establishing the formal framework toward quantitative studies of nucleon tomography and the physics encoded in the QCD EMT form factors.\footnote{In some literature, they have been referred to as ``gravitational form factors.'' We will avoid this terminology in this review for two reasons: (i) their connection to actual gravitational phenomena is rather indirect, and (ii) unlike electromagnetic form factors, they cannot be measured through a corresponding gravitational scattering experiment.} Major advances have been made in exclusive-scattering measurements, theoretical precision, lattice QCD calculations, and phenomenological analyses.

Following its first observations in 2001, DVCS has become the primary precision channel for constraining GPDs because of its clean final state and known BH reference amplitude. Measurements span the small-$x_B$ regime at HERA, intermediate $x_B$ at HERMES and COMPASS, and the valence region at JLab~\cite{H1:2009wnw,ZEUS:2008hcd,HERMES:2012gbh,COMPASS:2018pup,CLAS:2018bgk,JeffersonLabHallA:2015dwe}. Recent 12~GeV JLab cross sections and polarization observables, including neutron data~\cite{CLAS:2022syx,JeffersonLabHallA:2022pnx,CLAS:2024qhy}, have substantially strengthened Compton form factor (CFF) constraints and global GPD analyses~\cite{Kumericki:2016ehc,Cuic:2020iwt,Guo:2025muf}.

For DVMP, HERA measurements remain the principal inputs at high $Q^2$, providing complementary sensitivity to gluon and flavor structure~\cite{H1:2009cml,ZEUS:2007iet}. JLab and COMPASS have broadened the polarization observables and longitudinal--transverse separations available across light-meson channels~\cite{JeffersonLabHallA:2016wye,CLAS:2019uzc,COMPASS:2019fea,COMPASS:2024hvm}. Interpretation is more challenging than for DVCS because leading-power collinear factorization applies to longitudinal meson production, while transverse amplitudes and power corrections can be sizable~\cite{Collins:1996fb,Goloskokov:2007nt,Cuic:2023mki}. Future programs at the planned Electron--Ion Collider (EIC) at Brookhaven National Laboratory in the U.S.~\cite{AbdulKhalek:2021gbh} and the proposed Electron--Ion Collider in China (EicC)~\cite{Anderle:2021wcy, Xiao:2026tbs} will further extend the precision measurements of DVCS and DVMP to the high-$Q^2$ and small $x_B$ region.

Near-threshold $J/\psi$ production has recently emerged as a probe of large-skewness gluon GPDs and EMT form factors of the nucleon. High-statistics GlueX and $J/\psi$-007 measurements, together with CLAS12 results, have expanded the threshold coverage for the proton target~\cite{GlueX:2019mkq,Duran:2022xag,GlueX:2023pev,007:2026dow,Chatagnon:2026qsv} with recent efforts extending to the neutron target~\cite{Tyson:2026gnd}. The amplitude admits a GPD-moment expansion whose leading terms are governed by gluon EMT form factors~\cite{Hatta:2018ina,Guo:2021ibg,Guo:2023qgu}, motivating their phenomenological extractions and lattice-QCD comparisons~\cite{Guo:2023pqw,Hackett:2023rif}, though relativistic corrections and competing reaction mechanisms remain under study~\cite{Du:2020bqj,JointPhysicsAnalysisCenter:2023qgg,Blask:2025jua}.

The experimental advances have been accompanied by improved theoretical control. Beyond the established next-to-leading-order (NLO) description, next-to-next-to-leading-order (NNLO) coefficient functions are now available for major DVCS channels and pion DVMP, while two-loop GPD evolution and partial three-loop results mark progress toward fully consistent NNLO analyses~\cite{Braun:2017cih,Braun:2020yib,Braun:2022bpn,Braun:2025noa,Chen:2026vff}. Complementary developments in conformal-partial-wave methods, dispersion relations, and threshold resummation help organize amplitudes and stabilize phenomenological analyses~\cite{Mueller:2005ed,Kumericki:2019ddg,Schoenleber:2022myb}. At the moderate $Q^2$ of current experiments, systematic finite-$t$ and target-mass corrections, together with new analyses of genuine twist-three factorization, improve control over power-suppressed effects~\cite{Braun:2020zjm,Braun:2025xlp,Schoenleber:2024ihr}. Finally, doubly virtual Compton scattering (DDVCS) and hard exclusive $2\to3$ processes offer new opportunities for the $x$ dependence of GPDs~\cite{Belitsky:2002tf,Guidal:2002kt,Qiu:2022bpq,Qiu:2022pla}. These advances enable more consistent comparisons with data, although perturbative and power-suppressed uncertainties remain important at present experimental scales.

On the nucleon-structure frontier, conventional lattice QCD has steadily improved calculations of GPD moments and generalized form factors, including higher moments and increasingly complete quark--gluon decompositions of the proton momentum, angular momentum, and EMT form factors~\cite{Bhattacharya:2023ays,Hackett:2023rif,Alexandrou:2026oks}. However, Euclidean lattice methods cannot directly access light-cone, $x$-dependent GPDs. LaMET addresses this limitation by computing nonlocal equal-time matrix elements in fast-moving hadrons and matching quasi-distributions to their light-cone counterparts~\cite{Ji:2013dva,Ji:2014gla,Ji:2020ect}.

LaMET studies have progressed from the first pion calculation at zero skewness to nucleon GPDs at nonzero momentum transfer and skewness~\cite{Chen:2019lcm,Alexandrou:2020zbe,Bhattacharya:2022aob,Chu:2025kew, Ding:2024saz}. Milestones include physical-pion-mass tomographic studies of the nucleon and pion~\cite{Lin:2020rxa,Lin:2021brq,Lin:2023gxz} and an NNLO reanalysis of zero-skewness nucleon GPDs incorporating renormalization-group and leading-renormalon resummation~\cite{Holligan:2023jqh}. More recently, threshold resummation has been developed for quasi-GPD matching at nonzero skewness~\cite{Holligan:2025baj}. These results establish the feasibility of direct lattice tomography, although nucleon calculations remain limited by finite boosts, statistical noise, disconnected contributions, and systematic uncertainties, especially for gluon GPDs.

GPD phenomenology has moved from model studies toward global analyses that combine experimental data with PDFs, elastic form factors, lattice-QCD inputs, theoretical constraints such as polynomiality and sum rules, and improved perturbative accuracy. Frameworks such as PARTONS have made the implementation and comparison of GPD models more systematic~\cite{Berthou:2015oaw}. Modern analyses increasingly use flexible double-distribution~\cite{Vanderhaeghen:1999xj,Goloskokov:2007nt}, conformal-partial-wave~\cite{Cuic:2023mki,Guo:2025muf}, or neural-network parameterizations~\cite{Moutarde:2019tqa,Grigsby:2020auv,Almaeen:2024guo,CaleroDiaz:2025luc,Xu:2026lko}. Simultaneous NLO DVCS-DVMP fits have begun to test GPD universality across multiple exclusive channels~\cite{Cuic:2023mki,Guo:2025muf}. 
A recent milestone was the first global analysis using ``GPD through universal momentum parametrization (GUMP)'' combining lattice-QCD calculations, PDFs, form factors, and DVCS as well as DVMP data to extract nonzero-skewness GPDs~\cite{Guo:2025muf}. Nonetheless, flavor separation, $E$ and polarized GPDs, sea-quark and gluon GPDs, and the $D$-term remain less well constrained, motivating further progress through precision measurements of exclusive processes at existing and future facilities, improved lattice-QCD constraints, and consistent global analyses that incorporate these complementary inputs.

Finally, GPD studies have substantially broadened our knowledge of nucleon structure, as anticipated three decades ago~\cite{Ji:1996ek}. Phenomenological studies of twist-two GPDs can provide a complete partonic description of transverse proton spin, whereas resolving the longitudinal spin decomposition at the density level additionally requires the far more difficult twist-three quark--gluon correlations~\cite{Ji:1996ek,Hatta:2012cs,Ji:2012sj,Ji:2020hii}.Through GPD sum rules, the form factors of the QCD EMT determine energy and scalar-field distributions and their associated radii, while the momentum-current distributions encode the forces acting on quarks and gluons~\cite{Polyakov:2018zvc,Shanahan:2018nnv,Ji:2021mtz,Hackett:2023rif,Ji:2025qax}.

This review focuses primarily on nucleon structure physics. We have made every effort to include GPD-related works published since 2005 and apologize for any relevant contributions that may have escaped our attention. We also note a recent workshop summary white paper on closely related topics, which provides a useful complementary reference~\cite{Boer:2025ixc}.

The remainder of this review is organized as follows. Section \ref{sec:2_gpd_basic} summarizes the key properties of GPDs and the physical insights they provide. Section \ref{sec:3_hard_proc} reviews recent theoretical progress on hard exclusive and diffractive processes, including higher-order perturbative calculations and near-threshold $J/\psi$ production. Section \ref{sec:4_lattice_qcd} discusses first principles lattice-QCD calculations of GPDs and related form factors. Section \ref{sec:5_exp_and_pheno} reviews experimental advances and global analyses of GPDs from experimental and lattice-QCD constraints, together with selected model studies. Finally, Section \ref{sec:6_conclude} presents our conclusions and future prospects.

\section*{Conventions and Notations}

Throughout this review, we adopt natural units by setting $\hbar=c=1$ and the metric convention that $g^{\mu\nu}=\mathrm{diag}(1,-1,-1,-1)$ and $\epsilon^{0123}=+1$. We define two dimensionless light-cone vectors as,
\begin{equation}\label{eq:lc_vectors}
    p^\mu=\Lambda\,(1,0,0,1)/\sqrt{2}\, ,\quad n^\mu=\Lambda^{-1}\,(1,0,0,-1)/\sqrt{2}\, ,
\end{equation}
in $(V^t,V^x,V^y,V^z)$ or $(V^0,V^1,V^2,V^3)$ order. The light-cone vectors satisfy $p^2=n^2=0$ and $p\cdot n=1$ with an arbitrary normalization $\Lambda$: a boost in the $z$ direction with $n^\mu\to e^{-\eta} n^\mu$ and $p^\mu\to e^{\eta}p^\mu$ leaves all physical expressions unchanged. We denote the space-time coordinates by $x^\mu=(x^0,x^1,x^2,x^3)=(t,x,y,z)$ and consider a frame in which the hadron moves along the $+z$ ($p^\mu$) direction conjugate to the $n^\mu$ direction.

For any four-vector $V^\mu$, we define its light-cone projection with $V^+=V\cdot n$, $V^-=V\cdot p$, and $V_\perp^\mu=g_\perp^{\mu\nu}V_\nu$ and the following decomposition,
\begin{equation}
    V^\mu=V^+p^\mu+V^-n^\mu+V_\perp^\mu\ ,
\end{equation}
with Einstein summation convention implied unless stated otherwise. The transverse tensors are $g_\perp^{\mu\nu}\equiv g^{\mu\nu}-p^\mu n^\nu-n^\mu p^\nu$ and $\epsilon_\perp^{\mu\nu}\equiv\epsilon^{\mu\nu\rho\sigma}p_\rho n_\sigma$, with $g_\perp^{11}=g_\perp^{22}=-1$ and $\epsilon_\perp^{12}=-\epsilon_\perp^{21}=1$ for the basis above.  In tensor notation, Greek indices $\mu,\nu,\rho,\sigma,\ldots$ denote Lorentz components, Latin indices $i,j,k,\ldots$ transverse components or color indices in fundamental representation, and $a,b,c,\ldots$ color indices in adjoint representation unless stated otherwise. Symmetrization and antisymmetrization are defined as $A^{(\mu\nu)}\equiv(A^{\mu\nu}+A^{\nu\mu})/2$ and $A^{[\mu\nu]}\equiv(A^{\mu\nu}-A^{\nu\mu})/2$. We also define the leading-twist projection operator $\mathbf{S}$, which extracts the symmetric and traceless part of a tensor:
$\mathbf{S}\,T^{\mu\nu\rho\cdots}\equiv T^{(\mu\nu\rho\cdots)}-\rm{all~trace~terms}$.

In QCD, quark fields are denoted by $\psi_q(x)$ with $\bar{\psi}_q\equiv\psi_q^\dagger\gamma^0$, where the flavor label $q$ and the fundamental color and Dirac indices will be suppressed unless needed. The gluon potential is denoted by $A_a^\mu$, with field-strength tensor $F_a^{\mu\nu}\equiv\partial^\mu A_a^\nu-\partial^\nu A_a^\mu+gf^{abc}A_b^\mu A_c^\nu$ and its dual $\widetilde F_a^{\mu\nu}\equiv\epsilon^{\mu\nu\rho\sigma}F_{a,\rho\sigma}/2$. We define the matrix-valued fields $A^\mu\equiv A_a^\mu T^a$, $F^{\mu\nu}\equiv F_a^{\mu\nu}T^a$, and $\widetilde F^{\mu\nu}\equiv\widetilde F_a^{\mu\nu}T^a$, where the $SU(3)$ generators satisfy $[T^a,T^b]=if^{abc}T^c$ and $\operatorname{Tr}(T^aT^b)=\delta^{ab}/2$. We introduce the covariant derivative $D^\mu\equiv\partial^\mu-igA^\mu$. Arrows on the covariant derivatives, $\overrightarrow D$ and $\overleftarrow D$, indicate the direction in which the derivative acts, with right action understood when no arrow is shown, and we define $\overleftrightarrow D_\mu\equiv(\overrightarrow D_\mu-\overleftarrow D_\mu)/2$. For all light-cone correlators, gauge links between separated fields are understood and will be suppressed unless otherwise specified.

\section{Basic of GPDs}
\label{sec:2_gpd_basic}

We begin by reviewing some basic properties of GPDs, most of which have been covered in previous reviews. We introduce the light-cone definitions of twist-two GPDs and their partonic interpretations, and then discuss the polynomiality conditions and their connections to form factors, particularly those of the energy-momentum tensor. We next turn to the spin structure of the nucleon and briefly discuss twist-three GPDs, which are necessary for establishing the complete longitudinal spin sum rule. We further highlight the physics encoded in the EMT form factors by considering mass and scalar-field distributions, as well as force densities. We then return to the impact-parameter-space interpretation of GPDs, followed by two subsections on perturbative evolution and conformal expansion, and on GPDs in extreme kinematic regimes, including small $x$ and large $\xi$.

\subsection{Definition and partonic interpretation of GPDs}\label{subsec:2_GPD_def}

GPDs, also known as off-forward parton distributions, are defined in terms of the off-forward matrix elements of light-cone correlators with $P'\not=P$. For quarks,
\begin{equation}\label{def:GPD}
    F_q^{[\Gamma]}= \int \frac{\mathrm{d} \lambda}{2\pi} e^{i x \lambda}\left.\Lbra{P'}\bar \psi_q\left(-\frac{z}{2}\right) \Gamma\psi_q\left(\frac{z}{2}\right)\Rket{P}\right.\,,
\end{equation}
with $z=z^-n$, $\lambda=z\cdot\bar P$, and $\bar P\equiv(P+P')/2$ the average nucleon momentum. Different Dirac matrices $\Gamma$ project out distinct spinor structures. In the infinite-momentum frame, $\Gamma=\gamma^+$, $\gamma^+\gamma^5$, and $i\sigma^{+i}$ define the leading-power (leading-twist) quark GPDs $F_q$, $\widetilde{F}_q$ and $F^i_{q,T}$, whereas the following correlators define the leading-twist gluon GPDs of different polarizations~\cite{Ji:1998pc},
\begin{eqnarray} \label{def:GPD2}
&F_g&= \frac{2}{\bar{P}^+}\int \frac{\mathrm{d}\lambda}{2\pi}\, e^{ix\lambda}\Lbra{P'} F^{+i}\left(-\frac{z}{2}\right) F^{+}_{\phantom{n}i}\left(\frac{z}{2}\right) \Rket{P} , \nonumber \\ 
&\widetilde{F}_g &= \frac{2i}{x\bar{P}^+}\int \frac{\mathrm{d}\lambda}{2\pi}\, e^{ix\lambda}\Lbra{P'} F^{+i}\left(-\frac{z}{2}\right) \widetilde{F}^{+}_{\phantom{n}i} \left(\frac{z}{2}\right) \Rket{P}, \nonumber\\
&F_{g,T}^{ij}& = \! \frac{-2}{\bar{P}^+}\!\!\int\! \frac{\mathrm{d}\lambda}{2\pi}\, e^{ix\lambda}\mathbf{S} \Lbra{P'} F^{+i}\left(-\frac{z}{2}\right) F^{+j}\left(\frac{z}{2}\right) \Rket{P},
\end{eqnarray}
all with $z=z^-n$ and $\lambda = z \cdot \bar{P}$.  Henceforth, the initial and final proton spin labels $S$ and $S'$ and the arguments of $F_q^{[\Gamma]}$ and $F_g$ are suppressed for simplicity.

\begin{figure*}[t]
    \centering
    \includegraphics[width=0.92\textwidth]{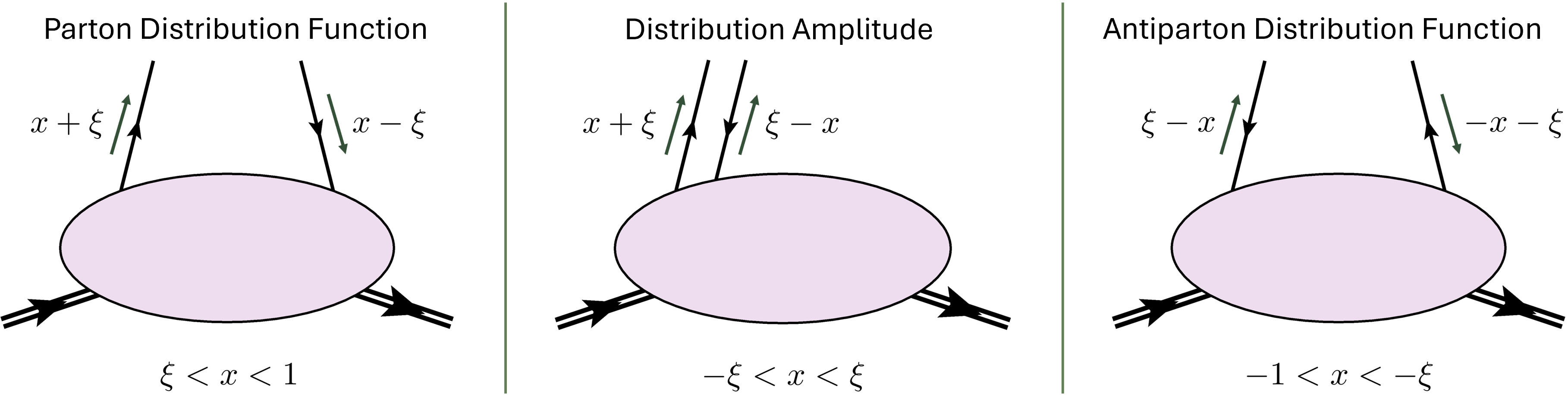}
    \caption{Partonic interpretation of quark GPDs as an example across kinematic regions. The regions $x>\xi$ and $x<-\xi$ generalize the parton and antiparton contributions of forward PDFs ($x>0$ and $x<0$), while the central region $-\xi<x<\xi$ corresponds to a DA-like configuration involving parton--antiparton pair production. All partonic momentum fractions are positive along the corresponding arrows, while their alignment relative to the fermion-number flow distinguishes particles from antiparticles.}
    \label{Fig:GPDparton}
\end{figure*}

Each off-forward matrix element above can be further decomposed into independent Dirac structures. For the leading-twist chiral-even sector, the parameterization reads~\cite{Ji:1996ek, Ji:1998pc}
\begin{align}
   F_{q,g}&=\bar U(P')\left[
   \gamma^+ H_{q,g}+\frac{i\sigma^{+\nu}\Delta_\nu}{2M} E_{q,g}\right] U(P)\ ,\label{def:HEq}\\
   \widetilde{F}_{q,g}&=\bar U(P')\left[
   \gamma^+ \gamma^5 \widetilde H_{q,g}+\frac{\Delta^+ \gamma^5 }{2M} \widetilde E_{q,g}\right] U(P)\ . \label{def:HEq-tilde}
\end{align}
And the chiral-odd ones read~\cite{Diehl:2001pm}
\begin{align}
\begin{split}
    F^i_{q,T}&=\bar U(P')\Bigg[
    H_{q,T}\,i\sigma^{+i}
    +\widetilde{H}_{q,T}
    \frac{2\bar{P}^{[+}\Delta^{i]}}{M^2}
    \\
    &\qquad\quad
    +E_{q,T}
    \frac{\gamma^{[+}\Delta^{i]}}{M}
    +\widetilde{E}_{q,T}
    \frac{2\gamma^{[+}\bar{P}^{i]}}{M}
    \Bigg]U(P)\ ,\label{def:HEqT}
\end{split}\\
\begin{split}
    F_{g,T}^{ij}
    &=\mathbf{S} \frac{\bar{P}^{[+}\Delta^{j]}}{M\bar{P}^+}
    \bar U(P')\Bigg[
    H_{g,T}\,i\sigma^{+i}
    +\widetilde{H}_{g,T}
    \frac{2\bar{P}^{[+}\Delta^{i]}}{M^2}
    \\
    &\qquad\quad
    +E_{g,T}
    \frac{\gamma^{[+}\Delta^{i]}}{M}
    +\widetilde{E}_{g,T}
    \frac{2\gamma^{[+}\bar{P}^{i]}}{M}
    \Bigg]U(P)\, .\label{def:HEgT}
\end{split}
\end{align}
These decompositions involve a set of scalar functions: the chiral-even GPDs $H$ and $E$ (unpolarized/vector) and $\widetilde H$ and $\widetilde E$ (polarized/axial-vector), together with the chiral-odd GPDs $H_T$, $\widetilde H_T$, $E_T$, and $\widetilde E_T$ (transversity). These eight functions, corresponding to eight quark/gluon-nucleon ``scattering amplitudes'', will be referred to collectively as the leading-twist GPDs. Each of them encodes different nonperturbative information associated with the corresponding Dirac spinor structures for a spin-$1/2$ nucleon. 
 
Besides the average nucleon momentum $\bar P$ and the momentum transfer $\Delta\equiv P'-P$, we define two additional scalar quantities, the skewness parameter $\xi\equiv -\left(n\cdot\Delta\right)/(2\,n \cdot \bar{P})$, and the invariant momentum transfer $t\equiv\Delta^2$. Then these leading-twist GPDs depend solely on $(x,\xi,t)$ resulting from Lorentz symmetry. Furthermore, they can always be defined as even functions of $\xi$ by virtue of charge, parity, and time-reversal symmetry, and it is therefore sufficient to consider $\xi>0$ only.

In a symmetric frame where $(k_1+k_2)/(P+P')=x$ with $k_1$ and $k_2$ the momenta of the two partons, one finds $k_1^\mu=(x+\xi)\bar P^+  \, p^\mu$ and $k_2^\mu=(x-\xi)\bar P^+ \, p^\mu$ neglecting power-suppressed components. This provides an intuitive partonic interpretation of GPDs. In the infinite momentum frame, partons predominantly propagate along the $+$ light-cone direction (the $+z$ axis). Fields associated with negative momentum fraction $-x$ are understood as antipartons carrying momentum fraction $x>0$. By contrast, modes genuinely moving in the $-z$ direction at the speed of light carry vanishing rather than negative $+$ momentum; such contributions are known as zero modes in light-front quantization, where the theory is formulated at fixed light-front time to elucidate partonic structure. This picture immediately suggests the partonic interpretation of GPDs~\cite{Diehl:2000xz}, as illustrated in Fig. \ref{Fig:GPDparton}.

\begin{figure}[t]
    \centering
    \includegraphics[width=0.41\textwidth]{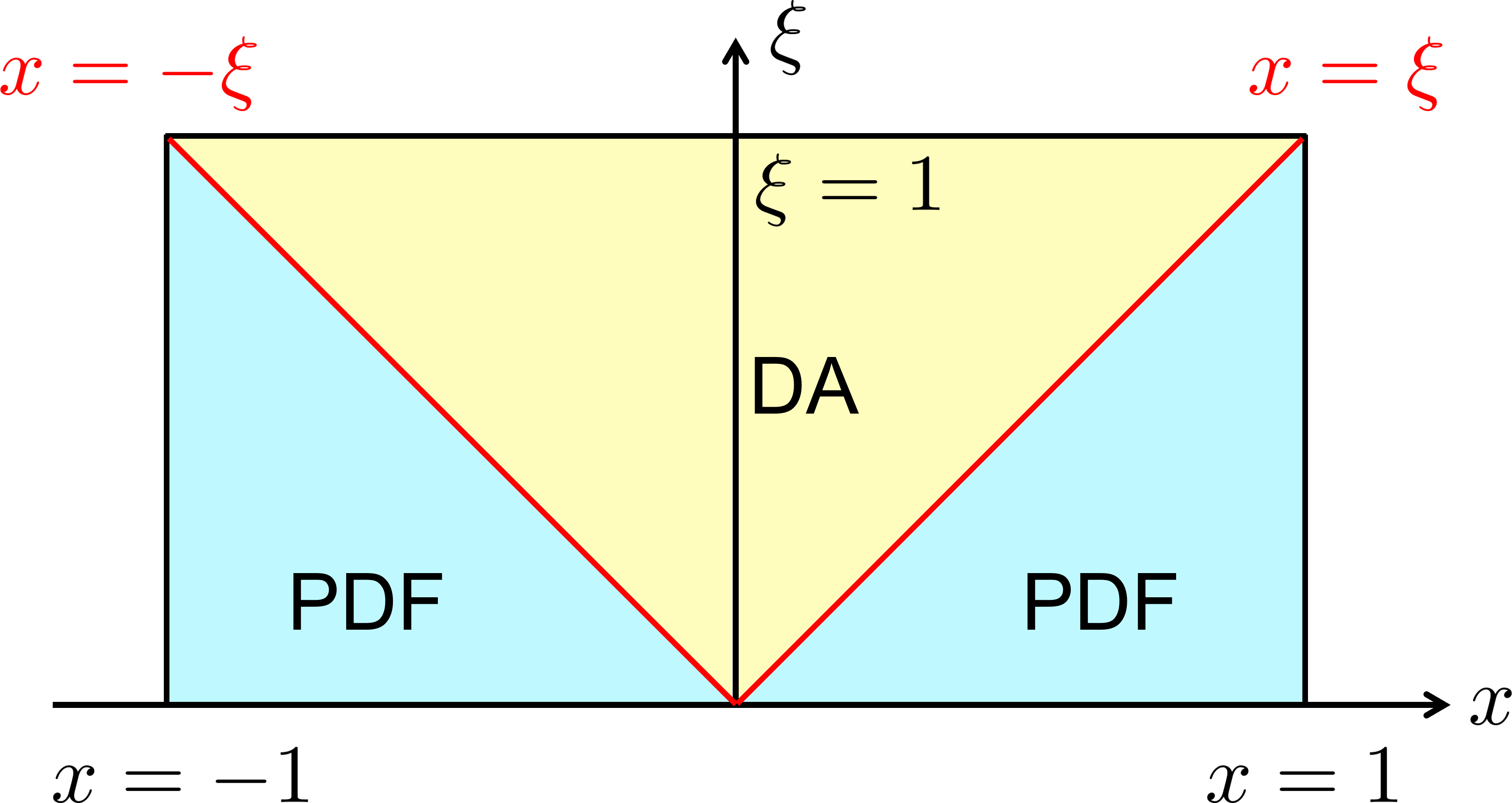}
    \caption{Phase space structure of GPDs in the $(x,\xi)$ plane. Generally, GPDs have support on $-1\le x\le1$, and $-1\le\xi\le1$ (we consider only positive $\xi$ here), and approach PDFs in the $\xi,t\to0$ limit. For non-zero skewness ($\xi$), the middle region $-\xi<x<\xi$ is DA-like, whereas the $x>\xi$ and $x<-\xi$ correspond to the parton and antiparton PDFs, respectively, as discussed above. GPDs are continuous but not necessarily smooth at the crossover lines $|x|=\xi$, where their partonic interpretation is unclear.} 
    \label{fig:gpdxxi}
\end{figure}

In the zero-skewness limit, the partonic interpretation of GPDs reduces to that of PDFs: the regions $x>0$ and $x<0$ describe partons and antipartons carrying momentum fraction $|x|$, respectively. For nonzero skewness, these domains generalize to $x>\xi$ and $x<-\xi$, which are therefore referred to as PDF-like regions.\footnote{In some literature, these have been referred to as the ``DGLAP regions,'' but this terminology obscures their underlying physical content. The same comment applies to the complementary region.} Their QCD evolutions resemble that of the PDFs, namely the Dokshitzer--Gribov--Lipatov--Altarelli--Parisi (DGLAP) equations~\cite{Gribov:1972ri, Dokshitzer:1977sg, Altarelli:1977zs}.
By contrast, the central region $-\xi<x<\xi$, which arises only at nonzero skewness, admits a DA interpretation in terms of parton--antiparton pair production. This DA-like region evolves according to equations analogous to the Efremov--Radyushkin--Brodsky--Lepage (ERBL) equations~\cite{Efremov:1978rn, Lepage:1980fj}. Thus, GPDs interpolate between two distinct types of partonic structure, unifying PDF-like and DA-like dynamics within a single framework. A ``phase space'' structure of GPDs on the $(x,\xi)$ plane is illustrated in Fig. \ref{fig:gpdxxi}. The transition lines $x=\pm\xi$, where one of the parton momentum fractions vanishes, are particularly subtle: GPDs have derivative discontinuities across
them, just like the ordinary PDFs have derivative discontinuities at $x=0, \pm 1$  where either the spectator or active parton has zero momentum fraction. 

\subsection{Polynomiality condition and EMT form factors}

\label{subsec:2_polynomiality}
Another important aspect of GPDs is their intimate connection to nucleon form factors, which characterize the internal structure of the nucleon by encoding the spatial distributions associated with its charge, axial charge, and other currents. Most importantly, GPDs provide the only known means of phenomenologically accessing the quark and gluon EMT form factors, which contain essential information on the distributions of mass, spin, momentum currents, and internal forces. Within the GPD framework, these form factors naturally emerge as moments in the longitudinal momentum fraction $x$, thereby establishing a direct link between partonic structure and electroweak form factors. 

Resulting from the Lorentz symmetry, the Mellin moments of GPDs satisfy the polynomiality constraints that they are finite-order polynomials in $\xi$~\cite{Ji:1998pc}:
\begin{align}\label{eq:quark_H_polynomiality}
\begin{split}
    \int_{-1}^{1} \mathrm{d}x\,
  x^{n-1}  &H_q(x,\xi,t)
 =
  \sum_{i=0}^{m}
  (2\xi)^{2i} A_q^{(n,2i)}(t)
  \\ & \qquad\quad + \operatorname{Mod}(n+1,2)\,
  (2\xi)^n C_q^{(n)}(t)\ ,
\end{split}
 \\\label{eq:quark_E_polynomiality}
 \begin{split}
     \int_{-1}^{1} \mathrm{d}x\,
  x^{n-1} &E_q(x,\xi,t)
  =
  \sum_{i=0}^{m}
  (2\xi)^{2i} B_q^{(n,2i)}(t)
  \\&\qquad\quad -
  \operatorname{Mod}(n+1,2)\,
  (2\xi)^n C_q^{(n)}(t)\ ,
 \end{split}
\end{align}
where $m=\left\lfloor (n-1)/2\right\rfloor$ denotes the largest integer less than or equal to $(n-1)/2$, while $\operatorname{Mod}(n+1,2)$ equals $0$ for odd $n$ and $1$ for even $n$. Here, the renormalization-scale and scheme dependence are suppressed with analogous relations holding for gluon GPDs as well as for other species, e.g., helicity form factors $\widetilde A_{q,g}^{(n,k)}(t)$ and $\widetilde B_{q,g}^{(n,k)}(t)$ for helicity GPDs $\widetilde H_{q,g}$ and $\widetilde E_{q,g}$ and tensor form factors $A_{q,g}^{T,(n,k)}(t)$, $B_{q,g}^{T,(n,k)}(t)$, $\widetilde {A}_{q,g}^{T,(n,k)}(t)$, and $\widetilde {B}_{q,g}^{T,(n,k)}(t)$ for transversity GPDs $H^T_{q,g}$, $E^T_{q,g}$, $\widetilde H^T_{q,g}$, and $\widetilde E^T_{q,g}$. 

The polynomiality condition is closely related to the short-distance operator product expansion (OPE) that the bilocal light-cone operators can be expanded into towers of local, gauge-invariant twist-two operators,
\begin{equation}
  \mathcal O_q^{\mu_1\cdots\mu_n}
  =
  \mathbf{S}\,\bar\psi_q\,
  \gamma^{\mu_1}
  i\overleftrightarrow D^{\mu_2}
  \cdots
  i\overleftrightarrow D^{\mu_n}
  \psi_q\ ,
\end{equation}
whose nonforward matrix elements are parametrized as
\begin{align}
&\left\langle P'
\left|
\mathcal O_q^{\mu_1\cdots\mu_n}
\right|
P
\right\rangle\nonumber\\
={}&
\mathbf{S}\bar U(P')\Bigg[\gamma^{\mu_1}
\sum_{i=0}^{m}
A_q^{(n,2i)}(t)\,
\Delta^{\mu_2}\cdots\Delta^{\mu_{2i+1}}
\bar P^{\mu_{2i+2}}\cdots\bar P^{\mu_n}
\nonumber\\
&+
\frac{i\sigma^{\mu_1\alpha}\Delta_\alpha}{2M}
\sum_{i=0}^{m}
B_q^{(n,2i)}(t)
\Delta^{\mu_2}\cdots\Delta^{\mu_{2i+1}}
\bar P^{\mu_{2i+2}}\cdots\bar P^{\mu_n}
\nonumber\\
&+
\operatorname{Mod}(n+1,2)\,
C_q^{(n)}(t)\,
\frac{1}{M}\,
\Delta^{\mu_1}\cdots\Delta^{\mu_n} \Bigg]U(P)\ .
\label{eq:quark_ff_decomposition}
\end{align}
These form factors are collectively referred to as the generalized form factors. %

Three groups of generalized form factors that are of particular interest will be highlighted here. First, the lowest ($x^0$) moments of quark vector GPDs $H_q$ and $E_q$ reproduce the Dirac and Pauli form factors, $F_{1,q}(t)$ and $F_{2,q}(t)$, respectively. They parameterize the hadronic matrix element of the electromagnetic current as,
\begin{align}
\langle P'|J^q_\mu(0)|P\rangle 
&=\bar U(P')\!\left[\gamma_\mu\,F_{1,q}(t)+\frac{i\sigma_{\mu\nu}\Delta^\nu}{2M}\,F_{2,q}(t)\right]\!U(P)\, \notag\\
J_\mu^q(x) &= \bar\psi_q(x)\gamma_\mu\psi_q(x)
\label{eq:JF1F2}
\end{align}
with $J^q_\mu(x)$ the quark electromagnetic current. 
In particular, the electric form factor $G_E(t)=F_1(t)+t/(4M^2)F_2(t)$ encodes the nucleon’s spatial charge distribution and thereby provides direct information on its charge radius; see~\textcite{Gao:2021sml} for a recent review.

Similarly, the lowest ($x^0$) moments of the axial-vector GPDs $\tilde{H}_q$ and $\tilde{E}_q$ yield the axial and pseudoscalar form factors, $G_A(t)$ and $G_P(t)$ defined through
\begin{align}
\langle P'|J^q_{5\mu}(0)|P\rangle
&=\!\bar U(P')\!\!\left[\gamma_\mu\gamma_5G_{A,q}(t)
+\frac{\Delta_\mu\gamma_5}{2M}G_{P,q}(t)\right]\!\!U(P)\notag\\
J^q_{5\mu}(x) &= \bar\psi_q(x)\gamma_\mu\gamma_5\psi_q(x)\ ,
\label{eq:J5GAGP}
\end{align}
which are associated with the nucleon’s axial charge and chiral structure. More specifically, $G_A(t)$ characterizes the distribution of axial charge, offering essential information on how quark spin contributes to the nucleon spin. Its forward limit, $G_A(0)$, the axial charge, gives the strength of the nucleon coupling to the axial current and plays an important role in beyond-standard-model physics as well; see, for instance,~\textcite{MINERvA:2023avz, Jang:2023zts} for recent progress from experimental programs and lattice QCD simulations.

More importantly, the second Mellin moments of the vector GPDs yield the EMT form factors~\cite{Ji:1996ek},
\begin{align} \label{eq:2_Hqpoly}
 \int_{-1}^{1}\text{d}x\,x\,H^q(x,\xi,t)
 &=A_q(t)+4\xi^2C_q(t)\, ,
 \\ \label{eq:2_Eqpoly}
 \int_{-1}^{1}\text{d}x\,x\,E^q(x,\xi,t)
 &=B_q(t)-4\xi^2C_q(t)\, .
\end{align}
with analogous relations for gluons. These form factors parameterize matrix elements of the QCD EMT. For a spin-$\frac{1}{2}$ nucleon, the decomposition reads~\cite{Ji:1996ek}
\begin{align}
    &\langle P'|T^{(\mu\nu)}_{f}|P\rangle =\bar U(P') \Big[A_{f}(t)\gamma^{(\mu}\bar P^{\nu)}  +B_{f}(t)\frac{\bar P^{(\mu}i\sigma^{\nu)\alpha}\Delta_\alpha}{2M} \nonumber\\
     &\quad+ C_{f}(t)\frac{\Delta^\mu\Delta^\nu-g^{\mu\nu}\Delta^2}{M}+\bar{C}_{f}(t)M g^{\mu\nu}\  \Big]U(P) \  , \label{eq:emtmatrix}
\end{align}
where $f=\{q,g\}$. It is important to note that the quark and gluon EMTs are not separately conserved and therefore contain the $\bar C_{q,g}(t)$ form factors, which cancel in the total EMT, $\bar C_q(t)+\bar C_g(t)=0$. The individual $\bar C_{q,g}(t)$ are not directly accessible through leading-twist GPDs and depend on the prescription used to separate the trace anomaly into quark and gluon components. With the standard  anomaly operator, they can be related to the corresponding nucleon scalar form factors~\cite{Ji:2026lyj,Yang:2026wzi}.

The EMT form factors encode key properties of the nucleon associated with its energy/mass, spin, and momentum current. In particular, $A_{q,g}(0)$ gives the total quark or gluon momentum fraction, while $[A_{q,g}(0)+B_{q,g}(0)]/2$ enters the spin sum rule as the quark or gluon angular momentum~\cite{Ji:1996ek}. Defining the total EMT form factors by
\begin{equation}
F(t)=F_q(t)+F_g(t), \qquad F={A,B,C},
\end{equation}
conservation of momentum and angular momentum implies $A(0)=1$ and $B(0)=0$. The form factor associated with the QCD Hamiltonian density, which is a linear combination of $A(t),B(t),C(t)$, characterizes the spatial distribution of internal energy~\cite{Ji:2021mtz}. Because direct gravitational scattering from a microscopic nucleon is experimentally infeasible, GPDs provide the only known phenomenological access to these quantities through Eqs.~\eqref{eq:2_Hqpoly} and \eqref{eq:2_Eqpoly}, via measurements of hard exclusive processes, most notably DVCS~\cite{Burkert:2018bqq} and exclusive heavy-quarkonium production~\cite{Guo:2021ibg}.

\subsection{Spin structure of the nucleon, twist-three GPDs}
\label{subsec:2_spin_t3GPD}
The significance of GPDs to the knowledge of the nucleon spin structures was first addressed in~\textcite{Ji:1996ek}, which showed that the total angular momentum of quarks and gluons are related to the gauge-invariant EMT form factors that can be accessed via GPD sum rules,
\begin{equation}
    J_{q/g} = \frac{1}{2} \left[A_{q/g}(0) + B_{q/g}(0) \right]\ ,
\end{equation}
where and hereafter we take $\hbar=1$.
Correspondingly, the angular momentum operator in QCD has the following expression,
\begin{align}
    \begin{split}
    \boldsymbol{J}_{\rm QCD}= \int \text{d}^3 \boldsymbol{r} \Big[ &\psi^\dagger_q \frac{\boldsymbol{\Sigma} }{2}\psi_q +
\psi^\dagger_q (\boldsymbol{r} \times i\boldsymbol{D}) \psi_q  \\&+  \boldsymbol{r}\times(\boldsymbol{E}_a\times\boldsymbol{B}_a)\Big]\,.
\end{split}
\label{eq:sec2_amgi}
\end{align}
where the first line corresponds to the quark angular momentum operator $J_q$ (summed over flavors) that can be further decomposed into the quark spin $\Delta\Sigma$ and the kinetic OAM $L_{q}$. Then the three-term sum rule for the nucleon helicity can be written as
\begin{equation}
\frac{1}{2}\Delta \Sigma  + L^z_q + J_g = \frac{1}{2}\,, \label{eq:sec2_jisumrule}
\end{equation}  
with $J_q=\frac{1}{2}\Delta \Sigma  + L^z_q $. Because all three terms are defined through local matrix elements of gauge-invariant operators, this sum rule is independent of the choice of frame and gauge. Note, however, that the helicity does not grow with momentum and 
the kinetic quark OAM $L_q^z$ involves a covariant derivative, the individual terms in the sum rule do not have a simple partonic interpretation.

On the other hand, \textcite{Jaffe:1989jz} considered the following decomposition,
\begin{align}
\begin{split}
    \boldsymbol{J}_{\rm QCD} = \int \text{d}^3 \boldsymbol{r}\Big[&\psi^\dagger_q\frac{\boldsymbol{\Sigma}}{2} \psi_q
+ \psi^\dagger_q (\boldsymbol{r} \times i\boldsymbol{\partial})\psi_q  \\
&+ \boldsymbol{E}_a\times \boldsymbol{A}_a + E^i_a(\boldsymbol{r}\times \boldsymbol{\partial})A_{ia}\Big] \ ,
\end{split}
\label{eq:sec2_amcanonical}
\end{align}
and derive the four-term sum rule for a longitudinally polarized nucleon in the infinite-momentum frame
\begin{equation}
\frac{1}{2} \Delta \Sigma+ \ell_q+ \Delta G  + \ell_g= \frac{1}{2} \ , \label{eq:sec2_jaffe-manohar}
\end{equation}  
where $\Delta G$ represents gluon helicity, and $\ell_q$ and $\ell_g$ denote the canonical OAM of quarks and gluons, respectively.

It is worth emphasizing that the Jaffe--Manohar sum rule is distinct from the Ji sum rule: $J_q \neq \frac{1}{2}\Delta\Sigma+\ell_q$ and $J_g \neq \Delta G+\ell_g$. The Jaffe--Manohar sum rule is most transparent in the infinite-momentum frame and in light-cone gauge, where each term admits a  partonic interpretation. The physical meaning of its manifest gauge dependence has nevertheless been the subject of extensive discussion and controversy~\cite{Hoodbhoy:1998bt,Hoodbhoy:1999dr,Chen:2008ag,Ji:2012gc,Hatta:2011zs,Wakamatsu:2013voa,Leader:2013jra,Hatta:2011ku,Hatta:2012cs}. A gauge-invariant extension can be constructed by specifying the Wilson-line path in the relevant partonic correlators~\cite{Hatta:2011ku,Hatta:2012cs}. This was subsequently connected to the Weizs\"acker--Williams description of the gauge fields generated by relativistic particles~\cite{Ji:2020ena}. In this framework, the quark and gluon helicity contributions are given by moments of the corresponding helicity PDFs and can be constrained through processes such as polarized DIS, whereas the canonical OAM contributions are related to moments of twist-three GPDs.

To make this more explicit, we introduce twist-three GPDs through the notations of~\textcite{Meissner:2009ww}:
\begin{eqnarray}
F_{q}^{[\gamma^j]}
 &=&\frac{M}{\bar P^+}\,
 \bar U(P')\Bigg[
 i\sigma^{+j}H_{2T}^q+\frac{2\bar P^{[+}\Delta^{j]}}{M^2}\widetilde H_{2T}^q \nonumber
 \\ \label{eq:t3GPD} &&\quad+\frac{\gamma^{[+}\Delta^{j]}}{M}E_{2T}^q
+\frac{2\gamma^{[+}\bar P^{j]}}{M}
 \widetilde E_{2T}^q\Bigg]U(P)\, ,\\
     F_{q}^{[\gamma^j\gamma^5]}
 &=&\frac{-i\epsilon_T^{ij}M}{\bar P^+}\,
 \bar U(P')\Bigg[
 i\sigma^{+j}H'{}_{\!\!2T}^q+\frac{2\bar P^{[+}\Delta^{j]}}{M^2}\widetilde{H}'{}^q_{\!\!2T} \nonumber
 \\&&\quad+\frac{\gamma^{[+}\Delta^{j]}}{M}E'{}^q_{\!\!2T}
+\frac{2\gamma^{[+}\bar P^{j]}}{M}
 \widetilde{E}'{}^{q}_{\!\!2T}\Bigg]U(P)\ .
\end{eqnarray}
The eight scalar functions form a complete basis of chiral-even GPDs at twist three. Hereafter, we focus on the OAM in the quark sector, while a consistent treatment applies to the gluon sector as well.

One specific combination of particular relevance to longitudinal OAM is ~\cite{Penttinen:2000dg,Kiptily:2002nx,Courtoy:2013oaa,Rajan:2017cpx}
\begin{equation}
G_L^q(x,\xi,t)\equiv \widetilde E_{2T}^q(x,\xi,t)  +H_q(x,\xi,t)+E_q(x,\xi,t)\ ,
\end{equation}
whose forward second moment obeys
\begin{equation}
L_q^{\mathrm{Ji}}
=\int_{-1}^{1}\text{d}x \,x\,G_L^q(x,0,0) \ .
\end{equation}
This shows that kinetic OAM can be represented through the moment of twist-three GPDs. The $xG_L^q(x,0,0)$ should not, however, be interpreted as the partonic OAM density, where additional $x$-dependent contributions from equation-of-motion, total-derivative, and genuine quark--gluon correlation terms whose relevant moments vanish could enter as well~\cite{Hatta:2012cs,Ji:2012sj,Leader:2013jra, Rajan:2017cpx}. 

To look closer into the partonic OAM density, the three-parton correlators that emerge at twist three become relevant. Schematically, they have the forms
\begin{align}
\mathcal O_{q,\,D}^\beta(\lambda,\mu)
&=
\bar\psi_q\left(-\frac{\lambda n}{2}\right)
\gamma^+\,
i\overleftrightarrow D_{\perp}^{\beta}(\mu n)\,
\psi_q\left(\frac{\lambda n}{2}\right)\, ,
\\
\mathcal O_{q,\,F}^\beta(\lambda,\mu)
&=
\bar\psi_q\left(-\frac{\lambda n}{2}\right)
\gamma^+\,
gF^{+\beta}(\mu n)\,
\psi_q\left(\frac{\lambda n}{2}\right)\, ,
\end{align}
which define the $D$- and $F$-type GPDs, respectively~\cite{Hatta:2011ku,Hatta:2012cs,Ji:2012sj}. %
The $D$-type correlator contains a covariant derivative and is therefore naturally related to the kinetic OAM appearing in the Ji decomposition. The $F$-type correlator contains an explicit gluon field strength and describes genuine quark--gluon correlations, which may be interpreted as the color force or torque acting on the active quark~\cite{Burkardt:2012sd}. The two correlators represent complementary descriptions of twist-three dynamics that are crucial to obtain the partonic OAM~\cite{Hatta:2011ku,Hatta:2012cs,Ji:2012sj,Guo:2021aik,Lorce:2011kd,Burkardt:2012sd,Engelhardt:2017miy,Engelhardt:2020qtg}. We also refer the reader to~\textcite{Ji:2020ena} for further details.

Experimentally, twist-three GPDs enter the scattering amplitudes with $1/Q$ suppression, particularly through longitudinal-to-transverse photon-helicity amplitudes and the associated subleading azimuthal harmonics in DVCS~\cite{Belitsky:2000vx,Anikin:2000em,Radyushkin:2000ap,Kivel:2000fg,Belitsky:2001ns,Aslan:2018zzk, Guo:2022cgq}. Similar to the twist-two case, physical observables generally constrain combinations of several vector and axial-vector twist-three GPDs, and their determination will therefore require a comprehensive set of inputs as well. This will be discussed in Sec.~\ref{sec:3_kpc} in more detail.

The situation is qualitatively different for transversely polarized nucleons, since transverse angular momentum transforms very differently under longitudinal boosts. Moreover, transverse polarization breaks azimuthal symmetry, so the transverse angular momentum receives a contribution from the center-of-mass motion; see, for example,~\textcite{Ji:2020hii}. The key observation is that transverse angular momentum is enhanced under a longitudinal boost and thus behaves as a leading-twist quantity. By identifying and subtracting the center-of-mass contribution, one can define the intrinsic transverse angular momentum of the nucleon, whose expectation value is $\langle P,S^x|J^x|P,S^x\rangle = \gamma /2$ with $\gamma$ the nucleon Lorentz boost factor. This intrinsic quantity then provides a basis for defining the partonic transverse angular-momentum density in the infinite-momentum frame. In this limit, the intrinsic spin contribution is suppressed by the large boost~\cite{Ji:2020hii}. Consequently, at leading twist, the partonic transverse angular-momentum density coincides with the corresponding OAM density and can be expressed entirely in terms of leading-twist GPDs~\cite{Hoodbhoy:1998yb,Ji:2012sj,Ji:2012vj}:
\begin{align}
    J_{q}^\perp(x) &= \frac{x}{2}(H_{q}(x)+E_{q}(x))\, , \\
    J_{g}^\perp(x) &= \frac{x}{2}(H_{g}(x)+E_{g}(x))\, ,
    \label{Eq:angden}
\end{align}
with the overall boost factor $\gamma$ removed so that the transverse spin sum rule is normalized as $J_q+J_g=1/2$.

Furthermore, a careful treatment of the subleading-twist component of the intrinsic partonic transverse angular momentum recovers the transverse quark and gluon spin contributions, leading to a four-term transverse spin sum rule analogous to the Jaffe--Manohar helicity decomposition~\cite{Guo:2021aik}:
\begin{equation}
\frac{1}{2}\Delta\Sigma_T(\mu)
+\Delta G_T(\mu)
+\ell_q^T(\mu)
+\ell_g^T(\mu)
=\frac{1}{2}\, .
\end{equation}
Each term is the transverse counterpart of that in the Jaffe--Manohar sum rule and admits a partonic interpretation. In principle, these contributions can be constrained through moments of twist-three PDFs and GPDs, although their definitions and phenomenological extraction involve subtleties analogous to those encountered in the longitudinal case discussed above.

\subsection{Mass and scalar distributions, confinement forces}
\label{subsec:2_mass_scalar_force}

The energy density operator is given by the $T^{00}$ component of QCD EMT, whose integral generates the QCD Hamiltonian: $H=\int \text{d}^3x\, T^{00}$.  This allows one to define the
energy/mass form factor $G_m(t) \propto \langle P'| T^{00}(0) |P\rangle $ with proper normalization such that~\cite{Ji:2021mtz},
\begin{equation}
\label{def:Gmt}
    G_m(t)
=
M\left[
A(t)+\frac{t}{4M^2}B(t)-\frac{t}{M^2}C(t)
\right] \  ,
\end{equation}
that characterizes the spatial distribution of the internal energy contributing to the nucleon mass.
It can be further decomposed 
into quark and gluon contributions through the corresponding 
form factors, 
$G_m=G_m^q +G_m^g$. Individual quark and gluon components, however, also contain the nonconserved form factors
$\bar C_{q,g}$, which cancel only in the sum, while the separation of
the trace anomaly into quark and gluon contributions depends on the
adopted operator prescription~\cite{Yang:2026wzi}.

On the other hand, the trace of the EMT defines a scalar field $T^\mu_{~ \mu}$.  Although the corresponding scalar form factor $G_s$ is a twist-four quantity,  
and difficult to measure in high-energy experiments directly, 
the total scalar form factor is related to the twist-two form factors~\cite{Ji:2021mtz} 
\begin{equation}
G_s(t)
=
M\left[
A(t)+\frac{t}{4M^2}B(t)-\frac{3t}{M^2}C(t)
\right],
\end{equation}
due to the EMT conservation. In the chiral limit, the scalar form factor $G_s(t)$ is associated with the scalar color field response generated by the presence of the quarks and may therefore be interpreted as the modification of the QCD vacuum condensate induced by the nucleon. This is analogous to the vacuum-energy difference represented by the bag constant in the MIT bag model~\cite{Chodos:1974je}. Nevertheless, further splitting it into quark and gluon contributions $G_s^{q,g}$ does not generally admit an equally straightforward physical interpretation.

Within the Breit-frame convention, the scalar and mass densities are defined by the three-dimensional Fourier transforms
\begin{equation}
\epsilon_{s,m}(\boldsymbol r) =
\int\frac{\mathrm{d}^3\boldsymbol\Delta}{(2\pi)^3} e^{-i\boldsymbol\Delta\cdot\boldsymbol r}
G_{s,m}(-\boldsymbol\Delta^2) \, ,
\end{equation}
satisfying $\int \text{d}^3\boldsymbol r\, \epsilon_{s,m}(\boldsymbol r)=G_{s,m}(0)=M$ with $A(0)=1$. The corresponding scalar and mass radii are
\begin{align}
\label{eq:scalar_radii}
\langle r_s^2\rangle
&=
6\left.\frac{\mathrm{d}A(t)}{\mathrm{d}t}\right|_{t=0}
-\frac{36}{2M^2}C(0),
\\\label{eq:mass_radii}
\langle r_m^2\rangle
&=
6\left.\frac{\mathrm{d}A(t)}{\mathrm{d}t}\right|_{t=0}
-\frac{12}{2M^2}C(0).
\end{align}
It is important to note that these definitions are subject to several caveats~\cite{Miller:2007uy,Jaffe:2020ebz}. Experimental and lattice-QCD studies of the nucleon mass and scalar distributions will be discussed in later sections.

 The physics of these momentum current distributions, however, has recently been carefully examined~\cite{Ji:2025qax}. More direct characterizations of the forces acting on quarks have instead been developed in terms of the divergence of momentum-current density~\cite{Ji:2025gsq,Ji:2026lyj}.

The $C/D$ form factor has attracted considerable attention in recent years because of the interpretation of the momentum current in terms of pressure and shear forces acting on quarks and gluons inside the nucleon~\cite{Polyakov:2002yz,Polyakov:2018zvc}, particularly following phenomenological extractions of the quark contribution from DVCS data~\cite{Burkert:2018bqq,Burkert:2023wzr}. However, because the color force is not short-ranged within the nucleon, interpreting its interior in terms of a fluid or continuous medium may be questionable~\cite{Ji:2025qax}. An alternative perspective is provided by the observation that the divergence of the quark momentum current defines the force density acting on quarks, thereby offering a new way to investigate the long-standing problem of quark confinement~\cite{Ji:2026lyj}. To establish a well-defined spatial description, Ref.~\cite{Ji:2026lyj} formulates the transverse force through the divergence of the EMT in the infinite-momentum frame, where a transverse position variable can be consistently defined in accordance with relativity and the uncertainty principle.

Since quarks exchange energy and momentum with the gluons, the quark contribution to the QCD EMT is not separately conserved. Its four-divergence thus relates to the color-Lorentz force density exerted on the quarks by the chromoelectric and chromomagnetic fields:
\begin{equation}
\mathcal{F}_q^{\,i}(x)
\equiv
\partial_\mu T_q^{\mu i}(x)
=
g\,\bar{\psi}(x)\gamma_\mu F^{\mu i}(x)\psi(x)\, .
\end{equation}
In the infinite-momentum frame, the force density becomes proportional to the transverse gradient of the two-dimensional spatial transform of $G_{s,q}$ since,
\begin{equation}
\left\langle P'\left|T_q^{\mu\nu}\right|P\right\rangle_{\mathrm{nc}}
=
\bar U(P')
\left[
-\frac{1}{4}g^{\mu\nu}\,G_{s,q}(t)
\right]
U(P),
\end{equation}
where the subscript ``nc'' denotes the nonconserved part. 
The two-dimensional Fourier transform is defined by
\begin{equation}\label{eq:Gsq_Fourier}
G_{s,q}(\bm r_\perp)
=
\int\frac{\text{d}^2\bm\Delta_\perp}{(2\pi)^2}
e^{-i\bm\Delta_\perp\cdot\bm r_\perp}
G_{s,q}(-\bm\Delta_\perp^2)\, .
\end{equation}
Therefore, the color-Lorentz force density reads,
\begin{eqnarray}
    \bm{\mathcal F}_q(\bm r_\perp)&=&\frac{1}{4}\bm\nabla_\perp G_{s,q}(\bm r_\perp)+\mathcal O\!\left(\frac{1}{(P^+)^3}\right)\ , \nonumber\\
    &=& \widehat{\bm r}_\perp
\frac{1}{4}
\frac{\text{d}G_{s,q}(r_\perp)}{\text{d}r_\perp}
+
\mathcal O\!\left(\frac{1}{(P^+)^3}\right)\, ,
\end{eqnarray}
where rotational symmetry in the transverse plane is assumed in the second line, with $r_\perp=|\boldsymbol{r}_\perp|$ and $\widehat {\bm r}_\perp$ the unit vector in the transverse plane.

The transverse proton and neutron Dirac densities are
\begin{equation}
F_1^N(\bm r_\perp)
=
\int\frac{d^2\bm\Delta_\perp}{(2\pi)^2}
e^{-i\bm\Delta_\perp\cdot\bm r_\perp}
F_1^N(-\bm\Delta_\perp^2),
\quad N=p,n.
\end{equation}
Dividing the color-Lorentz force in the transverse plane $\bm{\mathcal F}_q(\bm r_\perp)$ by the total transverse quark density, approximated using the proton and neutron Dirac form factors,  
\begin{equation}
\rho_q(\bm r_\perp)
\simeq
3\left[
F_1^p(\bm r_\perp)+F_1^n(\bm r_\perp)
\right],
\end{equation}
yields the transverse force per quark,
\begin{equation}
\bm F_q(\bm r_\perp)
\equiv
\frac{\bm{\mathcal F}_q(\bm r_\perp)}
{\rho_q(\bm r_\perp)}
=
\frac{(1/4)\bm\nabla_\perp G_{s,q}(\bm r_\perp)}
{3\left[
F_1^p(\bm r_\perp)+F_1^n(\bm r_\perp)
\right]}\, .
\end{equation}
The same force-density construction, when applied to nonrelativistic
systems such as the hydrogen atom, reproduces the well-known Coulomb force.

\subsection{Impact parameter space distributions}\label{sec:2_IPS}

Besides their connection to nucleon form factors, GPDs directly encode the spatial distributions of quarks and gluons within the nucleon. Analogous to how the three-dimensional Fourier transform of electromagnetic form factors provides a non-relativistic picture of charge and magnetization distributions, GPD yields the impact-parameter-space distribution of partons~\cite{Burkardt:2000za,Burkardt:2002hr}. Since partons predominantly move along the $z$-axis, a simple coordinate-space interpretation requires setting $\xi=0$, thereby eliminating longitudinal momentum transfer. A subsequent two-dimensional Fourier transform in the transverse $(x, y)$ plane then provides the transverse spatial distributions of quarks and gluons. This approach, commonly referred to as nucleon tomography, allows one to probe the nucleon’s spatial structure while minimizing relativistic effects from longitudinal recoil~\cite{Miller:2010nz, Miller:2018ybm, Lorce:2020onh}.

More explicitly, in a light-front frame with $\xi=0$, the longitudinal momentum transfer
vanishes and $t=-\boldsymbol{\Delta}_\perp^2$. The two-dimensional Fourier
transform of the unpolarized GPD $H_q$ then defines the
impact-parameter-dependent quark distribution,
\begin{equation}
q(x,\boldsymbol b_\perp)
=
\int \frac{\mathrm{d}^2\boldsymbol{\Delta}_\perp}{(2\pi)^2}
\,e^{-i\boldsymbol{\Delta}_\perp\cdot\boldsymbol b_\perp}
H_q\left(x,0,-\boldsymbol{\Delta}_\perp^2\right),
\end{equation}
where $\boldsymbol b_\perp$ is the transverse distance between the active
quark and the nucleon’s transverse center of mass. For fixed $x$,
$q(x,\boldsymbol b_\perp)$ has a genuine probability-density interpretation
and describes the transverse spatial distribution of quarks carrying
longitudinal momentum fraction $x$. 

\textcite{Burkardt:2005hp} further showed that, for a transversely polarized nucleon, the GPD $E_q$ induces a sideways distortion of the transverse spatial distribution, thereby linking nucleon polarization to partonic orbital motion. Later, the importance of removing center-of-mass contributions from the transverse angular momentum was emphasized by \textcite{Ji:2020hii}. After subtracting these extrinsic contributions, the intrinsic transverse-space quark distribution in a transversely polarized nucleon takes the form~\cite{Guo:2021aik}
\begin{equation}
\label{eq:sec2_transspacedist}
\begin{split}
    q_{\rm{Intrin}}^{\perp}(x,\boldsymbol b_\perp)&=\int \frac{\text{d}^2\boldsymbol \Delta_\perp}{(2\pi)^2} e^{-i \boldsymbol{\Delta}_\perp\cdot \boldsymbol b_\perp} \Big[H_q(x,-\boldsymbol \Delta_\perp^2)\\
    &\quad\qquad+\frac{i\Delta_y}{2M}\left(H_q+E_q\right)(x,-\boldsymbol \Delta_\perp^2)\Big].
\end{split}
\end{equation}
In particular, the contribution proportional to $H_q$ in the second term emerges only after removing the center-of-mass contribution. The impact-parameter distribution $q_{\rm{Intrin}}^{\perp}(x,\boldsymbol b_\perp)$ therefore characterizes the displacement of partons in the transverse space relative to the nucleon’s actual center of mass, which only depends on $H_q$.

Then the transverse angular-momentum density reads
\begin{equation}
    \label{eq:sec2_amdist}
    J^{\perp}_{q}(x,\boldsymbol b_\perp)=  \left(\vec{b}_\perp \times  x \vec{P}^z \right)^\perp q_{\rm{Intrin}}(x,\boldsymbol{b}_\perp) \,.
\end{equation}
Integrating $J^{\perp}_{q}(x,\boldsymbol b_\perp)$ over $\boldsymbol b_\perp$ reproduces the transverse angular-momentum density $\gamma J^{q}_\perp(x)$ carried by quarks with longitudinal momentum fraction $x$, with an analogous relation for gluons (see Eq. (\ref{Eq:angden})). A further integration over $x$ yields the total quark and gluon transverse angular momenta, $J_{q}^\perp$ and $J_{g}^\perp$, respectively. Here, the Lorentz factor $\gamma$ has been displayed explicitly, while $J_{q,g}^\perp(x)$ are defined with the boost factor removed so that their sum rule satisfies $J_q^\perp+J_g^\perp=1/2$. This framework shows that fundamental nucleon properties, including its spin and magnetic moment, can be consistently described in terms of intrinsic transverse-space distributions.

\subsection{RG scale evolution and conformal expansion}
\label{subsec:2_evolution_conf}
Experimentally, GPDs can be accessed through hard exclusive processes within the framework of QCD collinear factorization~\cite{Collins:2011zzd}, which generally states that the relevant amplitudes factorize into perturbatively calculable coefficient functions (CFs) and universal nonperturbative hadronic matrix elements with sufficiently large hard scales. In particular, collinear factorization has been well established for DVCS~\cite{Ji:1996nm,Collins:1998be} and DVMP~\cite{Radyushkin:1996ru, Collins:1996fb} to be further discussed in the following Sec.~\ref{sec:3_hard_proc}.

As a consequence of factorization, GPDs acquire a dependence on the factorization scale $\mu$, which separates short-distance dynamics contained in the CFs from long-distance hadronic structure. This scale dependence is governed by perturbative renormalization group (RG) evolution equations that generalize the familiar DGLAP evolution of parton distribution functions and ERBL evolution of meson distribution amplitudes. Since GPDs depend on both the momentum fraction $x$ and skewness $\xi$, their evolution encompasses the PDF regions, $|x|>\xi$, and the DA region, $|x|<\xi$. The corresponding kernels follow from the renormalization of nonlocal light-ray operators~\cite{Muller:1994ses}, or equivalently their towers of local twist-two operators, and preserve fundamental constraints such as polynomiality of Mellin moments.

The complete twist-two evolution kernels are known through two loops in both the nonsinglet and singlet sectors~\cite{Belitsky:1998vj,Belitsky:1998gc,Belitsky:1999hf,Braun:2014vba,Braun:2019qtp}. Conformal symmetry provides a particularly efficient organization of the GPD evolution in terms of conformal moments and is widely employed in phenomenological analyses~\cite{Kumericki:2007sa, Kumericki:2009uq, Guo:2025muf}. At three loops, the nonsinglet off-forward kernel is partially available~\cite{Braun:2017cih,Braun:2021tzi,Ji:2023eni}, whereas the singlet~\cite{Braun:2022byg} and quark transversity~\cite{Manashov:2024fcd} kernels are less known with the gluon transversity kernel remaining entirely undetermined, constituting an important missing ingredient for fully consistent NNLO GPD evolution.

Beyond the leading twist, the evolution is considerably more involved due to mixing among operators of equal twists. Results are presently available mainly at one-loop accuracy~\cite{Braun:2009vc,Braun:2009mi,Ji:2014eta} in which case the evolution equations are equivalent to a Heisenberg open spin chain model at the large $N_c$ limit~\cite{Braun:1998id,Derkachov:1999ze}. Further progress in higher-twist evolution will be important for a systematic control over power corrections at the moderate hard scales characteristic of current fixed-target experiments, as will be discussed in Sec. \ref{sec:3_kpc}.

One prominent feature of the twist-two GPDs is that their one-loop evolution kernels are diagonalized by the Gegenbauer polynomials, just as the leading-twist DAs, as a direct consequence of the (canonical) $SL(2,R)$ collinear conformal symmetry~\cite{Braun:2003rp}. It is therefore convenient to construct GPDs in the Gegenbauer basis. Taking $F$ as a GPD representative, we can write~\cite{Mueller:2005ed,Kirch:2005tt,Manashov:2005xp},
\begin{align}\label{eq:GPD_conformal_expansion}
    F(x,\xi,t,\mu) = \sum_{j=0}^\infty (-1)^j p_j(x,\xi){\cal F}_j(\xi,t,\mu)\, .
\end{align}
The ${\cal F}_j(\xi,t,\mu)$ are called partial wave amplitudes, also known as conformal moments of GPD, defined as GPD projections onto Gegenbauer polynomials:
\begin{equation}
\mathcal{F}_{j}(\xi,t) \equiv \xi^{j} \frac{\Gamma\!\left(3/2\right)\Gamma(j+1)}{2^j \Gamma\!\left(3/2+j\right)} \int_{-1}^{1} \mathrm{d}x \, C_{j}^{3/2}\!\left(\frac{x}{\xi}\right) F(x,\xi,t)\,,
\end{equation}
where quark GPDs are implied and factorization scale $\mu$ will be suppressed unless stated otherwise. The extra normalization factor ensures that they reduce to the Mellin moments of corresponding PDFs in the forward $\xi\to0$ limit. Analogous definitions hold for gluon GPDs as well, with Gegenbauer polynomials $C_{j}^{5/2}(z)$ of weight $5/2$, rather than the $3/2$~\cite{Mueller:2005ed}.

The function basis $p_j(x,\xi)$ reads,
\begin{align}
p_j(x,\xi) &= \xi^{-j-1}f_j\left(\frac x\xi\right)\, ,\\
f_j(z) &= \theta(1-|z|)\frac{2^j\Gamma(5/2+j)}{\Gamma(3/2)\Gamma(3+j)}(1-z^2)C_j^{(3/2)}(-z)\, .\notag
\end{align}
Nevertheless, the formal sum in Eq.~\eqref{eq:GPD_conformal_expansion} is generally divergent and must be defined through analytic resummation. This can be achieved using a Sommerfeld--Watson transformation, which, after analytic continuation, recasts the series as a Mellin--Barnes contour integral as,
\begin{align}
    F(x,\xi,t) = \frac i{2}\int^{c+i\infty}_{c-i\infty}\frac{\text{d}j\,p_j(x,\xi)}{\sin(\pi j)}{\cal F}_j(\xi,t)\, .
\end{align}
This establishes the theoretical foundations for the perturbative evolution and global analysis of GPDs based on conformal moments; see Sec.~\ref{subsec:5c} for details.

\subsection{GPDs in extreme regimes: small $x$ and large $\xi$}

At small Bjorken $x$, GPDs predominantly probe the sea-quark and gluon structure of the nucleon. This regime has been extensively explored through DVCS and exclusive vector-meson production at HERA. The rapid growth of the cross sections with the photon--proton center-of-mass energy $W$ reflects the rise of singlet-quark and gluon distributions, while their $t$ dependence constrains the transverse spatial distribution of small-$x$ partons~\cite{Kumericki:2009uq,Jones:2013pga,Flett:2019pux,Kowalski:2006hc}. A crucial feature in this regime is the rapidly increasing gluon density, which eventually enhances nonlinear gluon-recombination effects beyond linear DGLAP/BFKL evolution. These effects are described by BK/JIMWLK evolution and naturally organized within the color-glass-condensate (CGC) framework~\cite{Balitsky:1995ub,Kovchegov:1999yj,Jalilian-Marian:1996mkd,Gelis:2010nm}. Impact-parameter-dependent dipole and saturation approaches have provided successful descriptions of HERA DVCS and vector-meson data and have been extensively applied to high-energy exclusive production~\cite{Iancu:2003ge,Bartels:2002cj,Kowalski:2006hc,Rezaeian:2012ji,Mantysaari:2021ryb,Mantysaari:2022kdm}. 

More recently, increasing attention has focused on establishing the connection between this high-energy description and GPDs, including their small-$x$ evolution, finite-skewness structure, and representation in terms of Wilson-line operators~\cite{Hatta:2022bxn,Kovchegov:2025yyl}. Nevertheless, the simultaneous small-$x$ and small-$\xi$ regime remains theoretically subtle, since the underlying small-$x$ dynamics can depend nontrivially on the relative magnitude of $x$ and $\xi$, particularly between the $x\gtrsim\xi$ and $x\ll\xi$ regions. Such developments connect GPD-based tomography with complementary descriptions of hadron and nuclear structure at small $x$, including dipole and CGC approaches, with important applications to exclusive processes at the EIC and in ultraperipheral collisions~\cite{Hatta:2016dxp,Hatta:2017cte,Klein:2019qfb,Mantysaari:2020axf,Mantysaari:2022ffw}.

Another important kinematic regime that has attracted increasing interest is the large-skewness region. As $\xi\to1$, the DA-like region, $|x|<\xi$, becomes dominant, suggesting that the behavior of GPDs approaches that of the DA in this limit. Lately, exclusive near-threshold production processes, most notably heavy-quarkonium photoproduction and deeply virtual leptoproduction of vector mesons, have been extensively discussed for accessing large-skewness GPDs~\cite{Hatta:2018ina,Hatta:2019lxo,Guo:2021ibg,Guo:2023qgu,Guo:2025jiz,Hatta:2025vhs}. Near threshold, the momentum transfer is necessarily large and predominantly longitudinal, corresponding to large skewness. In this regime, the hard amplitudes can be organized in a DA-like expansion in moments of the GPDs. The lowest moments are directly related to the EMT form factors, thereby connecting measurements of near-threshold exclusive productions and the mass, angular-momentum, and mechanical structure of the nucleon.

Near-threshold $J/\psi$ photoproduction has emerged as the principal process for pursuing this connection. Measurements at Jefferson Lab by GlueX, $J/\psi$-007, and CLAS12 have substantially improved the available near-threshold cross-section data~\cite{GlueX:2019mkq,Duran:2022xag,GlueX:2023pev,007:2026dow,Chatagnon:2026qsv}, motivating phenomenological studies of gluon EMT form factors and the associated mass and mechanical distributions of the proton~\cite{Mamo:2019mka,Mamo:2021krl,Guo:2021ibg,Guo:2023pqw}. Related ideas have also been explored in near-threshold exclusive leptoproduction of the $\phi$ meson~\cite{Hatta:2025vhs}. Although the quantitative interpretation remains subject to perturbative, heavy-quark, relativistic, and possible competing production mechanisms~\cite{Du:2020bqj,Strakovsky:2019bev,JointPhysicsAnalysisCenter:2023qgg,Blask:2025jua}, this program has established large-skewness GPDs as an important frontier for accessing the gluonic EMT form factors. More details will be discussed in the Sec. \ref{sec:5:subsec_HeavyQuark}.

\section{Hard Exclusive Processes}

\label{sec:3_hard_proc}

The internal structure of the proton and neutron has traditionally been studied in two complementary ways: through electron diffraction (or elastic scattering)~\cite{Hofstadter:1956qs}, which maps the spatial charge distribution while leaving the nucleon intact, and through deep-inelastic scattering~\cite{Bloom:1969kc,Breidenbach:1969kd}, which probes the momentum distributions of its constituents as the nucleon breaks up. It was therefore particularly exciting when hard exclusive processes such as DVCS~\cite{Ji:1996ek,Ji:1996nm} were proposed to probe entirely new correlations between the spatial and momentum distributions of quarks and gluons, which are indispensable for understanding the angular momentum structure of the nucleon. DVCS is much more practical 
than an earlier proposal of two-virtual photon process in~\textcite{Muller:1994ses}, and its 
QCD factorization was quickly confirmed ~\cite{Radyushkin:1997ki,Ji:1998xh,Collins:1998be} that the final-state real photon does not introduce extra infrared singularities. Beyond DVCS, \textcite{Ji:1996ek} also pointed out that hard exclusive meson production can also probe GPDs, which was followed up
by calculations~\cite{Radyushkin:1996ru} and factorization proof~\cite{Collins:1996fb}. The two virtual photon case~\cite{Muller:1994ses}
has been studied decade late as Double DVCS or DDVCS~\cite{Pire:2011st,Mueller:2012sma}, and has a potentially-powerful 
GPD resolution capability although experimentally much more demanding. 

In reactions such as DVCS and DVMP, one has a diffractive scattering combined with highly energetic emissions in the center-of-mass frame, i.e., 
    a highly-virtual photon resolves individual quarks or gluons while the highly exclusive final state preserves information about the recoiling hadron. The QCD factorization framework separates the short-distance parton scattering process from the GPDs encoding the long-distance nonperturbative nucleon structure. As already emphasized in earlier sections, GPDs correlate parton momentum, transverse position, and spin, thereby enabling three-dimensional imaging of the hadron and giving access to its energy-momentum tensor, angular momentum, and internal mechanical properties.

\subsection{Deeply virtual Compton scattering}\label{sec:3_DVCS}

Regarded as the ``golden'' channel for experimental determination of GPDs, DVCS plays a pivotal role in the JLab 12~GeV, EIC and EicC experiment agenda. As the name suggests, it is a photon-hadron scattering process that reads,
\begin{align}
\gamma^*(q)+ N(P) \longrightarrow \gamma(q')+ N(P')\, ,
\label{DVCSprocess-BMP}
\end{align}
where the initial virtual photon $\gamma^*(q)$ is typically emitted from a high-energy electron $e\to e+\gamma^*$.

Since the DVCS process involves particles traveling in multiple directions, there are ambiguities in choosing the reference frame, particularly in defining the longitudinal plane required to construct the twist expansion. Each frame choice provides unique insights into the different facets of GPDs and the DVCS reaction. In the pioneering work of~\textcite{Ji:1996nm}, the longitudinal plane is spanned by the average nucleon momentum $\bar P$, hence treating the nucleon momenta in a symmetric setting,  and the initial photon $q$, with the two light-cone vectors in Eq.~\eqref{eq:lc_vectors} within the longitudinal plane subsequently defined. This frame offers a direct connection to the DIS frame and is also where the LP factorization theorem for DVCS was originally proved historically. 
Another notable frame is the symmetric Compton frame, where the average momenta of hadron $\bar P$ and photon $\bar q=(q+q')/2$ define the longitudinal plane. It also offers a direct, intuitive connection to the DIS frame, with the momentum transfer $\Delta$ entirely transverse as discussed in Sec.~\ref{subsec:2_GPD_def}. 
The Breit frame, also known as the ``brick-wall'' frame, is defined by the vanishing energy component of the momentum transfer, i.e., $\Delta^0=0$. In this frame, the target hadron recoils symmetrically, as if elastically bouncing off a hard brick wall.
The advantage of the Breit frame is that it allows us to acquire a clean non-relativistic physical interpretation for the hadron's mass, scalar, and electric form factor $G_{m}(t),G_s(t), G_e(t)$ as the Fourier transforms of the corresponding distribution in ${\bm r}$ inside the hadron, as discussed in Sec.~\ref{subsec:2_mass_scalar_force}. A dedicated discussion of all possible frame choices has been made in \textcite{Guo:2021gru}.

\begin{figure}[t]
    \centering
    \includegraphics[width=0.483\textwidth]{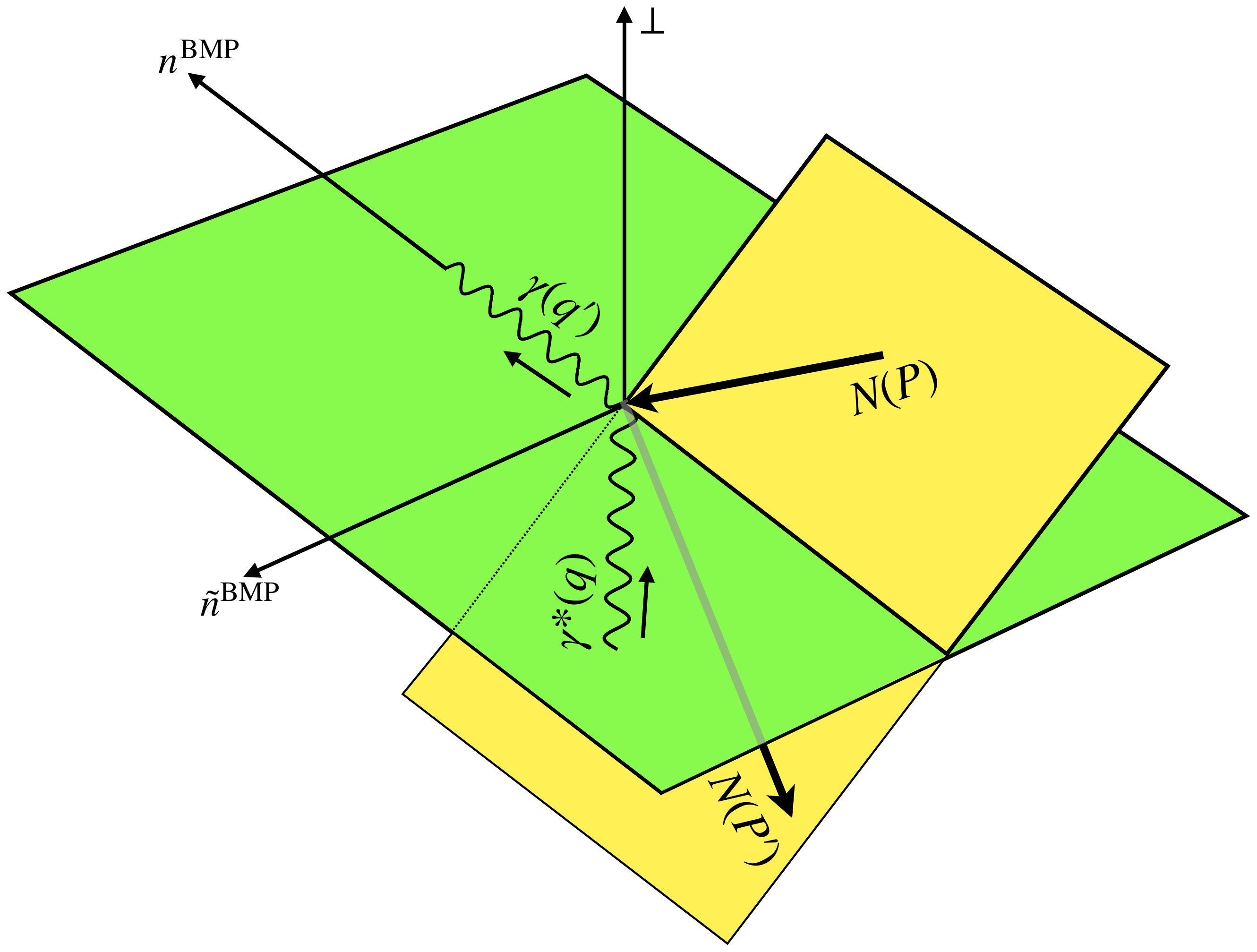}
    \caption{\raggedright Reference frame for the DVCS process in the BMP frame. The main characteristics of this frame are that the momentum transfer $\Delta$ is purely longitudinal, i.e., in the $(n^{\rm BMP},\tilde n^{\rm BMP})$ plane.}
    \label{fig:sec3:BMP}
\end{figure}

\begin{figure}[t]
    \centering
    \includegraphics[width=0.483\textwidth]{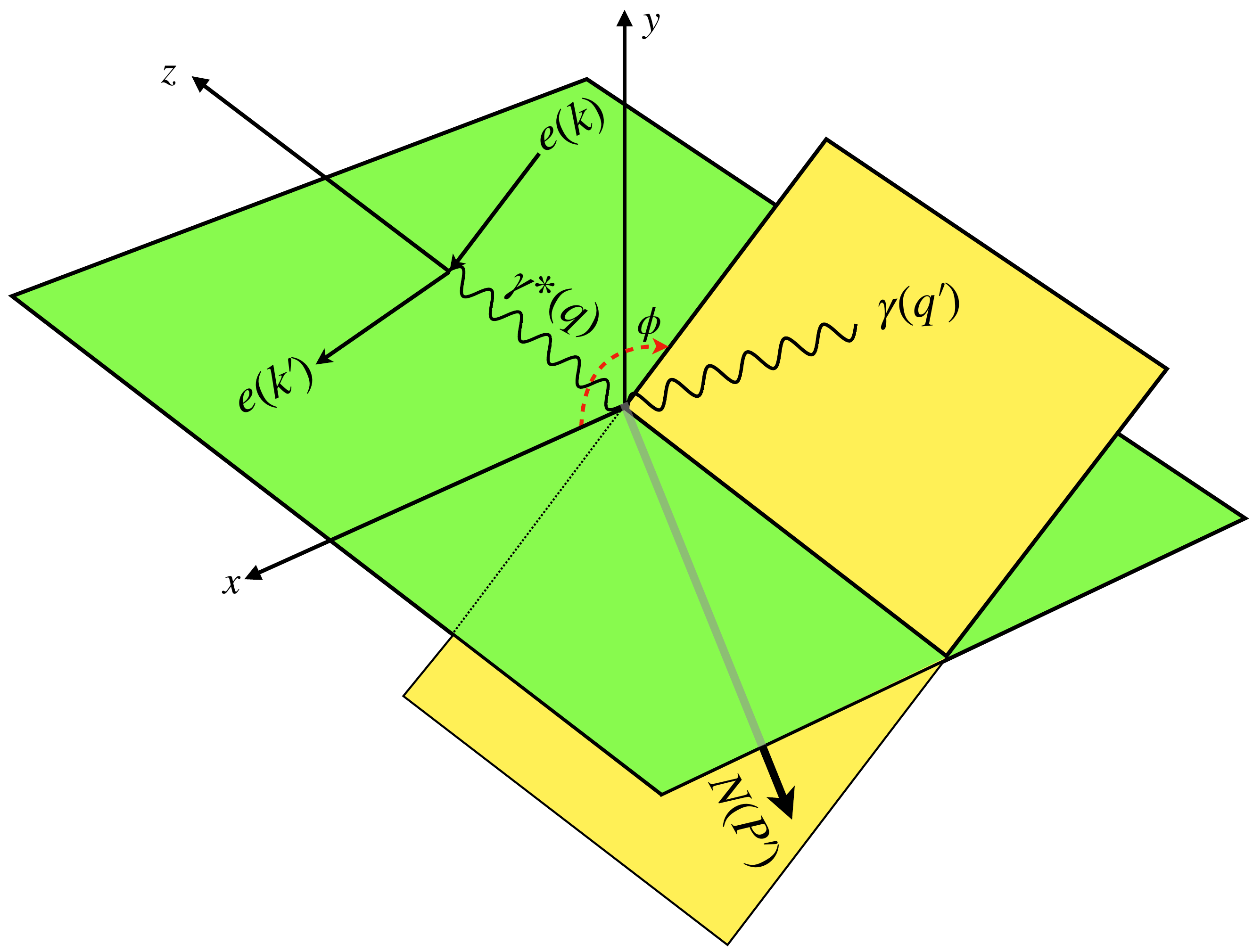}
    \caption{\raggedright Target frame, also known as the laboratory frame for lepton--nucleon scattering $e(k)N(P)\to e(k')N(P')\gamma(q')$, where the initial nucleon is at rest. The azimuthal angle $\phi$ is between the lepton plane and the hadron plane defined by the momenta of the outgoing nucleon ($P'$) and real photon ($ q'$) together. Experimental analyses often adopt the Trento convention, $\phi\to2\pi-\phi$~\cite{Bacchetta:2004jz}.}
    \label{fig:sec3:target}
\end{figure}

Two other special reference frames, known as the BMP frame (see Fig.~\ref{fig:sec3:BMP})~\cite{Braun:2012hq,Braun:2012bg} and the target/laboratory frame (see Fig.~\ref{fig:sec3:target}), prove to be particularly useful. While the former is better suited for deriving the higher-twist contributions, the latter is more convenient for experimental analysis. We adopt the BMP frame throughout this section, unless stated otherwise, for its convenience in perturbative calculations. It should be emphasized, however, that all frames are equivalent at LP approximation. 
The scattering amplitude in the BMP frame can be readily defined as,
\begin{align}
\label{Amunu-def}
 \mathcal{A}_{\mu\nu}(q,q',P) &=i\! \int\!\! d^4 x\, e^{-i(z_1q-z_2 q')\cdot x }\notag\\
&\quad\times\langle P'|T\{J_\mu(z_1x)J_\nu(z_2x)\}|P\rangle,
\end{align}
where $J_\mu(z_1x)$ and $J_\nu(z_2x)$ are the electromagnetic currents in Eq.~\eqref{eq:JF1F2} induced by quark fields, $z_1,z_2$ are real numbers such that without loss of generality 
$z_1-z_2=1$. The BMP frame~\cite{Braun:2012hq,Braun:2012bg} uses the photon momenta, $q$ and  $q'$, to define a 
longitudinal plane spanned by the two light-like vectors
\begin{align}
n_{\rm BMP}^\mu=q'^\mu\,, \qquad \tilde n_{\rm BMP}^\mu=-q^\mu+(1-\tau)\, q'^\mu\,,
\end{align}
where $\tau= t/(Q^2+t)$ with $Q^2=-q^2$ and $Q\equiv\sqrt{Q^2}$.  %
In this convention, the light-cone vector $n_{\rm BMP}$ enters the GPD definitions, hence playing the role of $n$ in Eqs.~\eqref{def:GPD} and~\eqref{def:GPD2}. The resulting differences in using different light-cone vectors, or more generally different reference frames, in defining the leading-twist GPDs and DVCS amplitude can be systematically studied, which we shall discuss in detail in Sec.~\ref{sec:3_kpc}. 
In the BMP frame, momentum transfer to the target 
$$\Delta^\mu = P'^\mu-P^\mu= q^\mu-q'^\mu\,, %
$$
is purely longitudinal and both the initial and final state nucleons have nonzero transverse momentum components,
\begin{align}\label{Pperp}
\bar P^\mu&=\frac12\left(P+P^\prime\right)^\mu=\frac{1}{2\xi}\left(\tilde n^\mu_{\rm BMP}-\tau n^\mu_{\rm BMP}\right) + \bar P_{\perp}^{\mu}\,, \notag\\
\bar P^+ &= \frac12n_{\rm BMP}\cdot(P+P')\, .
\end{align}
The skewness parameter $\xi$ is defined using the unique lightlike vector $q'$ as
\begin{align}
\xi \equiv \xi_{\rm BMP} &= -\frac{\Delta\cdot q'}{2\bar P\cdot q'}
= \frac{x_{ B}(1+t/Q^2)}{2-x_{ B}(1-t/Q^2)}\, ,
\label{xiBMP}
\end{align}
where the Bjorken-$x$ is defined as $x_{ B}=Q^2/(2P\cdot q)$ 
and $|\bar P_\perp|^2$ can  be written in terms of kinematic invariants as
\begin{align}
 |\bar P_\perp |^2 = \frac{1-\xi^2}{4\xi^2} (t_{\rm min}-t)\,, && t_{\rm min} = -\frac{4M^2\xi^2}{1-\xi^2} \,.
\end{align}
The main feature of the BMP frame/convention is that the photon polarization vectors take on simple forms that %
satisfy the normalization condition $\varepsilon^+\cdot\varepsilon^{-} = -1$\,,
$(\varepsilon^0)^2 = +1$, see~\textcite{Braun:2012hq} for their explicit expressions. 
The vectors $\varepsilon^\pm_\mu$ form  
a basis %
in the transverse plane perpendicular to both $q$ and $q'$, whereas $\varepsilon^0_\mu$
is a unit vector in the longitudinal plane %
orthogonal to the initial  
photon momentum $q$. The full DVCS amplitude can consequently be parameterized as,
\begin{align}
\!\mathcal{A}_{\mu\nu} ={} &\varepsilon^+_{\mu} \varepsilon^-_{\nu} \mathcal{A}^{++}
+\varepsilon^-_{\mu} \varepsilon^+_{\nu} \mathcal{A}^{--}
+\varepsilon^0_{\mu} \varepsilon^-_{\nu} \mathcal{A}^{0+}\notag\\
&+\varepsilon^0_{\mu} \varepsilon^+_{\nu} \mathcal{A}^{0-}\!
+\!\varepsilon^+_{\mu} \varepsilon^+_{\nu} \mathcal{A}^{+-}\!
+\!\varepsilon^-_{\mu} \varepsilon^-_{\nu} \mathcal{A}^{-+} \,,
\label{Amunu}
\end{align}
where we have neglected a term proportional to $q'_\nu$ because it doesn't contribute to any physical observables due to the Ward identity. Parity symmetry further imposes constraints on the helicity amplitudes by reversing both photon and nucleon helicities, which we have made implicit because they are irrelevant for perturbative QCD calculations. We can choose four amplitudes 
\begin{align}
  {\cal A}^{\pm\pm}\, ,\quad {\cal A}^{0+}\, ,\quad {\cal A}^{+-}\, ,
\end{align}
as representatives to illustrate perturbative results because amplitudes ${\cal A}^{0+}$ (${\cal A}^{+-}$) and ${\cal A}^{0-}$ (${\cal A}^{-+}$) share exactly the same perturbative Wilson CFs
\footnote{In case the target is spin zero, the  identities ${\cal A}^{++}={\cal A}^{--}\, ,{\cal A}^{0+} = {\cal A}^{0-}$ hold decreasing the number of independent scalar helicity functions to three.}. 
These scalar helicity amplitudes can be organized as power series in $\sqrt{-t}/Q\, , \Lambda_{\rm QCD}/Q$ and $M/Q$ (all are collectively written as $1/Q$ below), which %
read schematically,
\begin{align}
{\cal A}^{\pm\pm} &= {\cal A}_{0}^{\pm\pm} + {\cal O}(1/Q)\, ,%
\qquad
{\cal A}^{0+} = \frac{{\cal A}_1^{0+}}{Q} + {\cal O}(1/Q^2)\, , %
\notag\\
{\cal A}^{+-} &= %
{\cal A}_{0}^{+-} + {\cal O} (1/Q) %
\, .\label{eq:A_Qexpansion}
\end{align}
The above formulas indicate that at the LP approximation ${\cal O}(1/Q^0)$, only the helicity preserving ${\cal A}^{\pm\pm}$ and the double-helicity flip amplitude ${\cal A}^{+-}$ survive, allowing us to access the various twist-two GPDs experimentally. 
The helicity-conserving nucleon DVCS amplitudes ${\cal A}_0^{\pm\pm}$ can be reorganized into vector (${\mathbb V}$) and axialvector (${\mathbb A}$) channels as, 
\begin{align}\label{eq:VA_decomposition}
    {\mathbb V} &= \frac12\left({\cal A}^{++} + {\cal A}^{--}\right)\, ,\quad
    {\mathbb A} = \frac12\left({\cal A}^{++} - {\cal A}^{--}\right)\, .
    \end{align}

At the LP approximation, the DVCS amplitude can be computed in the collinear factorization framework, which is well established to all orders in perturbation theory~\cite{Radyushkin:1997ki,Collins:1998be,Ji:1998xh,Bauer:2002nz}, as a convolution between hard CFs calculable using perturbative QCD and all the twist-two GPDs defined in Eqs.~\eqref{def:HEq},~\eqref{def:HEq-tilde}, and \eqref{def:HEgT}. %
Note that the quark transversity GPDs in Eq.~\eqref{def:HEqT} are inaccessible at the leading power for spin-1/2 targets by selection rules in the hard scattering. This provides a unique opportunity to probe the transversity gluon distribution (also known as the maximal-helicity/double-flip sector) inside nucleons without quark ``contamination''. It should also be emphasized that the same CFs apply to computing DVCS cross-sections for both spin-1/2 and spin-0 hadron targets (e.g., pions), after appropriate adjustments to the fundamental properties of the targets (e.g., parity and spin).

The DVCS tensor amplitude in Eq.~\eqref{Amunu-def} cannot be accessed directly in the experiment. In the leptoproduction of a real photon off a hadron, $e(k)+N(P)\to e(k')+N(P')+\gamma(q')$, the outgoing photon can be radiated either from the hadron, as described by the DVCS amplitude $\mathcal{T}_{\rm DVCS}\propto \epsilon_\mu^{\lambda_1}\epsilon_\nu^{*\lambda_2}{\cal A}^{\mu\nu}$ carrying the GPD information, or by bremsstrahlung off the incoming or outgoing lepton, known as the Bethe--Heitler (BH) amplitude $\mathcal{T}_{\rm BH}$. %
Sharing the same final state, the two mechanisms add coherently, so that the total scattering amplitude-absolute-squared $|{\cal T}|^2$, which is proportional to the differential cross-section, splits into three pieces,
\begin{align}
\left|\mathcal{T}\right|^2=\left|\mathcal{T}_{\rm BH}\right|^2+\left|\mathcal{T}_{\rm DVCS}\right|^2+\mathcal{I}\,,\quad \mathcal{I}=\mathcal{T}_{\rm DVCS}\mathcal{T}_{\rm BH}^\ast+\mathrm{c.c.}\, .
\label{eq:BHdecomp}
\end{align}
The BH amplitude carries no imaginary part at the Born approximation in quantum electrodynamics; its only nonperturbative input is the matrix element of the EM current between off-forward hadronic states, as defined in Eq.~\eqref{eq:JF1F2} for a nucleon target. %
With $\mathcal{T}_{\rm BH}$ fixed at LO, the interference term ${\cal I}$ in Eq.~\eqref{eq:BHdecomp}, which in contrast to the quadratic $|\mathcal{T}_{\rm DVCS}|^2$ is linear in the DVCS amplitude, turns the BH process into a reference function against which both the magnitude and the phase of the DVCS amplitude %
can be read off, enabling a ``holographic'' determination that the DVCS-squared term alone cannot provide. This interferometry is, however, double-edged: over much of the JLab and EIC phase space, the BH term dominates the cross section, so any high-quality extraction of GPDs requires an accurate subtraction of the BH contribution and therefore precise knowledge of Dirac ($F_1$) and Pauli ($F_2$) nucleon form factors. The same structure also persists in the timelike-photon and lepton-pair reactions %
(see Sec.~\ref{sec:3_DDVCS}) %
whose signals likewise ride on the BH background%
~\cite{Berger:2001xd,Belitsky:2002tf,Guidal:2002kt}. The meson-production channels (see Sec.~\ref{sec:3_DVMP}) are, however, free of the BH-type ``contamination''. 
We note that the real and imaginary parts of the DVCS amplitude induced by twist-two GPDs can be related via dispersion relations~\cite{Diehl:2007jb}, whose consistency with a direct evaluation of the amplitude is a direct consequence of the polynomiality condition of the Mellin moments of the GPDs.

To confront experimental data, it is usually more convenient to work in the target frame. In particular,~\textcite{Belitsky:2010jw, Belitsky:2012ch} constructed a kinematical-singularity-free parameterization of the DVCS amplitude in the target frame, allowing DVCS data to be described for arbitrary initial-photon virtuality. 
In practice, $\mathcal{I}$ is disentangled in the target/laboratory frame through its dependence on the azimuthal angle $\phi$ between the leptonic and hadronic planes as shown in Fig.~\ref{fig:sec3:target} and through the opposite behavior of the two amplitudes under a reversal of the lepton charge, $\mathcal{T}_{\rm DVCS}\to-\mathcal{T}_{\rm DVCS}$ while $\mathcal{T}_{\rm BH}$ remains the same, so that a combined analysis of beam-charge, beam-spin, and target-spin asymmetries reveals, in principle, the full DVCS amplitude. %
Since the only source of the azimuthal angular dependence is the set of scalar products among the external momenta, each of the three terms in~\eqref{eq:BHdecomp} can be written as a finite Fourier series in $\phi$~\cite{Belitsky:2001ns} after factoring out contributions of the leptonic propagators. For an electron scattering off a nucleon $e(k)N(P)\to e(k')N(P')\gamma(q')$ in the target frame, the interference, which is our primary source of information for the full DVCS amplitude, reads\footnote{The formulas take on similar forms for a polarized target.}
\begin{align}\label{eq:DVCSharmonics}
    \mathcal{I}&=\frac{\pm e^6}{x_{\rm B}\,y^3\,t\,\mathcal{P}_1(\phi)\mathcal{P}_2(\phi)}\notag\\
    &\qquad\quad\times\left[c_0^{\mathcal I}+\sum_{n=1}^{3}\left(c_n^{\mathcal I}\cos n\phi+s_n^{\mathcal I}\sin n\phi\right)\right],
\end{align}
where $y =(P\cdot q)/(P\cdot k)$ is the fractional lepton energy loss. The overall sign of ${\cal I}$ follows the lepton charge, and $\mathcal{P}_{1}(\phi) = (k-q')^2/Q^2, {\cal P}_2(\phi)=(k'+q')^2/Q^2$ are proportional to the denominators of the two lepton propagators in the BH process. The $|\mathcal{T}_{\rm BH}|^2$ and $|\mathcal{T}_{\rm DVCS}|^2$ terms are similar finite sums, truncating at $n=2$ (for details, see e.g.,~\textcite{Belitsky:2000vk,Belitsky:2000gz,Belitsky:2001ns,Belitsky:2001yp,Diehl:1997bu,Kivel:2001rw}). The harmonic coefficients are well organized by twist: the leading interference harmonics $c_1^{\mathcal I},s_1^{\mathcal I}$ start at the LP, encoding the leading-twist coefficient ${\cal A}_0^{\pm\pm}$ in Eq.~\eqref{eq:A_Qexpansion}; $c_2^{\mathcal I},s_2^{\mathcal I}$ start at next-to-leading power (NLP), i.e., ${\cal O}(1/Q)$ induced partially by twist-three GPDs; and the highest harmonics $c_3^{\mathcal I},s_3^{\mathcal I}$ again receive LP contributions from ${\cal A}^{+-}_0$ as shown in Eq.~\eqref{eq:A_Qexpansion} where the twist-two gluon transversity GPDs of a spin-1/2 target provide the nonperturbative input (see Sec.~\ref{sect:3_singlet} for details). 
This twist separation is what makes DVCS a quantitative probe of GPDs. With sufficiently complete measurements across different beam/target polarization configurations, the $\cos n\phi$ and $\sin n\phi$ harmonics enable a full reconstruction of the complex DVCS amplitude. Combined with additional inputs from complementary processes and constraints from polynomiality, evolution, forward limits, form factors, and lattice results, ultimately, a more realistic construction of the twist-two GPDs can be achieved via the collinear factorization theorem, as discussed in the following subsections.

\subsubsection{The flavor-nonsinglet channel}

Among the leading-twist contributions to DVCS, the flavor-nonsinglet sector is the simplest as it does not involve quark--gluon mixing under evolution. While the one-loop vector and axialvector CFs have been known for nearly three decades~\cite{Ji:1997nk,Mankiewicz:1997bk,Belitsky:1997rh,Ji:1998xh}, the complete two-loop CFs have become available only recently~\cite{Braun:2020yib}, confirming earlier estimates of higher-order contributions beyond one-loop based on conformal partial wave analysis~\cite{Mueller:2005nz,Kumericki:2007sa}.

Following the power expansion~\eqref{eq:A_Qexpansion} 
and flavor decomposition, we can write the LP contributions as,
    \begin{align}\label{eq:A_VA}
    {\mathbb V}_0 = {\mathbb V}_{0,\rm NS} + {\mathbb V}_{0,\rm S}\, ,\qquad
    {\mathbb A}_0 = {\mathbb A}_{0,\rm NS} + {\mathbb A}_{0,\rm S}\, ,
\end{align}
where NS and S stand for flavor-nonsinglet and singlet channel, respectively. 
 For ${\mathbb V}_{0,\rm NS}$ and ${\mathbb A}_{0,\rm NS}$, which are the topics of this subsection, the LP factorization theorem dictates that
\begin{align}\label{eq:fac}
    &{\mathbb V}_{0,\rm NS}(\xi,t,Q^2)  \\
    &\quad= \sum_{q}e_q^2\int^1_{-1}\frac{dx}{\xi}\, C_{\rm NS}(x/\xi,Q^2/\mu_F^2)F_{q,\rm NS}(x,\xi,t,\mu_F)\, ,\notag
\end{align}
where $\mu_F$ is the factorization scale, $e_q$ is the electric charge of the quark flavor $q$. 
$C_{\rm NS}$ is the perturbatively calculable nonsinglet vector CF. 
The integration variable $x$ denotes the momentum fraction carried by the parton participating in the hard scattering. 
$F_{q,\rm NS}(x,\xi,t,\mu)$ is the nonsinglet-projected $F_q$ defined explicitly in Eq.~\eqref{def:HEq} and carries a nonzero flavor quantum number, hence forbidding mixing with gluon GPDs under evolution to all orders. %
For example, the flavor-nonsinglet matrix elements
 $F_{q,\rm NS}$ in the three-flavor QCD with massless $q=u,d,s$ quarks read, 
\begin{align}\label{eq:NS-proj}
    F_{q,\rm NS} &=F_q-F_{q,\rm S}\, ,\quad F_{q,\rm S} = \frac13(F_u+F_d+F_s)\, .
\end{align}
The factorization formula for the nonsinglet axialvector contribution takes exactly the same form as in Eq.~\eqref{eq:fac} with the simple replacements ${\mathbb V}\mapsto{\mathbb A}$, $C\mapsto \widetilde C$, and $F\mapsto \widetilde F$, so that the axialvector GPDs generate the axialvector contribution. %

It is convenient to introduce the so-called Compton form factors (CFFs) which are convolutions of the CFs with the corresponding GPDs. For the flavor nonsinglet vector case, Eqs.~\eqref{def:HEq} and~\eqref{eq:fac} imply the CFFs take the form, 
\begin{align}
&\mathcal H_{\rm NS}(\xi,t,Q^2) \label{CFF-H}
\\
&\quad=\sum_q e_q^2\int_{-1}^{1}\frac{dx}{\xi}C_{\rm NS}(x/\xi,Q^2/\mu_F^2)H_{q,\rm NS}(x,\xi,t,\mu_F)\,,%
\notag
\end{align}
and similarly for ${\cal E}_{\rm NS}$ with $H\mapsto E$.  The corresponding expressions for the axialvector CFFs follow intuitively with $C\mapsto \widetilde C$, and ${\cal H}\mapsto \widetilde{\cal H},\widetilde{\cal E}$ in combination with $H\mapsto \widetilde H, \widetilde E$, respectively.
The LP flavor-nonsinglet vector contribution can therefore also be written as,
\begin{align}\label{eq:V=CFF}
          {\mathbb V}_{0,\rm NS}(\xi,t,Q^2)&=\bar U(P')\bigg[
   \gamma^+ {\cal H}_{\rm NS} (\xi,t,Q^2)\\
   &\qquad\quad+\frac{i\sigma^{+\nu}\Delta_\nu}{2M} {\cal E}_{\rm NS}(\xi,t,Q^2)\bigg] U(P)\, ,\notag
\end{align}
so as ${\mathbb A}_{0,\rm NS}$ with evident adjustments to the corresponding CFFs and the Dirac structures according to Eq.~\eqref{def:HEq-tilde}.

The exact NNLO, i.e., two-loop CFs $C_{\rm NS}$ and $\widetilde C_{\rm NS}$ in the $\widebar{\rm MS}$-scheme was first derived using conformal symmetry~\cite{Braun:2020yib,Braun:2021grd}, utilizing the previously known two-loop DIS coefficient function~\cite{Zijlstra:1992qd,Vermaseren:2005qc}, and the so-called conformal anomaly at two loops~\cite{Braun:2016qlg}. These results were subsequently confirmed in~\cite{Gao:2021iqq} using the conventional diagrammatic approach with dimensional regularization (DR). 
We refer to~\textcite{Braun:2003rp,Braun:2013tva,Braun:2018mxm} for details of the conformal framework.  The final two-loop CFs in the $x$-space can be written in terms of the so-called harmonic polylogarithms (HPLs)~\cite{Remiddi:1999ew,Maitre:2005uu,Maitre:2007kp} with indices only 0 and 1. Recently, the LP NNLO CFs for all flavor and chirality channels in the Gegenbauer (conformal) moments (see Eq.~\eqref{eq:GPD_conformal_expansion}) have been derived~\cite{Braun:2025noa}, exploiting a new framework based on the conformal symmetry technique. The results are readily applicable in the extraction of GPDs via the GPDs-through-universal-moment-parameterization (GUMP) approach, see Sec.~\ref{sec:GUMP} for more phenomenological details.

An extra complication arises in the axialvector channel at the loop level due to the ambiguous $\gamma_5$ definition in $d=4-2\epsilon$ under DR. In the diagrammatic approach, it is necessary to introduce a so-called evanescent operator~\cite{Wang:2017ijn} to implement a finite renormalization transforming $\widetilde C_{\rm NS}$ from the Larin scheme~\cite{Larin:1993tq}, which comes directly from the Feynman diagram calculations, to the commonly used $\widebar{\rm MS}$, or ``naive dimensional regularization'' (NDR) scheme. In the conformal approach used in~\textcite{Braun:2021grd}, $\widetilde C_{\rm NS}$ is instead obtained from the already known vector CF $C_{\rm NS}$, supplemented by a Larin $\mapsto$ $\widebar{\rm MS}$ scheme change that is calculable order by order in $\alpha_s$. It can be shown that the different techniques used in these two approaches are equivalent. 

The phenomenological impact of the NNLO QCD corrections to DVCS in the nonsinglet channel was subsequently investigated. %
Since ${\mathbb V}_{0,\rm NS}$ is expected to be numerically dominated by the contribution from ${\cal H}_{\rm NS}$ in small $\xi$ region due to the $\sim \frac{\sigma^{+\nu}\Delta_\nu}{2M}$ suppression in the ${\cal E}_{\rm NS}$ sector visible in Eq.~\eqref{eq:V=CFF}, only NNLO corrections to ${\cal H}_{\rm NS}$ was analyzed. Adopting a simple $t$-independent GPD model %
and writing the resulting vector CFF as ${\cal H}_{\rm NS}(\xi)=R(\xi)e^{i\Phi(\xi)}$, the two-loop correction lowers the magnitude $R$ by about $10\%$ compared to the tree-level result throughout the range $0<\xi<0.5$ and is roughly a factor of two smaller than the NLO correction, whereas its effect on the phase $\Phi$ is much milder. 
Although this provides only a model-dependent benchmark rather than a general uncertainty estimate, its magnitude already exceeds the projected experimental precision at the JLab 12~GeV facility and the EIC. It is therefore essential to employ the complete NNLO vector CF, together with dedicated NNLO (three-loop) factorization scale variations for nonsinglet vector GPDs with realistic phenomenological modeling, to fully describe future high-precision data at the next-to-next-to-leading-logarithmic (NNLL) accuracy. Similar conclusions can also be made for the axialvector contributions.

One prominent feature shared by the quark CFs across all channels in DVCS is that they are singular at $x=\pm\xi$, as already visible in the tree-level nonsinglet result,
\begin{align}
C_{\rm NS}^{(0)}(x/\xi)=\frac{\xi}{\xi-x-i0}-\frac{\xi}{\xi+x-i0}\, ,
\label{eq:treeCFns}
\end{align}
with $\widetilde C_{\rm NS}^{(0)}$ taking the plus sign betwen the two fractions. 
Note that since $C_{\rm NS}$ ($\widetilde C_{\rm NS}^{(0)}$) is antisymmetric (symmetric) under $x\leftrightarrow-x$, only the antisymmetric (symmetric) part of the corresponding GPDs contributes to the nonsinglet vector (axialvector) CFFs. The simple poles sit at the partonic thresholds/boundary $x=\pm\xi$ (see Fig.~\ref{Fig:GPDparton} for a physical picture), where one of the Mandelstam invariants $ s\propto(1-x/\xi)Q^2$ and $ u\propto(1+x/\xi)Q^2$ of the hard parton--photon sub-processes vanish. %
The minus sign of the pole prescription $-i0$ is fixed by the contour deformation required in the proof of LP DVCS factorization and guarantees that the convolution integral is well-defined. In practice, it is often advantageous to analytically continue the convolution integral into the complex plane to tame the large numerical uncertainties, especially for CFs at higher orders. %

Beyond tree level, the CFs at the thresholds are further dressed by logarithmic enhancements $\sim\ln^k(1\mp x/\xi)$ visible in explicit higher-loop results. These threshold logarithms can be resummed to all orders in perturbation theory as shown in~\textcite{Schoenleber:2022myb}. 
In the limit $x\to\pm\xi$, each CF refactorizes into a (Sudakov) hard function times a collinear function, separating the large scale $Q^2$ from the vanishing Mandelstam variables $ s$ and $u$. The resulting renormalization-group equations are governed in part by the cusp anomalous dimension, allowing us to resum the series in $\ln(1\mp x/\xi)/(1\mp x/\xi)$ up to NNLL accuracy with currently available information. A preliminary numerical study based on a simple GPD model indicated only a modest threshold resummation effect on the quark CFF~\cite{Schoenleber:2022myb}; nevertheless, more comprehensive numerical analyses remain desirable.

\subsubsection{Flavor-singlet and transversity channels}\label{sect:3_singlet}

The flavor-singlet channel is significantly more challenging than its nonsinglet counterpart because of quark and gluon GPD mixing beyond tree level, which induces additional diagrams and more involved IR subtraction procedures. The gluon vector, axialvector, and transversity GPDs are defined from the matrix elements of the corresponding twist-two gluon operators as shown in Eqs.~\eqref{def:HEq},~\eqref{def:HEq-tilde}, and~\eqref{def:HEgT}.
The collinear factorization theorem implies that the LP contribution to the flavor-singlet vector channel takes the form,
\begin{align}
\hspace*{-1.8mm}\mathcal {\mathbb V}_{0,\rm S}&(\xi,Q^2,t)=\sum_q e_q^2\int_{-1}^{1}\frac{dx}{\xi} C_{\rm S}\left(\frac x\xi,\frac{Q^2}{\mu_F^2}\right) F_{q,\rm S}(x,\xi,t,\mu_F) \notag\\ 
&+\sum_q e_q^2\int_{-1}^{1}\frac{dx}{\xi^2} C_{g}(x/\xi,Q^2/\mu_F^2) F_{g}(x,\xi,t,\mu_F)\,, 
\label{CFF-Hs}
\end{align}
and similarly for the axialvector channel ${\mathbb A}_{0,\rm S}$ with the obvious replacements $F_{q,\rm S}\mapsto \widetilde F_{q,\rm S}$, $F_{g}\mapsto \widetilde F_{g}$, and $C_S\mapsto \widetilde C_S$, $C_{g}\mapsto \widetilde C_g$ above. 
The factorization for the transversity channel, however, becomes simpler for a spin-1/2 hadron target,
\begin{align}
   \hspace*{-1.8mm} {\cal A}_0^{+-}(\xi,Q^2,t) &= \sum_q e_q^2\int^1_{-1} \frac{dx}{\xi^2}C_{g,T}\left(\frac x\xi,\frac{Q^2}{\mu_F^2}\right) \notag\\
   &\qquad\qquad\qquad\times F_{g,T}^{ij}(x,\xi,t,\mu_F)\, ,%
\end{align}
 as the quark transversity GPDs cannot contribute due to angular momentum selection rules. 
The factorization formulas in terms of CFFs follow immediately from Eqs.~\eqref{def:HEq},~\eqref{def:HEq-tilde}, and~\eqref{def:HEgT}.

The singlet quark CFs $ C_{\rm S}$ and $\widetilde C_{\rm S}$ are identical to their nonsinglet counterpart up to ${\cal O}(\alpha_s)$. Starting from two loops, extra scattering channels induced by quark-gluon transition of the form $\bar qq\to gg\to\bar qq$ open up, giving rise to the so-called pure-singlet (PS) contribution. It is therefore common to write,
\begin{align}
    C_{\rm S}(x/\xi,Q^2/\mu_F^2) = C_{\rm NS}(x/\xi,Q^2/\mu_F^2) + C_{\rm PS}(x/\xi,Q^2/\mu_F^2)\, ,
\end{align}
and similarly for $\widetilde C_{\rm S}$, utilizing the nonsinglet CFs. 
Since gluons and photons do not couple directly, $C_g$ and $\widetilde C_g$ first contribute at $\mathcal{O}(\alpha_s)$ through quark box diagrams. 
The flavor-singlet vector hadronic matrix elements are given in Eq.~\eqref{eq:NS-proj} as $F_{q,\rm S}$. The axialvector and transversity counterparts and subsequently the corresponding singlet GPDs trivially follow. 

\begin{figure}[t]
    \centering
    \includegraphics[width=0.483\textwidth]{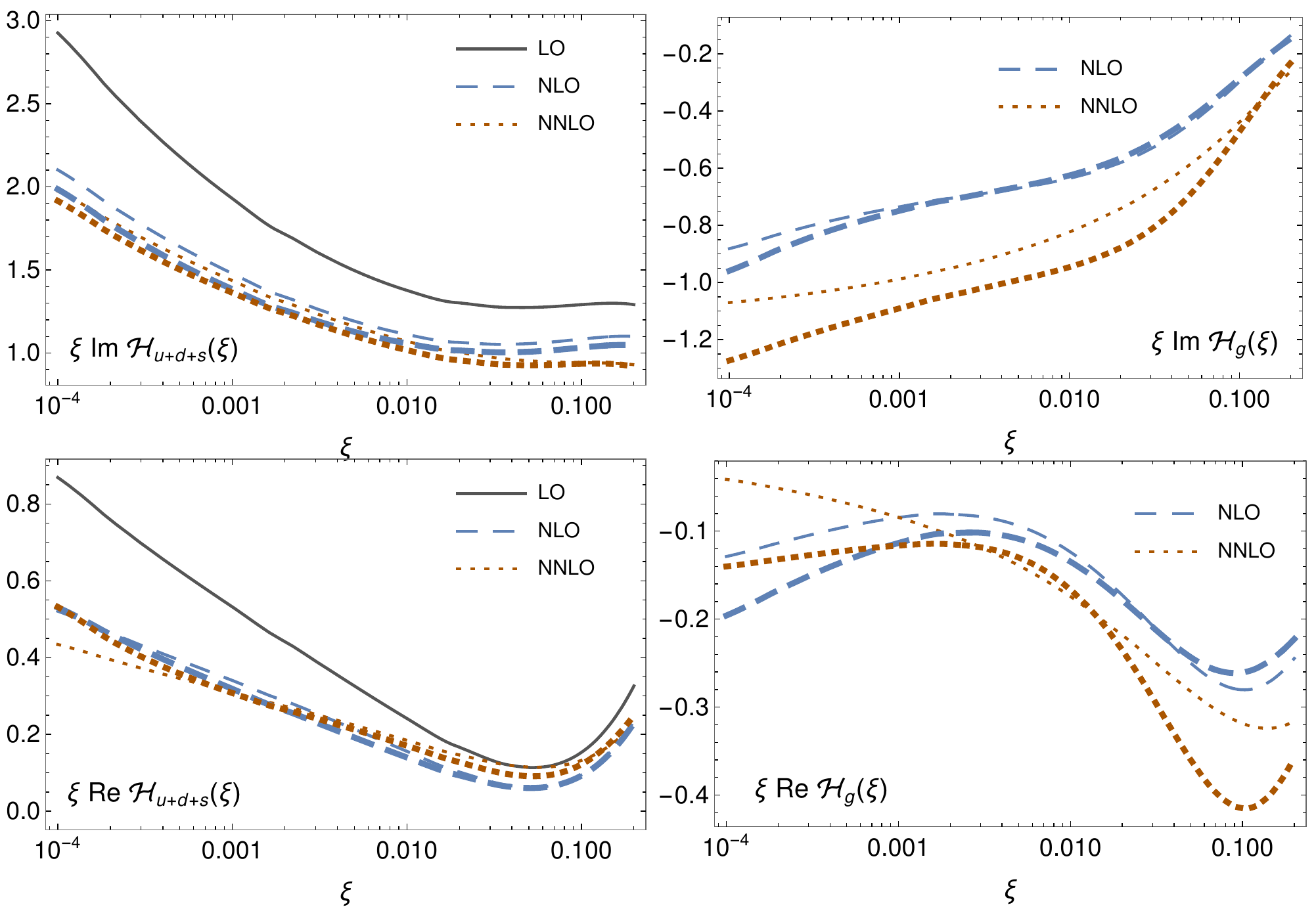}
\caption{\raggedright  Real and imaginary parts of quark and gluon CFFs in the singlet vector channel at $\mu^2 = Q^2 = 4\text{ GeV}^2, t = -0.1\text{ GeV}^2$ for GK-model GPDs normalized to HERAPDF20~\cite{H1:2015ubc} (thin lines) and ABMP16~\cite{Alekhin:2017kpj,Alekhin:2018pai} (thick lines, NLO and NNLO only) PDF fits at the appropriate $\alpha_s$ order. %
Adapted from~\textcite{Braun:2022bpn}.
}
\label{figH4GeVnew}
\end{figure}

The NLO flavor singlet and transversity CFs were obtained in~\textcite{Ji:1997nk,Belitsky:1997rh,Mankiewicz:1997bk,Hoodbhoy:1998vm,Belitsky:2000jk}, and with the modern technologies for diagrammatic calculations, the NNLO corrections to the singlet vector CFs were determined in~\textcite{Braun:2022bpn}. The two-loop singlet axialvector and transversity CFs were obtained soon after~\cite{Ji:2023xzk} in the Larin scheme. It is discovered that both the NLO and NNLO gluon CFs for vector and axialvector alike have opposite signs compared to the quark contributions, hence significantly reducing the total singlet contribution to the DVCS cross sections at moderate $Q^2$. Such cancellations are alleviated at large $Q^2$ due to the positive contribution induced by the quark-gluon mixing, eventually making the quark and gluon GPD contributions add up. By contrast, the quark and gluon contributions in DIS tend to add together. The distinct features of DVCS and DIS in the singlet channels imply that intuitions from DIS cannot be simply applied to DVCS. As in the nonsinglet channels, these two-loop CFs are written in terms of the HPL functions with indices 0 and 1. The phenomenological significance of the NNLO correction to the vector channel is analyzed using the Goloskokov-Kroll~\cite{Goloskokov:2006hr} (GK) GPD model. Due to the large gluonic content, the NNLO correction is much more prominent in the singlet channel as illustrated in Fig.~\ref{figH4GeVnew}, where the dominant CFF ${\cal H}_{\rm S}$ is plotted as a function of $\xi$. As a result, the NNLO corrections are indispensable for the precision determination of GPDs from measurements at the JLab 12~GeV facility and the EIC. Similar conclusions are also made for the axialvector and transversity channels.  %

\subsubsection{Kinematic power corrections} \label{sec:3_kpc}

As already discussed, the BMP frame/convention and the target frame are only two of several commonly used reference frames/conventions. Adopting a different frame to describe the DVCS process leads to different scalar functions of ${\cal A}^{0+}$, whereas ${\cal A}^{\pm\pm}$ and ${\cal A}^{+-}$, whose contributions start at the LP, are modified at the subleading order in the $\sqrt{-t}/Q$ and $M/Q$ expansion and beyond. Such rearrangements of the power series for the helicity amplitudes induced by the frame change, especially due to the ambiguous definition of the longitudinal plane in DVCS, 
are known as the kinematic higher-power corrections  (KPCs). %

Phenomenologically, %
since parton's transverse spatial information is probed by momentum transfer $\Delta$ through a Fourier transform as shown in Eq.~\eqref{eq:Gsq_Fourier}, a quantitative understanding of these finite-$t$ corrections is essential for hadron tomography. While increasing $|t|$ improves transverse resolution, it also amplifies the higher-power corrections for fixed $Q$, necessitating a systematic theoretical consideration.  %

From a theoretical perspective, KPCs are generated by the higher-twist descendants that are total derivatives of the twist-two local operators. They take on the schematic form of $\partial_\mu O^{\mu\mu_2\ldots\mu_N}$, $\partial_\mu\partial_\nu O^{\mu\nu\mu_3\ldots\mu_N}$, and $\partial^2O^{\mu_1\ldots\mu_N}\,,\cdots$. Sandwiched between off-forward nucleon states, a total derivative operator gives the momentum transfer, e.g., $\langle p'|\partial_\mu O|p\rangle=i\Delta_\mu\langle p'|O|p\rangle$. Their matrix elements are therefore determined by the same leading-twist GPDs, hence introducing no new nonperturbative functions, in contrast to contributions from genuine higher-twist GPDs with the twist-three ones defined in Eq.~\eqref{eq:t3GPD}. 
This implies that the higher-twist descendant operators that induce the KPCs can be disentangled from the genuine higher-twist operators by considering their distinct behaviors under QCD evolution, as the genuine higher-twist operators do not mix with the twist-two operators, albeit the reverse statement is false~\cite{Braun:2009vc,Ji:2014eta}. Therefore, by setting all the genuine higher-twist operators to zero at a given scale, they will never be generated at any other scale by the evolution. This approach was adopted in~\textcite{Braun:2011zr,Braun:2011dg,Braun:2012bg,Braun:2012hq}, pushing the study of KPCs to $1/Q^2$. The subsequent comprehensive phenomenological analysis showed significant KPCs in the intermediate photon virtuality range $Q^2\sim 1-5~{\rm GeV}^2$~\cite{Braun:2014sta}.

The inclusion of KPCs is also mandatory for theoretical consistency. The exact Compton tensor in Eq.~\eqref{Amunu-def} respects the electromagnetic Ward identities $q^\mu{\cal A}_{\mu\nu}=q'^{\nu}{\cal A}_{\mu\nu}=0$, whereas a leading-twist truncation satisfies these relations only at leading power. Translation invariance provides an equally useful diagnostic, where the translation identity 
$$e^{iz\widehat P\cdot x}{\cal A}_{\mu\nu}(z_1x,z_2x)e^{-iz\widehat P\cdot x} = {\cal A}_{\mu\nu}((z_1+z)x,(z_2+z)x)$$
with $\widehat P_\mu = -i\partial_\mu$  
is recovered only after all KPCs are summed over.

A systematic new treatment of KPCs at ${\cal O}(\alpha_s^0)$ was developed in~\textcite{Braun:2020zjm} using conformal theory techniques and the operator product expansion (OPE). In the OPE of two conserved electromagnetic currents $J_\mu(x_1)J_\nu(x_2)$, conformal symmetry and current conservation fix the Wilson coefficients of all local descendant operators up to common normalization constants, which can then be determined from the familiar transverse and longitudinal DIS CFs. The resulting local operators multiplying the Wilson CFs can be resummed and, after sandwiching it between the off-forward hadronic states, yields factorization formulas for KPCs, written as perturbative CFs convoluted only with the twist-two GPDs. The scalar-target result was obtained to ${\cal O}(1/Q^4)$ in~\textcite{Braun:2022qly}, and the calculation for a spin-$1/2$ target was subsequently completed to the same order in~\textcite{Braun:2025xlp}, obtaining schematically,
\begin{align}
    {\cal A}^{\lambda_1\lambda_2}_{\rm KPC} &= \frac{{\cal A}_{2,\rm KPC}^{\lambda_1\lambda_2}}{Q^2}+\frac{{\cal A}_{4,\rm KPC}^{\lambda_1\lambda_2}}{Q^4} +{\cal O}(Q^{-6})\, ,\notag\\
    {\cal A}^{0+}_{\rm KPC} &= \frac{{\cal A}_{1,\rm KPC}^{0+}}{Q}+\frac{{\cal A}_{3,\rm KPC}^{0+}}{Q^3} +{\cal O}(Q^{-5})%
\end{align}
with $\lambda_1\lambda_2\in\{\pm\pm,\pm\mp\}$. 
These higher KPCs restore EM gauge invariance up to corrections starting at $1/Q^5$.  Since only quark GPDs contribute at the tree level, the transversity GPDs have been absent from considerations of KPCs so far. It is worth pointing out that the technique established in~\textcite{Braun:2020zjm} can be generalized to compute the KPCs at higher orders of $\alpha_s$ as well. This investigation serves as a concrete testament that a finite twist truncation in the scattering amplitude does not commute with a Lorentz transformation.

A complementary derivation of KPC is formulated in~\textcite{Guo:2021gru} by examining how different light-cone bases and gauge choices for the final photon polarization affect the DVCS cross-section beyond the leading-twist expansion. It was found that, by keeping only the leading-twist CFFs, different light-cone conventions indeed give different twist-three helicity amplitudes. This convention dependence is not physical and shall be removed by the Wandzura--Wilczek component of twist-three GPDs as dictated by Lorentz symmetry and the equations of motion. In the special case of BMP frame/convention $\Delta_\perp=0$, so kinematic twist-three corrections of the form $|\Delta_\perp|/Q$ vanish, but other types of twist-three KPCs remain. It is confirmed that the finite twist truncation to the hadronic tensor~\eqref{Amunu-def} breaks EM current conservation beyond the resummed twist accuracy. This unphysical final-photon gauge dependence, which must disappear in the complete result, yields $\sim0.1\%$ differences at the cross-section level for kinematics tested in~\textcite{Guo:2021gru} and is much smaller than the
dependence induced by the choice of light-cone basis. By comparing results in different conventions (BKM~\cite{Belitsky:2001ns}, BMJ~\cite{Belitsky:2012ch}, and BMP~\cite{Braun:2011dg}), the study demonstrated explicitly how an incomplete twist expansion can generate intrinsic uncertainty in CFF extractions, and how the appropriate higher-twist terms convert convention-dependent formulas into covariant ones.

\begin{figure}[t]
    \centering
    \includegraphics[width=0.483\textwidth]{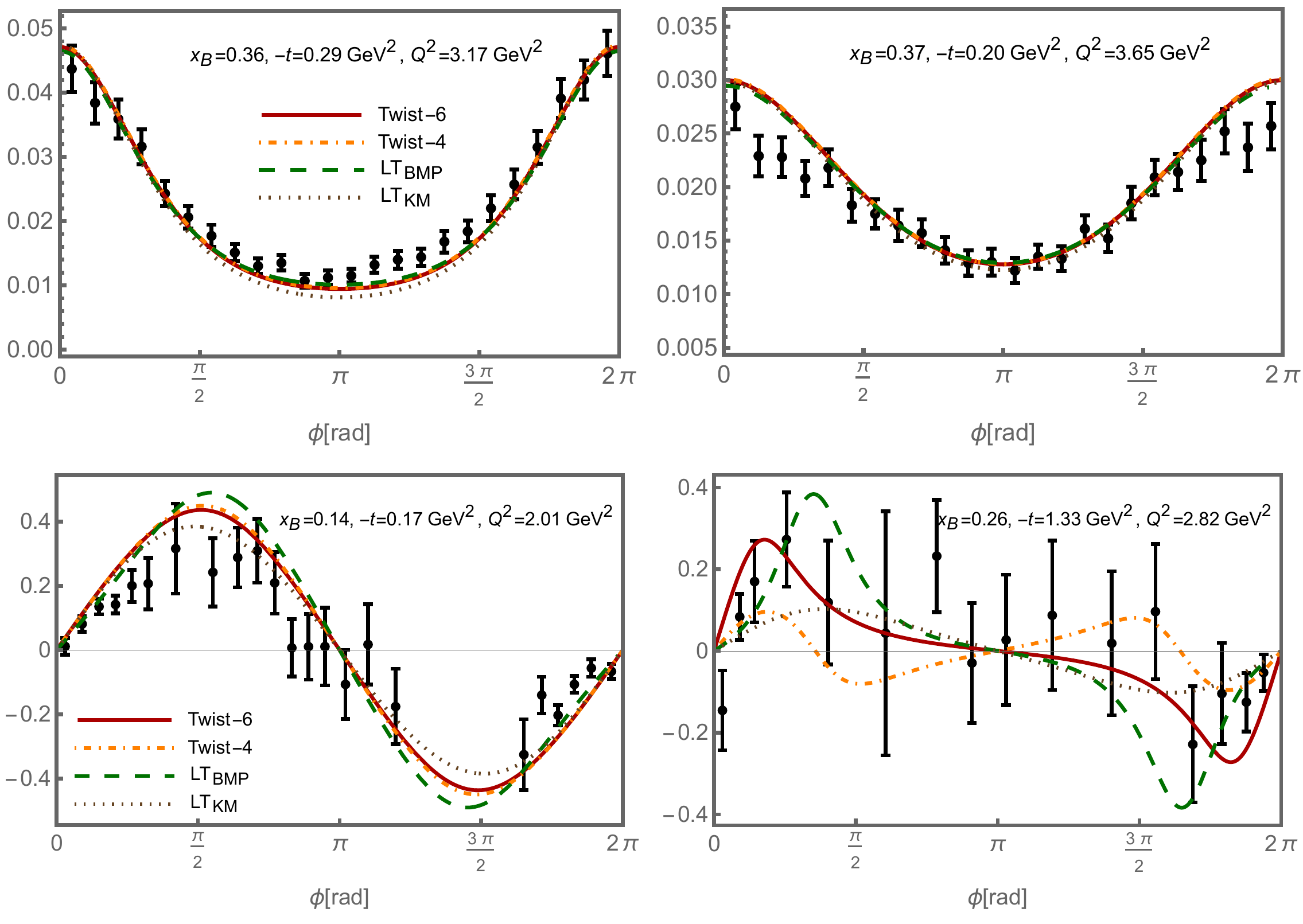}
\caption{Selected plots for total spin-averaged cross-section (upper panel) and beam-spin asymmetries (lower panel). The theory predictions of different twists are compared against the Jefferson Lab Hall A (upper)~\cite{JeffersonLabHallA:2022pnx} and CLAS12 10.6~GeV (lower) data set~\cite{CLAS:2022syx}. Adapted from~\textcite{Braun:2025xlp}.}
\label{CLAS2022}
\end{figure}

The numerical impact of kinematic power corrections depends strongly on the observable and on the ratio $|t|/Q^2$. The corrections to the spin-averaged Hall~A electroproduction total cross-sections, including the Bethe--Heitler contribution, are generally modest %
when $|t|/Q^2\lesssim0.2$. By contrast, the CLAS12 beam-spin asymmetries receive sizable corrections for $|t|/Q^2\gtrsim0.3$ and become very large near $|t|/Q^2\simeq0.5$. For most observables, the KPC expansion generally exhibits a convergence hierarchy for $|t|/Q^2\lesssim1/4$; beyond this range, the twist-five and twist-six terms can become comparable to the lower-twist contributions. It was found %
that organizing the KPCs in $1/(q\cdot q')=-2/(Q^2+t)$ instead of $1/Q^2$ improves convergence by effectively resumming additional higher-twist contributions.  The comparison with selected data of CLAS12 beam-spin asymmetries in Fig.~\ref{CLAS2022} illustrates that these corrections are important in experimentally accessible regions. They must therefore be included, together with a realistic estimate of genuine higher-twist effects, in precision extractions of GPDs from JLab, the EIC and EicC measurements. Recently, the dispersive analysis has been applied to KPC, showing that the twist-four KPCs carry nontrivial information of the nucleon EMT beyond that available at the twist-two level~\cite{Martinez-Fernandez:2025jvk}.

\subsubsection{Factorization of genuine twist-three correction}\label{sect:3_gt3}

The kinematic power corrections discussed above depend only on the leading-twist GPDs and carry no new nonperturbative information. At intermediate virtualities $2~{\rm GeV}^2\lesssim Q^2 \lesssim 10~{\rm GeV}^2$, however, the DVCS amplitude also receives potentially nonnegligible genuine power corrections of order $\sim\Lambda_{\rm QCD}/Q \sim M/Q \sim \sqrt{-t}/Q$ that demand additional nonperturbative input, namely the twist-three GPDs. They are parameterizations of the off-forward matrix elements of twist-three two-parton (e.g., Eq.~\eqref{eq:t3GPD}) and twist-three three-parton operators. The two-parton twist-three GPDs generally decompose into a Wandzura--Wilczek part, fixed by the twist-two GPDs related to KPCs, and a dynamical piece carrying genuinely new information~\cite{Kivel:2000fg,Aslan:2018zzk}, such as partons' orbital angular momentum and spin-orbit correlations inside their parent hadrons as discussed in Sec.~\ref{subsec:2_spin_t3GPD}. The inclusion of twist-three GPDs is also mandatory for a consistent description of DVCS observables---in particular, the beam- and target-spin asymmetries---away from the asymptotic regime.

 At the NLP, the DVCS amplitude excluding the KPCs has long been conjectured to factorize onto twist-three GPDs, with the tree-level coefficient functions known for two decades~\cite{Kivel:2000fg} and parts of the $\alpha_s^1$ correction computed in the Wandzura--Wilczek approximation~\cite{Kivel:2003jt}. A rigorous all-order justification was nonetheless missing. In addition to the dynamics already appearing in target-collinear directions at the LP, at NLP, the extra (anti)-collinear direction aligned with the real photon generates new soft modes that connect the two collinear directions. Such modes may then induce endpoint-divergent convolutions that spoil a clean collinear factorization onto the standard GPDs.

This gap was closed recently in~\textcite{Schoenleber:2024ihr}, where for the first time an all-order analysis of DVCS at NLP was performed using soft-collinear effective theory (SCET)~\cite{Bauer:2000yr,Bauer:2000ew,Bauer:2001yt,Beneke:2002ph}. Organizing the amplitude in the standard power-counting scheme $\lambda\sim\Lambda_{\rm QCD}/Q\sim\sqrt{-t}/Q\sim M/Q$, %
the factorization formula for the genuine twist-three amplitude in the doubly deeply virtual Compton scattering (DDVCS, see Sec.~\ref{sec:3_DDVCS}), $\gamma^{*}(q) + N(P)\to \gamma^{*}(q')+N(P')$, where both photons are far off shell, is shown to hold to all orders in the following form, confirming ealier conjectures,
\begin{align}
\mathcal{C}^{\lambda_1\lambda_2}_1 &= \sum_\ell \int\! dx\, C^{\lambda_1\lambda_2}_{\ell}(x,\xi)\,F_\ell^{(2)}(x,\xi,t)\label{eq:NLPfact}\\
&+ \sum_\ell \int\! dx_1 dx_2\, C^{\lambda_1\lambda_2}_{\ell}(x_1,x_2,\xi)\,F_\ell^{(3)}(x_1,x_2,\xi,t)
\notag\, ,
\end{align}
with $\lambda_1\lambda_2\in\{0\pm,\pm0\}$. $F_\ell^{(2)}$ and $F_\ell^{(3)}$ collect the two- (quark--quark and gluon--gluon with possible derivatives) and three-parton twist-three GPDs, respectively. %

For DVCS, the proper factorization requires some care as it depends on the polarization of the initial virtual photon,
\begin{align}
\mathcal{A}^{\lambda_1\lambda_2}\big|^{\rm BMP}_{\rm NLP} =
\begin{cases}
 \sum_\ell C^{(\ell)}\otimes F_\ell^{\rm tw\text{-}3}\, , & \lambda_1\lambda_2=0\pm\, ,\\[4pt]
\ C_H \times J\otimes\mathcal{S}\, , & \lambda_1\lambda_2=\pm\pm\, ,
\end{cases}
\label{NLPpol}
\end{align}
where $\otimes$ denotes the convolution in the momentum fractions, and the convolution in the first line represents Eq.~\eqref{eq:NLPfact}. In the BMP frame, the collinear twist-three GPDs contribute only when the initial virtual photon is longitudinally polarized, which enters the helicity amplitude $\mathcal{A}^{0\pm}$ whose LP term ${\cal A}_0^{0\pm}\sim (1/Q)^0$ simply vanishes as shown in Eq.~\eqref{Amunu}, and therefore the $\sim 1/Q$ contribution becomes dominant in this channel. By contrast, additional complications emerge when the initial virtual photon is transversely polarized. In addition to the anti-collinear modes in the real-photon direction, ultrasoft modes, which leave collinear momenta almost unchanged, appear. %
Such modes together with the dynamical chiral symmetry breaking induce an ``unusual'' nonzero soft function $\mathcal{S}$ that contributes to the DVCS amplitude at NLP through a hard function $C_H(Q^2)$ and a jet function $J$ as shown in Eq.~\eqref{NLPpol}. %
It should be emphasized, however, that the universality of the soft function ${\cal S}$ remains unknown. In summary, the longitudinal DVCS amplitude factorizes onto twist-three GPDs to all orders in $\alpha_s$, free of endpoint divergences. Whereas for the NLP contribution induced by the transversely polarized initial photon, factorization still holds as shown in the second line of Eq.~\eqref{NLPpol}, but the nonperturbative inputs do not organize into GPDs. These statements are frame-specific: in frames with $\Delta_\perp\neq0$, the contributions induced by twist-three GPDs and soft function ${\cal S}$ are redistributed to other helicity amplitudes at the NNLP, i.e., to ${\cal A}^{\pm\pm}$ (${\cal A}^{0+}$) %
without spoiling the NLP factorization formulas presented. %

The soft function ${\cal S}$ may be assessed nonperturbatively, for example, using light-cone sum rules. The result is expected to be sensitive to the strange-quark mass~\cite{Schoenleber:2024ihr}. The twist-three GPDs, on the other hand, can be determined through their leading-order effect on the longitudinal DVCS amplitude, potentially within the reach of the precision level from JLab 12~GeV and the EIC, provided that the longitudinal cross section $\sigma_L$ can be determined with sufficient precision. 

Before concluding the DVCS subsection, it is worth mentioning some DVCS-related reactions. The charged-pion DVCS as the subprocess of the Sullivan reaction $ep\to e\gamma\pi^+n$ was investigated in~\textcite{Chavez:2021koz}, where the proton emits a virtual pion, and the neutron acts as a spectator. Assuming one-pion-exchange dominance and neglecting off-shell effects in the pion GPDs, numerical analysis showed pronounced sensitivity to gluon contributions at NLO within EIC and EicC kinematics. For the spin-0 ${}^4\mathrm{He}$ nucleus, %
it was demonstrated that the ${}^4\mathrm{He}$ nuclear GPDs, constructed from the nucleon ones under the impulse-approximation, offer a good description of the measured coherent-DVCS beam-spin asymmetry~\cite{Fucini:2018gso}. 
Very recently, a comprehensive phenomenological analysis combining coherent-DVCS and elastic-FF data with NLO amplitudes, KPCs, and GPD evolution has led to nontrivial constraints on a nuclear GPD model~\cite{Martinez-Fernandez:2026zog}, yielding transverse spatial distributions of parton within that model. 
The GPDs for spin-one targets, e.g., the deuteron, have also been introduced~\cite{Berger:2001zb}. They are accessible through coherent DVCS and meson electroproduction, with more complex polarization tensor structures than in spin-1/2 targets. The one-body contributions to the deuteron EMT FFs have recently been calculated in the nonrelativistic impulse approximation~\cite{Cosyn:2026gyy}.

\subsection{Doubly virtual Compton scattering}\label{sec:3_DDVCS}

The DDVCS
$
\gamma^*(q)+ N(P) \longrightarrow \gamma^*(q')+ N(P') 
$
generalizes the DVCS reaction~\eqref{DVCSprocess-BMP} to the case where both photons are far offshell. It is accessed experimentally through exclusive lepton-pair electroproduction $e\,N\to e\,N\,\ell^+\ell^-$, where the timelike outgoing photon $q'^2>0$ materializes as the lepton pair. 
To better demonstrate the characteristics of the DDVCS process, let us first establish the kinematics of this reaction. %
Setting $\bar q=(q+q')/2$ with $Q^2\equiv -\bar q^2$, we define two scaling variables together with their ratio, 
\begin{align}
\xi = -\frac{\Delta\cdot \bar q}{2 \bar P\cdot\bar  q}\,, \quad \eta = \frac{Q^2}{2 \bar P\cdot \bar q}\,, \quad \omega \equiv \frac{\xi}{\eta}=\frac{q^2-q'^2}{q^2+q'^2-\Delta^2/2}\, .
\label{DDVCSkin}
\end{align} 
At LP, DVCS corresponds to $\omega=1$ (so that $\eta=\xi$), forward DIS to $\xi=0$, TCS to $\omega=-1$, and the physical DDVCS region of dilepton production to $\omega<-1$. DDVCS therefore provides a unified framework in which DVCS and TCS, together with other double-photon-nucleon processes, appear as special cases. 

The timelike Compton scattering (TCS), $\gamma(q)+N(P)\to\gamma^*(q')+N(P')$ gives us access to GPDs through exclusive lepton-pair photoproduction~\cite{Berger:2001xd}. For $q^2=0$ and the timelike $q'^2\gg M^2, -t$ establishing the hard scale, the LP amplitude factorizes into hard CFs and the same GPDs as DVCS, allowing us to test GPD universality.  Similar to DVCS, it was shown in~\cite{Berger:2001xd} that the BH-TCS interference term is separable from $|{\cal T}_{\rm TCS}|^2$ through the angular moments by flipping the lepton charge. 

The LP NLO CFs were calculated by~\cite{Pire:2011st} with $q'^2>0$ generating additional absorptive and $\pi^2$ terms compared to its DVCS counterparts, producing imaginary contributions from the whole $-1<x<1$ region. This is qualitatively different from DVCS, where the imaginary part of the amplitude comes entirely from the $|x|\geq \xi$ region. The numerical analysis of~\cite{Moutarde:2013qs} showed sizable gluon effects even at moderate energies, especially in the real part of CFFs. Since both DVCS and TCS admit LP collinear factorization and they are related by crossing symmetry, their LP amplitudes can be mapped to each other as explicitly demonstrated to the NLO~\cite{Mueller:2012sma}. This allows us to make predictions for TCS observables based on more abundant DVCS data~\textcite{Grocholski:2019pqj}. %
Beyond LP, the LO KPCs for the TCS process have been computed up to twist-four for the spin-0 target~\textcite{Martinez-Fernandez:2025gub} with numerical analysis showing sizable effects at moderate $q'^2$ based on the DD-inspired pion GPD model. The KPCs to nucleon TCS remain to be determined. The CLAs collaboration has recently demonstrated the TCS measurement is feasible~\cite{CLAS:2021lky}, see Sec.~\ref {sec:5_hard_ex_mea} for details.

For the general DDVCS process, it is convenient to define two light-cone vectors in the LP approximation ($t=0$): $n^\mu=q_1^\mu/q_1^2-q_2^\mu/q_2^2$ that enters the GPD definition, and $\tilde n^\mu = q_1^\mu-q_2^\mu=\Delta^\mu$. Together, they define the longitudinal plane. The $n^\mu$ then reduces to $n^\mu_{\rm BMP}$ at for $q_2^2=0$ after rescaling. 

The distinctive advantage of DDVCS comes from its greater kinematic freedom. The DVCS and TCS correspond to having a real photon in the final (initial) state. They detect the GPDs predominantly along the line $x=\pm\xi$, especially for the imaginary parts of the amplitude because of the tree-level CF structure (see e.g.,~\eqref{eq:DDVCS_Cperp}). The DDVCS, on the other hand, lifts this restriction~\cite{Belitsky:2002tf,Guidal:2002kt} by allowing $\eta$ and $\xi$ to vary independently within the physical kinematic domain so that the imaginary part of the LO DDVCS amplitude maps out the interior $x\neq\xi$ region of the GPDs by scanning the invariant mass of the lepton pair. %
Such a simple relation, however, is spoiled by higher-order corrections to the CFs.

Since both photons are highly virtual, the LP amplitude, induced by helicity-conserving processes, receives contributions from both the transverse channel ${\cal A}^{\pm\pm}$, and the longitudinal channel ${\cal A}^{00}$, as in the case of DIS. Taking the vector sector as a concrete example, it is convenient to decompose the DDVCS hadronic matrix element at LP as
\begin{align}
{\cal A}_V^{\mu\nu} &=\left(-g^{\mu\nu}+\frac{q'^\mu q^\nu}{q\cdot q'}\right){\cal F}_\perp
\\
&\quad+\frac{2\eta}{\bar P\cdot \bar q}\left(\bar P^\mu + \frac{q'^\mu}{2\eta} \right)\left(\bar P^\nu + \frac{q^\nu}{2\eta} \right)
\left({\cal F}_\perp + {\cal F}_L\right)\, ,\notag
\end{align}
where the arguments $ (\xi,\eta,\Delta^2,Q^2)$ of the transverse ${\cal F}_\perp$ and longitudinal ${\cal F}_L$ scalar amplitudes have been omitted for simplicity. 
The LP collinear factorization theorem for the transverse and longitudinal flavor-nonsinglet vector amplitudes reads,
\begin{align}
&\mathcal{F}_i(\xi,\eta,t,Q^2) \notag\\
&\quad= \sum_q e_q^2 \int_{-1}^{1}\frac{dx}{\xi}\, C_i\!\left(\frac{x}{\eta},\frac{\xi}{\eta},\frac{Q^2}{\mu^2}\right) F_{q,\rm NS}(x,\xi,t,\mu)\,,
\label{DDVCSfact}
\end{align}
where $i$ can be either transverse ($\perp$) or longitudinal ($L$). $F_{q,\rm NS}$ is the appropriate matrix element in Eq.~\eqref{def:HEq},  and  $C_i$ is the Wilson CF computable perturbatively in QCD. 
The tree-level CFs read 
\begin{align}\label{eq:DDVCS_Cperp}
    C_\perp^{(0)}(z,\omega)=\frac{\omega}{1-z}-\frac{\omega}{1+z}\, ,\quad 
    C_L^{(0)}=0
\end{align} 
with $z=x/\eta$. It is clear that the longitudinal FF ${\cal F}_L$ is a genuine radiative effect generated first at one-loop~\cite{Pire:2011st}, hence suppressed in perturbative QCD. The LP factorization formulas for other DDVCS channels, including the vector, axialvector sectors in both singlet and nonsinglet combinations and the gluon-transversity double-flip channel, follow as natural generalizations to the corresponding DVCS ones, with the NLO results available for a long time~\cite{Ji:1998xh,Pire:2011st,Mankiewicz:1997bk,Blumlein:1999sc,Hoodbhoy:1998vm}, and we refer to Refs.~\cite{Radyushkin:1997ki,Ji:1998xh,Collins:1998be,Bauer:2002nz,Chen:1997rc} for more detailed discussions. 
Beyond the LP, the factorization theorem has also been established as discussed in Sec.~\ref{sect:3_gt3}.

The tree-level poles of $C_\perp^{(0)}(z,\omega)$ at $z=\pm1$ can again be treated by introducing the $-i0$ prescription in the denominator. At higher orders, the CFs are again enhanced by the threshold logarithms as discussed in Sec.~\ref{sec:3_DVCS}, and their all-order resummation was worked out in~\textcite{Schoenleber:2024dvq}. The qualitatively new element is that near $z\to\pm1$, with the two photon virtualities being independent, the CF factorizes into a product of a collinear function multiplied by two Sudakov hard functions $H(-q^2,\mu)\,H(-q'^2,\mu)$, one associated with each photon, in comparison to the single hard function of the DVCS or TCS process with one real photon. %
The leading $z\to\pm1$ behavior of the two-loop DDVCS CF $C_\perp^{(2)}$ was then determined by its one-loop data through the threshold analysis. The result agrees with the explicit calculation of~\textcite{Braun:2024srt}, thereby providing a nontrivial mutual check.

Numerically, using the toy GPD model and the anticipated kinematics for the first measurements at JLab, 
the NNLO corrections turn out to be large, warranting more dedicated DDVCS phenomenological investigations. Moreover, extending the two-loop description to the flavor-singlet and axialvector channels, in parallel with the DVCS program, remains the principal outstanding task. 
Such studies will certainly be beneficial to the experimental programs targeting DDVCS at JLab%
~\cite{Boer:2024hol,Zhao:2021zsm}, %
EIC~\cite{AbdulKhalek:2021gbh}, and EicC~\cite{Anderle:2021wcy}. The projected sensitivities of these experiments to the underlying GPDs have been quantified in dedicated impact studies~\cite{Deja:2023ahc,Alvarado:2025huq}.

\subsection{Deeply virtual meson production}\label{sec:3_DVMP}

DVMP offers another indispensable probe of GPDs, complementary to the Compton processes discussed above~\cite{Radyushkin:1996ru,Collins:1996fb}. The generic reaction
$
\gamma^*(q)+ N(P) \longrightarrow M(p_M)+ N'(P')
$
replaces the outgoing real photon of DVCS by an exclusively produced meson $M$, e.g., a pseudoscalar ($\pi,K,\eta$) or a vector ($\rho,\omega,\phi,J/\psi,\dots$) state. 
Note that initial and final nucleon state $N(P)$ and $N'(P')$ are not necessarily the same as in the case of $\gamma^*p\to\pi^+n$, which instead constrains the so-called transition GPDs; see, for example, \textcite{Diehl:2024bmd} for detailed discussions.  
When the produced meson is light compared to the hard scale, as for the pion and kaon, its momentum $p_M$ may be taken lightlike, making the BMP frame directly applicable, with $\xi$, $t$ and $Q^2$ retaining their DVCS definitions; heavier states, however, require the meson mass to be kept, as shall be discussed in Sec.~\ref{sec:3_DVMP_heavy}. The distinctive feature of the DVMP process is its flavor sensitivity: while the DVCS amplitude probes only the $e_q^2$-weighted sum $\sum_q e_q^2 F_q$ over quark flavors at LO, the identified meson allows us to distinguish the flavor content inside the hadron $N$. At LP approximation, the accessible twist-two flavor GPDs are determined by the quantum numbers of the produced meson: pseudoscalar mesons probe the polarized axialvector GPDs $\widetilde H$ and $\widetilde E$, whereas vector mesons are sensitive to
the unpolarized vector GPDs $H$ and $E$, as well as to the small-$x$ gluon sector. The valence content of the meson $M$ then selects the relevant
flavor combination. For instance, with exact isospin symmetry, charged-pion production $\gamma^* p\to\pi^+ n$ isolates the nonsinglet nucleon GPD combination $\widetilde H^{u}-\widetilde H^{d}$ and $\widetilde E^{u}-\widetilde E^{d}$, giving access to information no single DVCS measurement can provide. Here, isospin symmetry allows one to relate the transition GPDs directly to the nucleon GPDs. DVMP is in this sense a crucial complement to DVCS for the flavor separation of GPDs.

The factorization theorem for DVMP carries an important qualification relative to DVCS: collinear factorization is established only for a \emph{longitudinally} polarized virtual photon~\cite{Collins:1996fb}, with the transverse amplitude being power-suppressed by $1/Q$ and factorization-breaking, so that separating the longitudinal cross-section $\sigma_L$ from its transverse counterpart $\sigma_T$ is a prerequisite for any definite GPD interpretation. In the vector-meson production channels, the leading-twist factorization requires, in addition, that the outgoing meson be longitudinally polarized as well. The LP DVMP scalar amplitude ${\cal A}_M$ then factorizes in terms of twist-two GPDs and the twist-two meson distribution amplitude as, %
\begin{align}
&\mathcal{A}_M(\xi,t,Q^2) \label{DVMPfact}
\\
&\quad=\int_{-1}^{1}\frac{dx}{\xi}\int_{0}^{1}du\,\phi_M(u)\,C\left(u,\frac{x}{\xi},\frac{Q^2}{\mu^2}\right)F_f(x,\xi,t,\mu)\, ,\notag
\end{align}
where $F_f$ is the twist-two hadronic matrix elements in Eqs.~\eqref{def:HEq}--\eqref{def:HEgT}, $\phi_M$ is the twist-two meson DA, and $u$ is the quark momentum fraction inside the meson. As in DVCS, the perturbative CF $C$ is singular at the threshold points $x=\pm\xi$, hence the convolution over $x$ is understood to follow the prescription $\xi\to\xi-i0$.

The NLO coefficient functions have long been known for charged-pion and light vector-meson production~\cite{Belitsky:2001nq,Ivanov:2004zv}, with subsequent calculations completing the flavor-singlet pseudoscalar channels~\cite{Duplancic:2016bge}. Their formulation in conformal Mellin space facilitates the consistent inclusion of radiative corrections and scale evolution in global GPD analyses~\cite{Muller:2013jur}. Substantial NLO corrections at fixed-target kinematics and particularly large effects at small $x_B$, nevertheless, underscore the importance of assessing perturbative stability~\cite{Diehl:2007hd}. In the latter regime, logarithmically enhanced contributions also motivate high-energy resummation~\cite{Ivanov:2007je, Flett:2024htj}.

Recently, complete NNLO coefficient functions have been obtained for longitudinal pion production, $\gamma^*_L p\to\pi^+ n$ and $\gamma^*_L p\to\pi^0 p$~\cite{Chen:2026vff}, extending the perturbative description beyond the NLO results available for more than two decades. The charged-pion coefficient function follows by analytic continuation from the two-loop hard-scattering kernel of the spacelike pion electromagnetic form factor~\cite{Chen:2023byr,Ji:2024iak}, whereas neutral-pion production requires an additional pure-singlet contribution that first appears at two loops. Numerical studies find substantial NNLO corrections, comparable to the NLO corrections at intermediate $Q^2$, motivating further investigations of perturbative convergence and its implications for GPD extractions. Comparisons with available charged-pion data from JLab favor the GK parametrization among the models examined, although this preference remains tentative, especially given that the data are limited in the relatively small $Q^2$ region and the theoretical uncertainties, including pion DA, remain sizable. We expect future DVMP measurements at EIC and EicC with much higher $Q^2$ values and luminosity to provide cleaner and more stringent tests of LP-DVMP factorization and to impose more quantitative constraints on both leading-twist DA and GPDs. 

On the other hand, with proper treatment of the endpoint singularities, the twist-three pion DA contribution to the DV$\pi$P process has been considered in~\cite{Duplancic:2023xrt}, showing reasonable agreement with the experimental data. 
The related wide-angle photo- and electroproduction of pions, kaons, and eta particles, induced by twist-two GPDs and meson DAs up to twist three, have been discussed in~\cite{Kroll:2021ecb,Kroll:2021zss,Kroll:2018uvl,Kroll:2019wug}. Here, the angle is between the initial and final momenta of the nucleon in the center-of-mass frame, with a large angle corresponding to a large $-t$ momentum transfer, see~\cite{Kroll:2017hym} for the wide-angle analysis of DVCS. While the issues of factorization and experimental measurements remain challenging, it nonetheless offers a potential route to accessing GPDs in the large $-t$ region and gluonic DA of the $\eta'$ meson. The subsequent generalization to the heavier baryon productions $\gamma p\to \pi\Delta$ in the wide-angle limit has also been discussed~\cite{Kroll:2025osx,Kroll:2022roq} where the transition GPDs $p\to\Delta$ are no longer determined from nucleon GPDs using isospin symmetry. 

The hard exclusive light meson production through weak interaction provides additional parity and flavor combinations, where both vector GPDs $H,E$ and axialvector GPDs $\widetilde H,\widetilde E$ can enter the amplitudes at LP. The charged- and neutral-current neutrino production of $\pi,K,\eta$ was investigated in~\cite{Kopeliovich:2012dr}, using flavor SU(3) to relate baryon-transition GPDs to nucleon ones. The contribution of gluon GPDs in pion and longitudinal-$\rho$ production was studied in~\cite{Pire:2017tvv}, showing that comparisons of charged- and neutral-meson production on proton and neutron targets help disentangle quark-flavor and gluon contributions. The NLO CFs are used to analyze the charged-current channels in~\textcite{Siddikov:2019ahb} with estimates for the twist-three effects and a proposal for testing the leading-twist dominance via the cross-section ratio between $\rho_L$ and $\pi$ production. 
Charmed-meson production is another interesting process, as it provides access to transversity GPDs. \textcite{Pire:2015iza} showed that charm-quark mass effects make azimuthal distributions in neutrino-induced $D$-meson production sensitive to chiral-odd quark GPDs. They subsequently proposed the $D^*\to D\pi$ decay angular distribution in neutrino-induced production as a probe of gluon transversity GPDs~\cite{Pire:2017yge}. For charged-current electron scattering,~\textcite{Pire:2021dad} obtained LO EIC cross-section predictions for both $e^-N\to\nu_e D_s^-N$ and $\nu_e D_s^{*-}N$ channels, probing gluon GPDs with strange-quark GPDs neglected.

Another class of reactions known as the exclusive (light)-meson-induced Drell-Yan processes $MN\to\gamma^*N'\to\ell^+\ell^-N'$~\cite{Berger:2001zn}, or ``crossed-channel counterpart of DVMP'', has been considered to the NNLO order with numerical studies showing sizable corrections~\cite{Jia:2026ufq}. In addition to TCS, these processes provide another independent test of GPD universality with timelike kinematics and are expected to be measured at J-PARC with its proposed high-energy pion and kaon beams, providing another independent test of GPD universality~\cite{Sawada:2016mao,Aoki:2021cqa}.

Vector-meson channels provide complementary access to the ordinary, non-transition unpolarized GPDs $H$ and $E$, with particular sensitivity to gluons at small skewness~\cite{Ivanov:2004zv,Goloskokov:2007nt}. These processes have therefore played a crucial role in phenomenological analyses as well~\cite{Cuic:2023mki,Guo:2025muf}. Extending these channels to NNLO would permit a more systematic assessment of perturbative uncertainties and strengthen their role in precision GPD extractions. Heavy vector mesons, such as $J/\psi$ and $\Upsilon$, couple to gluon GPDs exclusively at LO in the conventional heavy-quarkonium factorization framework~\cite{Ivanov:2004vd,Flett:2021ghh, Chen:2019uit}; these channels are discussed in the following subsection.

\subsection{Exclusive heavy quarkonium production}\label{sec:3_DVMP_heavy}

Among vector-meson channels, exclusive productions of heavy vector mesons, e.g., $J/\psi$ and $\Upsilon$ provide direct access to gluon structure~\cite{Ivanov:2004vd}. The heavy-quark mass supplies a hard scale even in photoproduction, allowing an expansion in $\Lambda_{\rm QCD}^2/(Q^2+M_V^2)$ and $|t|/(Q^2+M_V^2)$, supplemented by the NRQCD expansion in the squared relative heavy-quark velocity $v^2$~\cite{Bodwin:1994jh}. The kinematics can be parametrized as in the DDVCS frame (Sec.~\ref{sec:3_DDVCS}), with $M_V^2$ replacing the outgoing photon virtuality. At Born level, the color-singlet heavy-quark pair couples to the target through two-gluon exchange, so that only gluon GPDs contribute; quark GPDs first enter at NLO~\cite{Ivanov:2004vd,Chen:2019uit,Flett:2021ghh}. At leading order in the velocity expansion, quarkonium formation is described by an NRQCD matrix element proportional to the wave function at the origin, $\psi_V(0)$. Suppressing this normalization and other overall factors, the leading-power LO photoproduction amplitude takes the form
\begin{align}
{\cal A}_{\rm LO}^{\rm LP} (\xi,t) &\propto\alpha_s\int_{-1}^{1}\,\frac{dx}{2\xi}\left[\frac{1}{x+\xi-i0}-\frac{1}{x-\xi+i0}\right]\notag\\
&\qquad\times x F_g(x,\xi,t)\,.
\label{eq:GJpsi}
\end{align}
where $F_g$ is the gluon correlator defined in Eq.~\eqref{def:HEq}. 

While the perturbative coefficients have been calculated to NLO in both the photo-~\cite{Ivanov:2004vd} and electro-production~\cite{Chen:2019uit,Flett:2021ghh}, it was found that the small skewness at high collision energies enhances sensitivity to gluon GPDs but also generates large logarithmic corrections that can destabilize fixed-order predictions~\cite{Ivanov:2004vd,Jones:2015nna,Guo:2024wxy}. Recent calculations have matched the NLO collinear amplitude, including GPD evolution, to high-energy resummation in the double-logarithmic approximation, improving perturbative stability and the description of HERA photoproduction data~\cite{Flett:2024htj}. The resummed coefficient function has subsequently been extended to include the first relativistic corrections of order $v^2$~\cite{Nefedov:2026cgj}. These developments provide a more systematic basis for extracting small-skewness gluon GPDs while accounting for radiative and quarkonium-structure effects, while the need for improved theoretical accuracies still persists.

Ultraperipheral collisions at the LHC extend heavy-quarkonium photoproduction to high-energy photon--proton and photon--nucleus interactions, providing complementary constraints on small-$x$ gluon structure~\cite{ALICE:2023jgu,CMS:2023snh}. NLO calculations demonstrate the importance of radiative corrections and quark contributions in interpreting these measurements~\cite{Eskola:2022vpi}. Additionally, the color-glass-condensate framework provides a complementary description in terms of Wilson-line correlators, incorporating nonlinear small-$x$ evolution and gluon saturation. Recent advances include NLO quarkonium-production calculations with leading relativistic corrections and studies of nuclear saturation effects~\cite{Mantysaari:2021ryb,Mantysaari:2022kdm,Penttala:2024hvp}. Developing a unified framework that clarifies the connection between these dipole amplitudes and GPDs, see, for instance,~\textcite{Hatta:2017cte}, will be important for connecting high-energy gluon dynamics with GPD-based hadron tomography.

Recently, the heavy quarkonia production in the near-threshold region has attracted rising interest. Near the threshold, the restricted phase space enforces a large and predominantly longitudinal momentum transfer in the center-of-mass frame. This drives the skewness towards its maximal value, $\xi\to1$ in the heavy-quark limit without spoiling the LO factorization formula~\cite{Guo:2021ibg} in Eq.~\eqref{eq:GJpsi}. In practice, the finite mass of the produced heavy meson implies $\xi<1$. This effect can, in principle, be incorporated through finite-mass corrections, informing us to keep $\xi$ general in order to describe different near-threshold heavy-meson production processes.
Since the GPDs have support $|x|\leq1$ and $\xi$ is large in heavy-meson production, the hard CF can be expanded in powers of $ x/\xi$, organizing the real part of ${\cal A}_{\rm LO}$ into an asymptotic series of the gluon GPD moments,
\begin{align}
{\rm Re}\left[\mathcal A_{\rm LO}^{\rm LP}(\xi,t)\right]\!\propto\!\sum_{n=0}^{\infty}\frac{1}{\xi^{2n+2}}\int_{-1}^{1}\text{d}x x^{2n+1}\,F_g(x,\xi,t),
\label{eq:GJpsimom}
\end{align}
whereas the imaginary part ${\rm Im}\left[\mathcal A_{\rm LO}^{\rm LP}(\xi,t)\right]\propto F_g(\xi,\xi,t)$ is suppressed in the $x=\xi\to1$ endpoint limit.

The real parts of the amplitudes were originally argued to be dominated by the leading term in a Taylor expansion in $(x/\xi)^2$~\cite{Hatta:2021can,Guo:2021ibg}. Higher moments carry additional powers of $(x/\xi)^2$ and are therefore expected to be suppressed as $\xi\to1$, when gluon GPDs are believed to become concentrated in the DA-like region $|x|<\xi$. In typical near-threshold kinematics for $J/\psi$ production, $\xi\simeq0.6$, and higher-moment contributions are estimated to be $\sim25\%$ of the leading-moment contribution in an asymptotic GPD model with $\mu\sim M_V\to\infty$~\cite{Guo:2021ibg}. This relation was subsequently examined using an asymptotic expansion~\cite{Guo:2023qgu}, extended to NLO~\cite{Guo:2025jiz}, and explored in near-threshold $\phi$-meson electroproduction~\cite{Hatta:2025vhs}. Together with the polynomiality condition discussed in Sec.~\ref{sec:2_gpd_basic}, this relates measured near-threshold production cross sections directly to the spin-two gluonic gravitational form factors $A_g(t)$, $B_g(t)$, and $C_g(t)$ introduced in Eq.~\eqref{eq:emtmatrix}. Combined with the growing body of near-threshold $J/\psi$ production measurements across various experimental halls at JLab~\cite{Hatta:2018ina,Guo:2021ibg,GlueX:2019mkq,Duran:2022xag,GlueX:2023pev,007:2026dow,Tyson:2026gnd}, this framework offers rare direct access to the gluonic EMT form factors, which play a pivotal role in determining the nucleon's mass, spin, and mechanical structure (see Sec.~\ref{subsec:2_mass_scalar_force})~\cite{Hatta:2018ina,Guo:2021ibg,Sun:2021gmi,Guo:2023qgu,Guo:2025jiz,Hatta:2025vhs} together with the lattice QCD advancements~\cite{Hackett:2023rif}. 

On the other hand, we note that relativistic corrections and competing mechanisms, including open-charm coupled channels, final-state interactions, and hidden-charm resonances, could also be important and remain under active investigation~\cite{Du:2020bqj,Strakovsky:2019bev,JointPhysicsAnalysisCenter:2023qgg,Blask:2025jua}. Addtionally, the near-threshold production has also been studied using holographic QCD, QCD sum rules, the operator-product expansion, and phenomenological models~\cite{Mamo:2019mka,Mamo:2021krl,Gryniuk:2016mpk,Kharzeev:2021qkd,Sun:2021pyw,Sun:2021gmi}. 

The experimental status in both the near-threshold and high-energy regimes and the prospects for promoting heavy quarkonia to a standard element of global GPD extractions are discussed in Sec.~\ref{sec:5_exp_and_pheno}.

\subsection{Hard exclusive $2\to3$ processes}

\begin{figure}[t]
    \centering
    \includegraphics[width=0.483\textwidth]{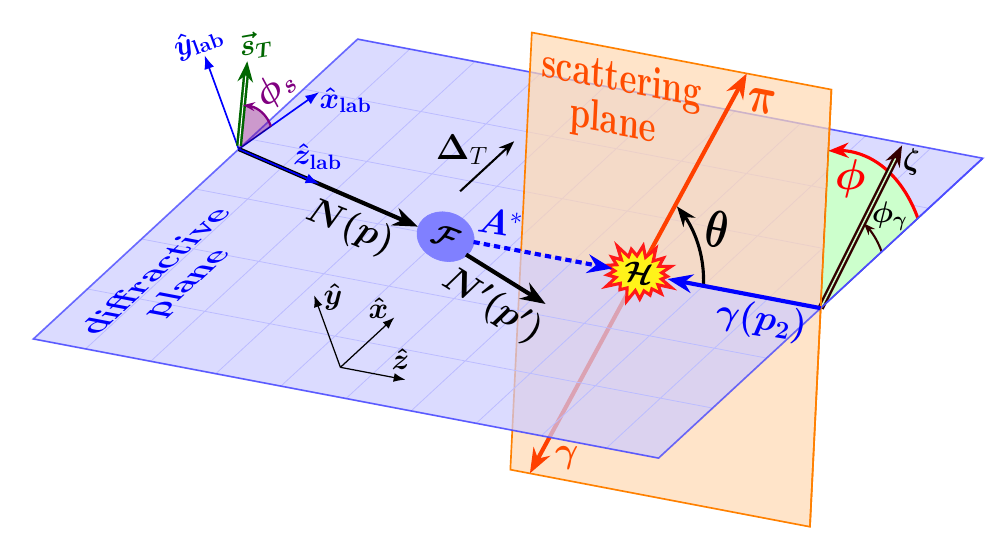}
    \caption{\raggedright Reference frame for the process $\gamma+p\to\gamma+p+\pi$, i.e., exclusive photoproduction of a photon-pion pair off the proton, suitable for describing $2\to3$ processes in general. From~\textcite{Qiu:2023mrm}.}
    \label{fig:sec3:SDHEP}
\end{figure}

The reactions considered so far are all of the $2\to2$ type, excluding the lepton from which the initial photon is emitted. A theoretically distinct, but more complex, class of reactions is provided by hard exclusive $2\to3$ processes, in which a diffractive scattering off the nucleon is accompanied by the exclusive production of a two-particle system of large invariant mass,
\begin{align}
A(p_A)+N(P)\longrightarrow X_1(k_1)+X_2(k_2)+N'(P')\, ,
\label{eq:SDHEPprocess}
\end{align}
with the pair $X_1$ and $X_2$ recoiling at large transverse momentum $q_T\gg\Lambda_{\rm QCD}$ compared to the final state nucleon. Here we have adopted a new frame for convenience where the momenta $P$ and $P'$ define the longitudinal plane (see Fig.~\ref{fig:sec3:SDHEP} for details). The incident particle $A$ and the produced pair may be photons (either real or virtual), mesons, or lepton pairs; representative cases are the photoproduction of a photon pair $\gamma N\to\gamma\,\gamma\,N'$, of a photon--meson pair $\gamma N\to\gamma\,M\,N'$, and the meson-induced reactions $\pi^\pm N\to\gamma\gamma\,N'$ and $\pi^\pm N\to\gamma\,\ell^+\ell^-\,N'$, with the last three also revealing the transition GPDs of an $N\to N'$ system~\cite{Pedrak:2017cpp,Duplancic:2022ffo,Duplancic:2018bum,Duplancic:2023kwe,Grocholski:2022rqj}. These exclusive reactions have been unified under the notion of a single-diffractive hard exclusive process (SDHEP)~\cite{Qiu:2022bpq,Qiu:2023mrm,Qiu:2024mny}: because of the off-forward kinematics, one nucleon is diffracted through a $t$-channel exchange probing the corresponding GPDs, while the remaining large-momentum-transfer of the $2\to2$ subprocess\footnote{e.g., $\gamma P\to\gamma M$ with $P$ being the appropriate parton state inside $N$ for producing the final state meson $M$.} sets the hard $q_T$ scale,
\begin{align}
    |k_{1T}| \sim |k_{2T}| = q_T \gg \sqrt{-t}\,, M\, , \Lambda_{\rm QCD}
\end{align}
so that the soft and hard dynamics are cleanly separated in rapidity. %
These reactions also permit leading-twist access to quark transversity (i.e., chiral-odd) GPDs. The Born amplitude for $\gamma_L^*p\to\rho_L^0\rho_T^+n$ with a large rapidity gap between the mesons was calculated in~\cite{Ivanov:2002jj}, connecting the chiral-odd quark GPDs to the chiral-odd transverse-$\rho$ DA. \textcite{Boussarie:2016qop} studied $\gamma N\to\gamma\rho^0N$ at large photon--meson invariant mass, where longitudinal and transverse $\rho$ polarizations select chiral-even and chiral-odd quark GPDs, respectively, and estimated LO cross sections for JLab 12~GeV kinematics.

An all-order LP factorization theorem was consequently established for amplitudes of several special SDHEP channels~\cite{Qiu:2022bpq,Qiu:2022pla}, taking the schematic form of a double convolution for the case of photon-meson production amplitude $\gamma N\to\gamma M N'$,
\begin{align}
\mathcal{A}_{\rm LP} (\xi,t,\theta,q_T^2)=\sum_{a,b}\int_{-1}^{1}\!\frac{dx}{\xi}&\int_{0}^{1}\!dz\;
 F_a(x,\xi,t,\mu)\,\label{eq:SDHEPfact}
\\
\times
&C_{ab}\!\left(x,\xi,z,\theta;\frac{q_T^2}{\mu^2}\right)
\phi_b(z,\mu)\, ,\notag
\end{align}
where $F_a$ is the GPD-defining matrix element in Eqs.~\eqref{def:HEq}--\eqref{def:HEgT}, and $\phi_b$ is the twist-two DA of the final state meson $M$ and $a,b$ run over all relevant parton species. In case of a double photon production, $\phi_b$ is absent, $C_{ab}$ simplifies into $C_{a}(x,\xi,\theta;\frac{q_T^2}{\mu^2})$, and Eq.~\eqref{eq:SDHEPfact} becomes a single convolution. %
While both DDVCS (see Sec.~\ref{sec:3_DDVCS}) and the SDHEPs enhance sensitivity to the $x$-dependence of GPDs over a wider kinematic domain, the latter provides a structurally distinct and complementary handle to DDVCS. %
 More specifically, the SDHEPs involve an independent, continuously measurable variable $\theta$ (see Fig.~\ref{fig:sec3:SDHEP}) beyond the invariant pair mass $M_{12}^2=(k_1+k_2)^2$.
 The difficulties in extracting GPDs from SDHEPs as an inverse problem are discussed quantitatively in Sec.~\ref{sec:5_exp_and_pheno}.

This factorization is not, however, universal across all partonic channels, and one prominent counterexample has recently been established. It was shown~\cite{Nabeebaccus:2023rzr,Nabeebaccus:2024mia} that the exclusive nucleon-photoproduction of a $\pi^0\gamma$ pair with large invariant mass %
breaks the collinear factorization at LP in the gluonic $t$-channel exchange. Assuming naive factorization in $\gamma N\to\gamma \pi^0 N$, the leading-order convolution already develops an endpoint-like divergence as a consequence of a Glauber pinch absent in DVCS, TCS, DDVCS and DVMP, where a (Glauber) gluon exchanged between a soft spectator of the pion and a collinear spectator of the nucleon is trapped in the Glauber region. Namely, both lightcone components of the Glauber gluon's momentum are pinched at the GPD singular points $x=\pm\xi$ that separate the GPD's PDF and DA regions, thereby obstructing the usual contour deformation away from these problematic points. The effect is unique to channels involving gluon GPDs. The quark channels remain free of Glauber pinches, so the factorization proof stands there. For example, $\pi^\pm\gamma$ and $\rho^{0,\pm}\gamma$ photoproduction are unaffected where gluon GPDs are absent. These channels, including $\gamma\eta$ production, have been phenomenologically studied at LO, showing sensitivity to the shape of the pion DA and gluonic content of the $\eta$ meson~\cite{Crnkovic:2025man}. By contrast, the closely related $\pi^0 N\to\gamma\gamma\,N$ reaction inherits the same gluon-channel obstruction. We note that the Glauber analysis is inherently process dependent, calling for more dedicated leading-region analyses of other gluon-GPD-related processes.

The resulting picture clarifies the LP factorization for families of SDHEPs with enhanced sensitivity to the $x$-dependence of GPDs. For the quark-driven channels, which incorporate a variety of photon--meson and dilepton final state production processes, the factorization is solid with the diphoton production already computed to NLO~\cite{Grocholski:2022rqj}. 
For the $\gamma N\to\gamma\pi^0N$ process, the Glauber
obstruction in the gluon GPD channel breaks the standard collinear factorization %
rather than causing a mere endpoint subtlety, and whether a modified factorization formula involving additional nonperturbative functions/effects beyond GPDs and DAs can restore predictability remains an open question. %
Exclusive dijet electroproduction $\gamma^* N\to q\bar q N, ggN$, where the $q\bar q$ quark pair or $gg$ gluon pair materializes into di-jets, provides another probe of GPDs when $Q^2$ and the jet transverse momentum are large, i.e., $Q^2,q_T^2\gg |t|,\Lambda_{\rm QCD}^2$~\cite{Braun:2005rg}. The LO calculation for the quark-pair channel follows collinear factorization with the subsequent phenomenological study~\cite{Chall:2026oes} showing increased sensitivity to valence GPDs when the nucleon target has a large fractional longitudinal-momentum loss. The recently updated LO analysis further included helicity GPDs~\cite{Pang:2026lsr}, the EM contributions from elastic FFs, and the gluon-pair dijet channel, which couples to the $C$-odd quark GPDs. The phenomenological analysis based on EIC kinematics shows an enhanced valence-quark contribution and a comparable cross-section between $u\bar u$ and gluon jets.

\section{GPDs and Their Moments from Lattice QCD}
\label{sec:4_lattice_qcd}
Lattice QCD
(LQCD), formulated by Wilson as a gauge theory on a discrete
Euclidean spacetime lattice, provides a first-principles framework for
studying nonperturbative QCD quantities~\cite{Wilson:1974sk}. Over the past decades, LQCD has enabled increasingly precise determinations of a broad range of hadronic observables, as summarized in the recent FLAG review~\cite{FlavourLatticeAveragingGroupFLAG:2024oxs}.  

GPDs are defined through light-cone correlations and therefore cannot
be accessed directly in Euclidean lattice QCD. Their Mellin moments, however, are given by matrix elements of local
operators and can therefore be calculated directly on the lattice. The corresponding generalized form factors also carry physical information
in their own right, as summarized in
Sec.~\ref{sec:2_gpd_basic}. Traditional lattice studies of GPDs have
therefore focused primarily on their Mellin moments and the associated generalized form factors. Under suitable regularity conditions~\cite{Zhang:2024djl}, the complete infinite sequence of Mellin moments determines the underlying GPD, for example through a Mellin--Barnes integral~\cite{Mueller:2005ed}. A finite
number of moments constrains its shape but does not determine it uniquely. 

On the lattice, however, reaching higher moments is particularly difficult for GPDs. Each additional moment requires more covariant
derivatives and introduces more independent generalized form factors.
Operator mixing, renormalization, and the numerical separation of these
form factors consequently become more complicated as the moment order increases. Lattice calculations using local operators have therefore focused mainly on the lowest few moments, with recent attempts to extend such calculations to the fourth moment~\cite{Alexandrou:2026tjs}. A few lower moments, however, remain insufficient for a model-independent reconstruction of the $x$-dependent GPD.

To overcome this limitation, one of the present authors proposed that light-cone distributions could be accessed from spatial correlations evaluated at equal Euclidean time in a hadron carrying a large momentum, an idea
subsequently formulated within large-momentum effective theory (LaMET)~\cite{Ji:2013dva,Ji:2014gla}. Several other approaches based on short-distance OPE have also been developed to fit models of PDFs with lattice data.  We refer to~\textcite{Cichy:2018mum,Lin:2025hka,Cichy:2026xrh,Ji:2022ezo} for more detailed discussions. Their ability to obtain
\(x\)-dependent GPDs, however, remains limited by the need for additional modelling. On the other hand, LaMET provides direct access to the \(x\) dependence and has so far been applied most widely to the direct calculations of GPDs.

In this section, we first review traditional lattice calculations of
GPD moments and form factors, then discuss the LaMET formulation and
its numerical results. We also give a brief discussion of other approaches to GPDs from lattice data.

\subsection{Lattice calculations of GPD moments}

The lowest GPD moments encode some of the most important aspects of hadron structure: how momentum and angular momentum are shared among quarks
and gluons, and how energy, momentum current, and mass are distributed inside a hadron. Here we first discuss the
extraction of GPD moments and generalized form factors and the
associated lattice systematics. We then review results for nucleon momentum, angular momentum, and mass distributions and radii, followed by recent attempts on
higher Mellin moments.

\subsubsection{From lattice matrix elements to physical form factors}
\label{sect4a1}
The road from bare lattice matrix elements to physical form factors is
convoluted and demanding. The local operators must be
renormalized, including all mixing allowed by the lattice symmetries,
while a suitable strategy is needed to disentangle the
form factors entering the moments. Excited-state contamination, discretization,
finite-volume, and unphysical pion-mass effects must also be controlled before
the resulting form factors can be interpreted as physical ones.

At finite lattice spacing, the Euclidean rotational symmetry \(O(4)\) is
reduced to the hypercubic group \(H(4)\). Consequently, a symmetric
traceless continuum operator generally decomposes into several
\(H(4)\) irreducible representations, enlarging the set of operators
allowed to mix under renormalization~\cite{Gockeler:1996mu}.
As the number of covariant derivatives increases, operator mixing becomes more complicated and can involve lower-dimensional operators with power-divergent coefficients~\cite{Capitani:2002mp}. For sufficiently high moments, such mixing
cannot in general be avoided by a suitable choice of lattice
representation. For moments higher than the fourth order, such mixing cannot in general be avoided by a suitable choice of lattice representation. For nonforward matrix elements, the operator basis must
further include total-derivative operators
~\cite{Gockeler:2006nb}. The resulting operator basis is commonly renormalized
nonperturbatively in an intermediate momentum-subtraction scheme and
then converted perturbatively to the \(\overline{\mathrm{MS}}\) scheme
~\cite{Martinelli:1994ty,Chetyrkin:1999pq,Gockeler:2010yr}. Gradient flow offers an alternative way to avoid the hypercubic
operator-mixing problem by taking the continuum limit at finite flow
time~\cite{Luscher:2010iy,Luscher:2011bx,Luscher:2013cpa}. 

Flavor-singlet moments pose an additional challenge. Beyond the well-known mixing between singlet-quark and gluon operators,
they require the lattice calculation of disconnected diagrams, in which
the operator insertion forms a closed quark loop that is not contracted
with the external hadron states. These contributions are considerably more difficult to compute and have
a poorer signal-to-noise ratio than the connected ones. Despite the many techniques developed to
reduce this noise~\cite{Dong:1993pk,Foley:2005ac,Bali:2009hu}, disconnected matrix elements remain
difficult to determine with comparable precision to connected ones. Gluon matrix elements
suffer from similarly poor statistical signals, for which 
smearing techniques and Wilson flow are commonly used to improve the signal~\cite{Albanese:1987ds,Hasenfratz:2001hp,Morningstar:2003gk,Luscher:2010iy}. The isovector, i.e., flavor-nonsinglet combination \(u-d\), which is free from disconnected diagrams and quark--gluon mixing in the isospin-symmetric limit, has therefore
been studied most extensively.

The number of generalized form factors grows with the moment order,
requiring an increasing number of independent operator and momentum
configurations for their separation. Techniques based on overdetermined
systems have been widely used for their extraction
~\cite{Hagler:2003jd}. At higher moments, however, operator choice entails a tradeoff between
renormalization and form-factor separation. Restricting to a smaller operator 
basis to reduce mixing can leave fewer independent kinematic
constraints, whereas a larger basis facilitates the extraction at the
cost of more complicated mixing. For particular choices of operators and kinematics, the
resulting linear system can become rank deficient.

Beyond these operator-level issues, reliable form factors require
control over discretization, finite-volume, quark-mass, and
excited-state effects
~\cite{FlavourLatticeAveragingGroupFLAG:2024oxs}. Discretization and
finite-volume effects are studied using multiple lattice spacings and
volumes. Lighter pion masses worsen the signal-to-noise ratio of nucleon
correlation functions~\cite{Parisi:1983ae,Lepage:1989hd}, so early lattice calculations were commonly
performed at heavier pion masses and extrapolated to the physical point. With increasing computational
resources, calculations at or near the physical pion mass have become
more common. At the same time, physical low-lying \(N\pi\) states are
closer to the nucleon ground state, making excited-state contamination
more important~\cite{Tiburzi:2009zp,Bar:2016jof}. A systematic control of excited-state contamination requires sufficiently large source--sink separations or variational methods based on the generalized eigenvalue problem (GEVP)~\cite{Blossier:2009kd,Luscher:1990ck}. In practice, most calculations employ multiple source--sink separations together with fitting strategies such as multistate fits or the summation method, although the extent to which excited-state contamination is fully controlled by these approaches remains under investigation.

Beginning in the early 2000s, lattice calculations of GPD moments
have progressively improved their control of the main systematic
uncertainties. Calculations spanning multiple lattice spacings,
volumes, and pion masses enabled increasingly controlled continuum,
infinite-volume, and physical-mass extrapolations
~\cite{Alexandrou:2011nr}. Later studies placed greater emphasis on
excited-state contamination~\cite{Bali:2018zgl,Alexandrou:2020sml},
while disconnected contributions and quark--gluon mixing were
incorporated in increasingly complete calculations
~\cite{Hackett:2023rif}. Physical-point continuum extrapolations have
also been achieved for selected generalized form factors
~\cite{Alexandrou:2022dtc,Alexandrou:2026oks}. A calculation
incorporating physical-mass, continuum, finite-volume, excited-state,
disconnected, and gluonic effects within a common analysis has also
been carried out~\cite{Yang:2018nqn}, although achieving this level of
systematic control remains computationally demanding. The rest of this
subsection presents representative lattice results organized according
to the physical information encoded in the GPD moments.

\subsubsection{Momentum and angular momentum sum rules}

The EMT form factors determine how the hadron momentum and angular
momentum are distributed among quarks and gluons. The forward values
\(A_{q,g}(0)=\langle x\rangle_{q,g}\) give the momentum
fractions carried by each parton species, while
\(J_{q,g}=\tfrac{1}{2}[A_{q,g}(0)+B_{q,g}(0)]\) quantifies the angular momentum partition.
For quarks, the kinetic orbital angular momentum in a longitudinally polarized nucleon follows from
\(L_q^{\rm kin}=J_q-\Delta q/2\), with
\(\Delta q = \widetilde A_q^{(1,0)}(0)\), as reviewed in
Sec.~\ref{subsec:2_spin_t3GPD}. In contrast to \(A\),
\(B\) is obtained from the (Pauli) GPD \(E\), which remains much less
constrained phenomenologically than \(H\). Lattice calculations can
therefore provide an important first-principles constraint on
\(B_{q,g}(0)\) and the resulting angular-momentum decomposition. Note that \(B_{q,g}(0)\) cannot be determined directly from a forward
matrix element and is instead obtained by fitting the nonzero-\(t\)
data and extrapolating to \(t=0\), commonly using a \(p\)-pole form,
\(F(t)=F(0)(1-t/m_p^2)^{-p}\), or a truncated \(z\) expansion,
\(F(t)=\sum_k a_k z(t)^k\).

Early lattice studies reported estimates of the \(u\)- and \(d\)-quark
contributions to the proton angular momentum
~\cite{Mathur:1999uf,Gockeler:2003jfa,LHPC:2007blg,Alexandrou:2011nr,Alexandrou:2013joa}. The flavor-resolved results
consistently found that \(J_u\) dominates, while \(J_d\) is small and
compatible with zero. The \(u\)- and \(d\)-quark orbital contributions showed a strong
cancellation, leaving their sum consistent with zero.
Most of these calculations neglected disconnected contributions,
although the available estimates generally found them to be small
~\cite{Mathur:1999uf}. Later isovector studies refined the extraction of these generalized
form factors through improved fitting strategies and detailed analyses
of excited-state and other lattice systematics
~\cite{Bali:2018zgl,Alexandrou:2019ali}.

A complete decomposition of the proton angular momentum requires
disconnected quark contributions as well as the gluon contribution.
An early quenched calculation by the \(\chi\)QCD Collaboration provided the first such determination
~\cite{Deka:2013zha}, followed by dynamical calculations at or near the
physical pion mass
~\cite{Alexandrou:2017oeh,Alexandrou:2020sml,Wang:2021vqy}. Across these studies, gluons were found to carry a substantial part of
the proton angular momentum, while the characteristic pattern
\(L_u<0\) and \(L_d>0\) persisted. These calculations were, however,
all based on a single lattice spacing.

\begin{figure}[t]
    \centering
\includegraphics[width=\columnwidth]{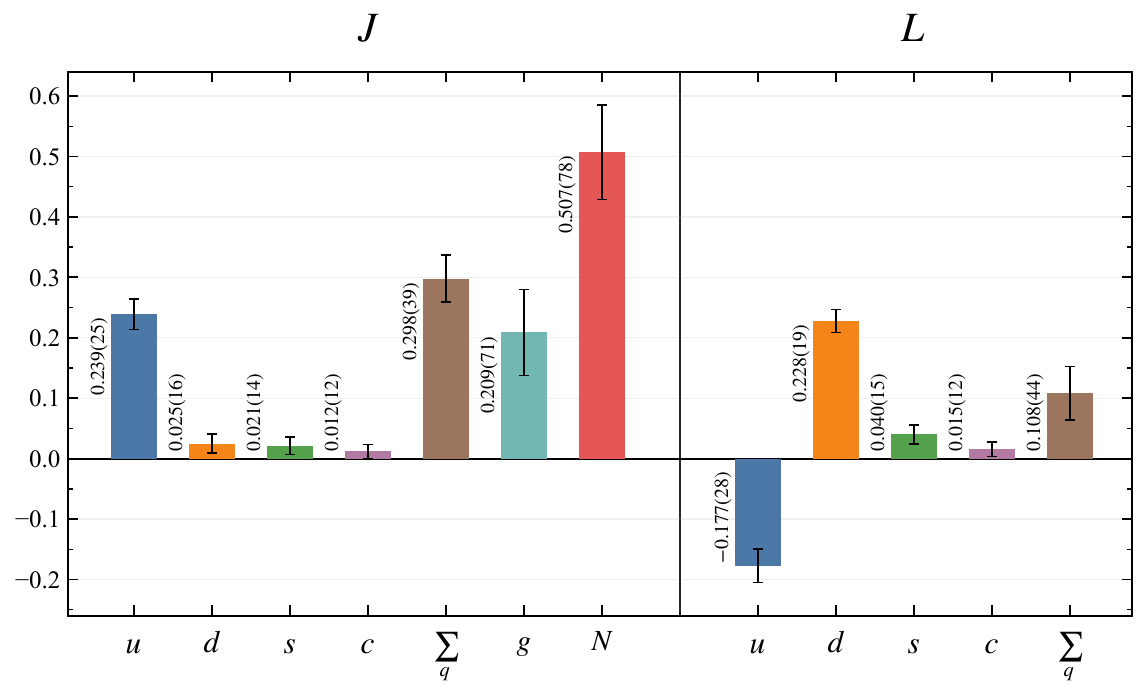}
\caption{
Decomposition of the proton angular momentum \(J\) and quark orbital
angular momentum \(L\) at the physical pion mass and in the continuum
limit, in the \(\overline{\mathrm{MS}}\) scheme at
\(\mu=2~\mathrm{GeV}\).
Data are taken from~\textcite{Alexandrou:2026oks}, with the error bars
showing the total uncertainties.}
\label{fig:JL_sumrule}
\end{figure}

The first complete decomposition of the proton momentum and angular
momentum was obtained by the Extended Twisted Mass Collaboration (ETMC)
at the physical point and in the continuum limit
~\cite{Alexandrou:2026oks}. The calculation included the \(u,d,s,c\) quark and gluon
contributions, with disconnected diagrams and
nonperturbative singlet--gluon mixing taken into account. Both the momentum and angular-momentum sum rules are satisfied
within uncertainties, with the corresponding total \(B(0)\) consistent
with zero. As summarized in
Fig.~\ref{fig:JL_sumrule}, quarks (gluons) carry about
\(60\%\) ($40\%$) of the proton's angular momentum.
The negative \(L_u\) and positive \(L_d\) largely cancel, in line with
the tendency already seen in earlier calculations. This is the most complete lattice determination of the proton momentum
and angular-momentum decomposition to date.

\subsubsection{Mass and momentum current decomposition}

Beyond the momentum and angular-momentum sum rules, the
\(t\) dependence of the EMT form factors probes the spatial
distribution of energy and momentum current inside the hadron.
Forward matrix elements of the EMT, including its trace,  provide
the basis for decompositions of the hadron mass~\cite{Ji:1994av}. The trace anomaly 
is a key piece of the mass components which arises from quantum anomalous breaking of classical scale symmetry, 
providing the necessary strong interaction scale for quark and gluon kinetic energy contributions~\cite{Ji:2021mtz}.  

\begin{figure}[b]
    \centering
\includegraphics[width=\columnwidth]{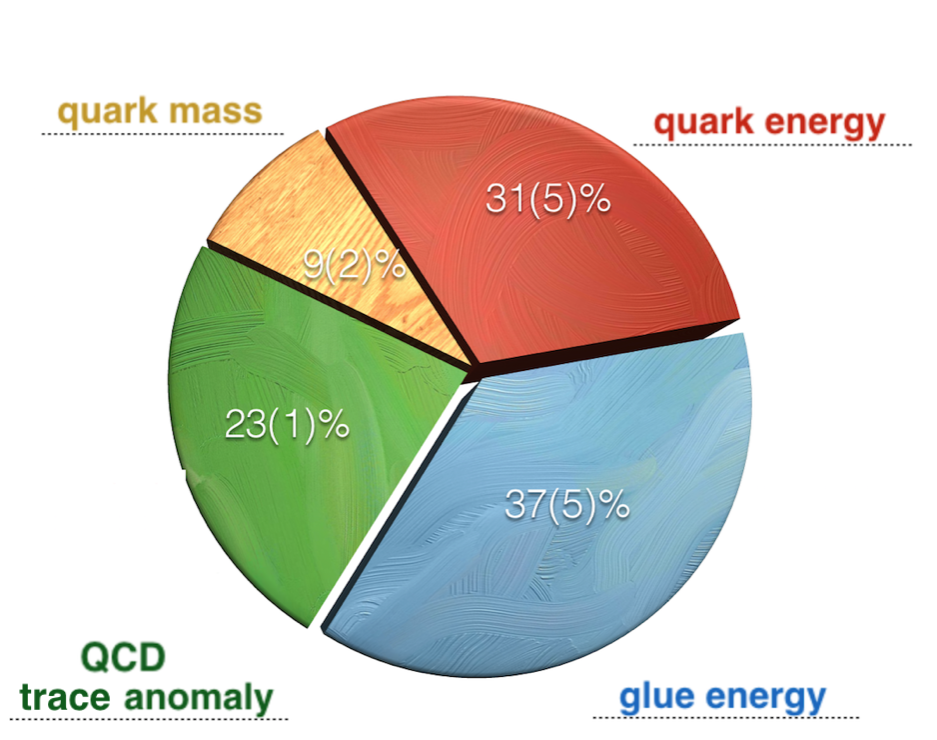}
\caption{\label{fig:proton_mass_decomposition}
Decomposition of the proton mass into the quark-mass, quark-energy,
gluon-field-energy, and QCD trace-anomaly contributions, which account
for \(9(2)\%\), \(31(5)\%\), \(37(5)\%\), and \(23(1)\%\), respectively.
The quark- and gluon-energy contributions are renormalized in the
\(\overline{\mathrm{MS}}\) scheme at \(\mu=2~\mathrm{GeV}\).
From~\textcite{Yang:2017erf}.}
\end{figure}

A pioneering lattice study of the proton mass decomposition was done
 with extrapolations to the physical pion mass and continuum limit~\cite{Yang:2018nqn}. The quark-energy, gluon-energy,
quark-mass, and anomalous contributions were found to be about
$31\%$, $37\%$, $9\%$, and $23\%$, respectively, as shown in Fig.~\ref{fig:proton_mass_decomposition}. However, their calculation required a renormalization of the total momentum fraction
and did not include an explicit calculation of the trace term. Their subsequent work provided the first direct lattice calculation of the
quark and gluon trace-anomaly, providing an explicit test of the mass sum rule~\cite{He:2021bof}.
The spatial structure of the trace has also been investigated through
the gluonic trace-anomaly density~\cite{He:2021bof,Wang:2024lrm}. 
A more recent calculation uses
gradient flow method to obtain the trace anomaly contribution~\cite{Bollweg:2026nnr}. The total scalar energy in the trace term, normalized to hadron masses, has been studied for light-, strange-, and charm baryons, finding a gluonic trace-anomaly contribution of roughly $1~\mathrm{GeV}$ with
only weak flavor dependence~\cite{Hu:2024mas}.

Several alternative schemes have been proposed
based on how the anomaly contribution to the mass is included in the original mass sum rule~\cite{Lorce:2017xzd,Hatta:2018sqd,Metz:2020vxd} . A discussion about these approaches can be found in~\cite{Yang:2026wzi}. 

\begin{figure}[t]
    \centering
\includegraphics[width=\columnwidth]{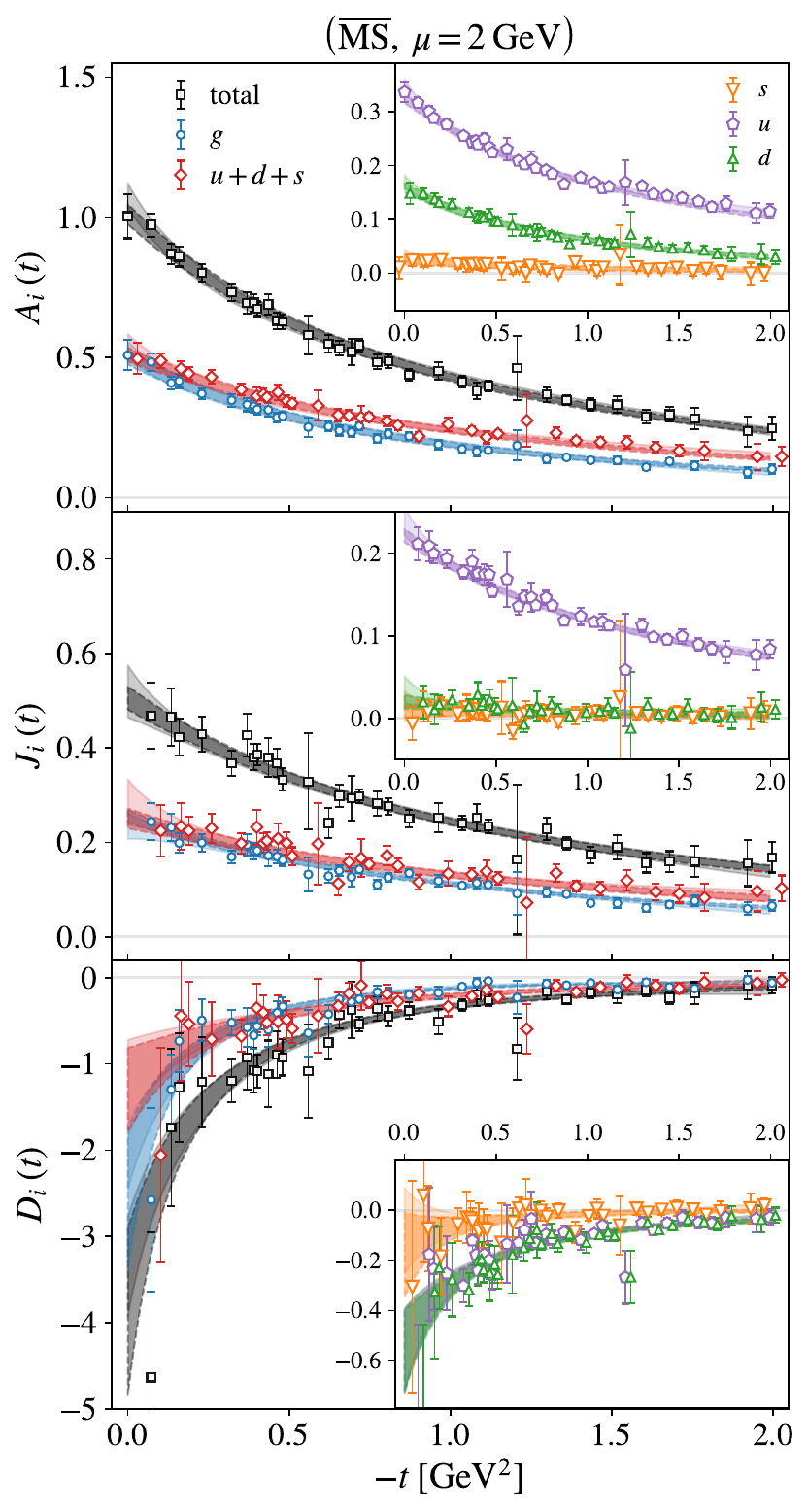}
\caption{\label{fig:proton_GFF}  The proton EMT form factors \(A(t)\), \(J(t)=[A(t)+B(t)]/2\), and \(D(t)=4C(t)\) as functions of \(-t\). The main panels show the total, gluon, and total-quark contributions, while the insets display the separate \(u\)-, \(d\)-, and \(s\)-quark contributions. Dark and light bands denote dipole and \(z\)-expansion fits, respectively. From~\textcite{Hackett:2023rif}.}
\end{figure}

A number of lattice studies went beyond the mass sum rule to investigate EMT form factors and the spatial energy distribution as well as the mass  radii. The total mass
radius is defined from the mass form factor \(G_m(t)\) introduced in Eq.~\eqref{def:Gmt}.
Early lattice calculations determined the quark generalized form
factors \(A_q(t)\), \(B_q(t)\), and \(C_q(t)\)
~\cite{LHPC:2003aa,Gockeler:2003jfa,LHPC:2007blg,
Bali:2018zgl,Alexandrou:2019ali}.
The gluon generalized form factors of the nucleon and pion were first calculated in~\textcite{Shanahan:2018nnv}, and were subsequently extended to
hadrons of different spin~\cite{Pefkou:2021fni}.
This was followed by a near-physical calculation of the pion with
a full quark--gluon decomposition~\cite{Hackett:2023nkr}.
Most recently, the individual \(u\)-, \(d\)-, \(s\)-quark and gluon
form factors of the proton were determined near the physical pion
mass~\cite{Hackett:2023rif}, providing the first flavor-resolved
determination of its mass and ``mechanical'' radii. The extracted \(A(t)\), \(J(t)\), and \(D(t)\) form factors are shown
in Fig.~\ref{fig:proton_GFF}. The gluon mass radius
was found to be larger than the corresponding quark radius, indicating
a more extended gluon distribution in the proton.  

Lattice determined the quark and gluon $C/D$ form factors have been used to construct distributions of momentum current density,
conventionally referred to as ``pressure and shear''
~\cite{Shanahan:2018pib,Shanahan:2018nnv}. The gluon contribution was found to dominate the proton shear
distribution. As discussed in
Sec.~II.D, however, such distributions are not literally ``pressure and shear''.

\subsubsection{Higher moments and construction of GPDs through ansatz}

\begin{figure}[t]
\centering
\includegraphics[width=\columnwidth]{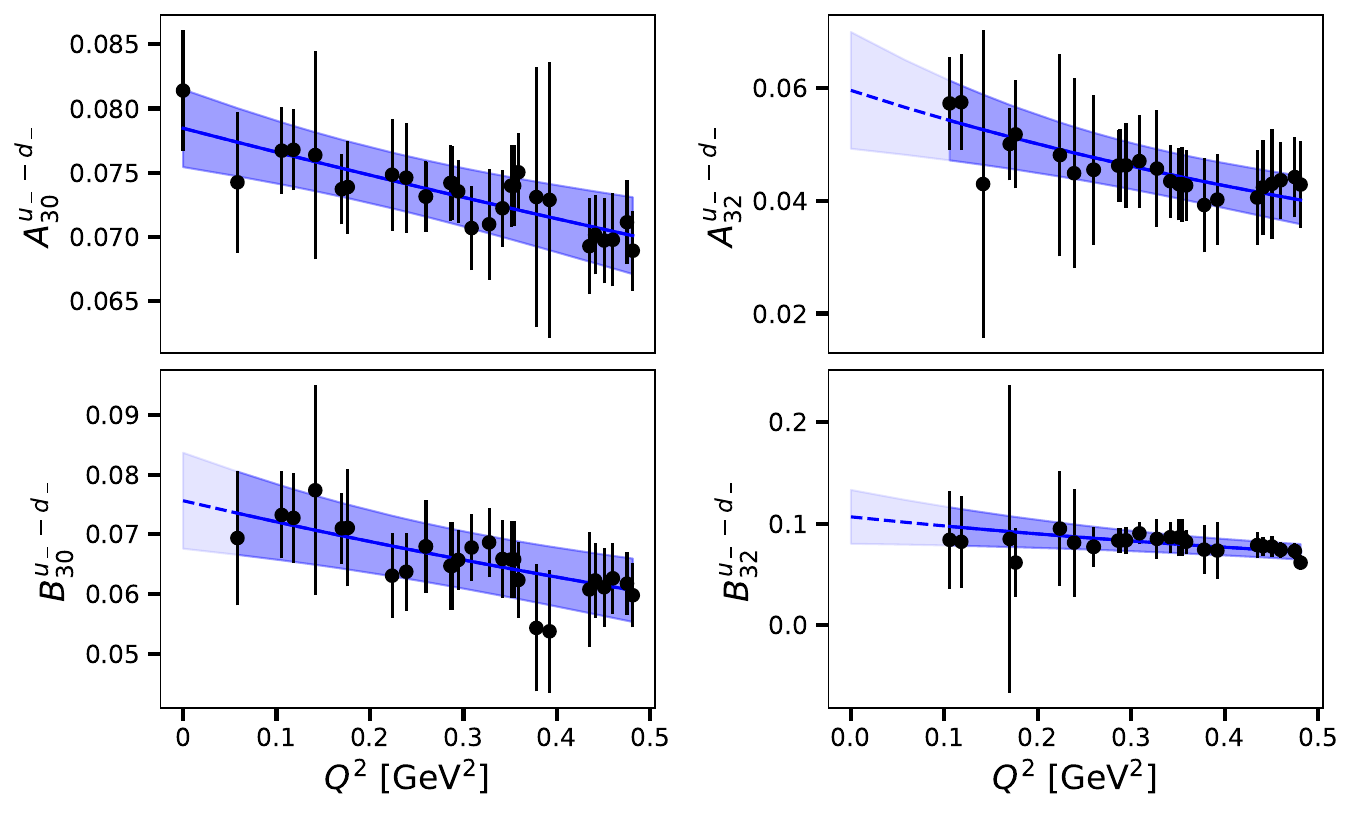}
\caption{\label{fig:higher_moments}
Isovector generalized form factors
$A^{(3,0)}_{u-d}(t)$, $A^{(3,2)}_{u-d}(t)$, $B^{(3,0)}_{u-d}(t)$, and $B^{(3,2)}_{u-d}(t)$
entering the third Mellin moments of the unpolarized nucleon GPDs.
The dashed curves and shaded bands show dipole fits and their
uncertainties, respectively. From~\textcite{Alexandrou:2026tjs}.}
\end{figure}

As discussed in Sec.~\ref{sect4a1}, direct calculations of higher GPD moments with
local operators become increasingly challenging. %
Early calculations by the LHPC and SESAM Collaborations reached the
third unpolarized quark Mellin moment at relatively heavy pion masses
~\cite{LHPC:2003aa,LHPC:2007blg}. Nearly two decades later, calculations at the physical pion mass
reached the third and fourth Mellin moments of the unpolarized quark
GPDs~\cite{Alexandrou:2026tjs}. 
The form factors entering the third moment are shown in
Fig.~\ref{fig:higher_moments}. Among the fourth-moment form factors,
only \(A^{(4,0)}(t)\) was resolved with statistical significance,
while the others were consistent with zero within the achieved
precision. Quark--gluon mixing was also neglected, leaving the
isovector combinations as the cleanest channels.

Recent studies using gradient flow have determined pion PDF moments up
to the sixth order, providing a promising demonstration of this
method to higher moments
~\cite{Francis:2025rya,Francis:2025pgf}. At finite flow time, gradient
flow provides a physical ultraviolet regulator that allows the
continuum limit to be taken prior to the UV renormalization, apart from the multiplicative
renormalization of the flowed quark fields. The restoration of \(O(4)\) symmetry in the continuum limit removes the \(H(4)\)
operator mixing, thereby removing a
major obstacle to higher-moment calculations
~\cite{Shindler:2023xpd}. Its
application to higher GPD moments, however, remains to be explored.

Higher moments can in principle also be accessed through short-distance factorization (SDF) of Euclidean correlation functions, avoiding the explicit construction of higher-derivative local operators. One realization is based on the off-forward Compton amplitude. Using a second-order
Feynman--Hellmann construction, the CSSM/QCDSF/UKQCD Collaborations
calculated the amplitude directly on the lattice and extracted GPD
moments through SDF, providing the first
lattice determination of the fourth moment at zero skewness
~\cite{CSSMQCDSFUKQCD:2021lkf}. A complementary strategy applies SDF to nonlocal quark or gluon bilinear matrix elements~\cite{Karpie:2018zaz}. Its extension to quasi-GPD matrix elements has
yielded unpolarized and axial-vector nucleon moments up to fifth order
~\cite{Bhattacharya:2023ays,Bhattacharya:2024wtg}, and more recently has been extended to the pion at nonzero skewness. Simultaneous fits constrained by polynomiality separated \(A^{(3,2)}(t)\), \(A^{(5,2)}(t)\), and \(A^{(5,4)}(t)\), providing the first lattice access to the skewness dependence of higher GPD moments~\cite{Gao:2025inf}.

The same SDF framework has also been used to access \(x\)-dependent GPDs~\cite{Radyushkin:2019owq}. It starts from the same nonlocal
matrix elements as the quasi-GPD used in LaMET, which will be discussed in the next
subsection. The distinction lies in
how the expansion is made in coordinate or momentum spaces, which are not equivalent~\cite{Ji:2022ezo}.
In SDF, the renormalized matrix element is expanded at short distances
in terms of Mellin moments, whereas LaMET exploits the full coordinate-space
information to construct a momentum-space quasi-distribution and subsequently performs
a large-momentum expansion to obtain the light-cone distribution at specific $x$ and $\xi$.
These two approaches expose different power corrections. The LaMET
expansion is controlled by the hadron momentum, whereas short-distance factorization requires
\(z^2\Lambda_{\mathrm{QCD}}^2\ll1\). 
SDF provides direct access to finite-range coordinate-space correlations, or equivalently, a few lower-order Mellin moments. A continuous GPD in \(x\), cannot be obtained from a correlation with a finite range of \(\lambda\), without introducing a model ansatz. The first such reconstruction obtained the zero-skewness isovector
nucleon GPDs \(H\) and \(E\) within pseudo-GPDs,
fitting the \(x\) dependence independently at each value of the momentum
transfer~\cite{Bhattacharya:2024qpp}. More recently, the HadStruc Collaboration performed a joint fitting of the \(x\)-, \(\xi\)-, and \(t\)-dependence of the isovector GPDs \(H\) and \(E\), using double distributions and multidimensional Gaussian-process regression (GPR)~\cite{Dutrieux:2026grg}.

\subsection{The LaMET route to \(x\)-dependent GPDs}

The Large-Momentum Effective Theory (LaMET) relates a quasi-distribution defined from equal-time correlations in a hadron carrying a large momentum to the corresponding light-cone distribution~\cite{Ji:2013dva,Ji:2014gla}. At large momentum,
the two share the same infrared physics, while their ultraviolet
difference can be treated perturbatively, allowing the light-cone
matrix elements to be recovered through a large-momentum
expansion~\cite{Ji:2024oka}. A given light-cone observable can be
accessed through different finite-momentum Euclidean quantities,
including conventional gauge-invariant quasi-distributions,
Coulomb-gauge quasi-distributions~\cite{Gao:2023lny}, and suitable
current--current correlators~\cite{Zhang:2026lle}. For GPDs, quasi-GPDs have received the most extensive theoretical
development and numerical study and are therefore the focus below. For a detailed review of LaMET, see \textcite{Ji:2020ect}.

In the following, we first introduce the construction of quasi-GPDs and
their large-momentum expansion, then review recent theoretical
developments, and finally discuss lattice calculations of GPDs within
LaMET.
\subsubsection{Quasi-GPDs and large-momentum expansion}

For quark GPDs, the corresponding equal-time bilocal operator reads
\begin{equation}
\widetilde{\mathcal O}_\Gamma^q(z)
=
\overline{\psi}_q\left(-\frac{z n_z}{2}\right)
\Gamma
W\left(-\frac{z n_z}{2},\frac{z n_z}{2}\right)
\psi_q\left(\frac{z n_z}{2}\right),
\label{eq:qgpd_op}
\end{equation}
where \(z^\mu=z n_z^\mu\), with
\(n_z^\mu=(0,0,0,1)\), and \(W\) is a straight Wilson line along the
\(z\) direction. Its renormalized nonforward matrix element is 
\begin{equation}
\widetilde h_\Gamma^q(z,P',P,\mu)
=
\langle P'|
\widetilde{\mathcal O}_\Gamma^q(z)
|P\rangle_\mu.
\label{eq:qgpd_mat}
\end{equation}
A tilde on \(\widetilde{\mathcal O}\), \(\widetilde h\), and
\(\widetilde F\) denotes a quasi quantity. The conventional notation
\(\widetilde H\) and \(\widetilde E\) for helicity GPDs is retained
and will be distinguished by context.

With \(\bar P=(P'+P)/2\) and \(\Delta=P'-P\), the longitudinal
momentum transfer is described by the quasi-skewness
\begin{equation}
 \widetilde\xi
 =
 -\frac{\Delta^z}{2\bar P^z}
 =
 \xi+
 \mathcal O\left(
 \frac{M^2}{(\bar P^z)^2},
 \frac{-t}{(\bar P^z)^2}
 \right).
 \label{eq:qgpd_skewness}
\end{equation}
For \(\bar{\bm P}_\perp=0\), the requirement that the transverse
momentum transfer be real gives
\begin{equation}
 |\widetilde\xi|
 \leq
 \frac{1}{2|\bar P^z|}
 \sqrt{
 \frac{
 -t\left[(\bar P^z)^2+M^2-t/4\right]
 }{
 M^2-t/4
 }
 } .
 \label{eq:qgpd_skewness_bound}
\end{equation}
Thus, \(\widetilde\xi\), \(t\), and \(\bar P^z\) cannot be chosen  completely 
independently. In the leading-power matching,
\(\widetilde\xi\) may be replaced by \(\xi\), with their difference
included among the finite-momentum corrections.

Early lattice calculations of quasi-GPDs mostly employed the symmetric
frame. A major practical improvement came with the introduction of the
asymmetric frame~\cite{Bhattacharya:2022aob}. These two frames are defined by the three-momenta of the hadron,
\begin{align}
\text{symmetric:}\quad
&\bm P=\bar{\bm P}-\frac{\bm\Delta}{2},
\qquad\qquad\,\,\,
\bm P'=\bar{\bm P}+\frac{\bm\Delta}{2},
\notag\\
\text{asymmetric:}\quad
&\bm P=(-\bm\Delta_\perp,P'^z-\Delta^z),
\,\,
\bm P'=(\bm0_\perp,P'^z).
\label{eq:qgpd_frames}
\end{align}
In the symmetric frame, varying \(t\) generally requires changing both
external momenta. In the asymmetric frame, the same sink momentum ($\bm P'$) can  be used
for several momentum transfers, substantially reducing the
computational cost and providing denser coverage in \(t\). 
While the asymmetric setup provides an efficient way to explore the
\(t\)-dependence, conventional quasi-GPD projections remain frame
dependent at finite momentum, reflecting kinematic power corrections similar to those discussed in
Sec.~\ref{sec:3_kpc}. A frame-independent definition can be
obtained by decomposing the matrix elements into Lorentz-invariant
amplitudes and constructing the quasi-GPD from appropriate combinations
of these amplitudes
~\cite{Bhattacharya:2022aob}.

The nonlocal operator must first be renormalized before Fourier transformation to obtain the quasi-distribution in momentum space. In lattice regularization, the Wilson-line self-energy
induces a linear power divergence in addition to logarithmic
divergences. The operator is nevertheless multiplicatively
renormalizable, with a renormalization factor independent of the
external hadron states~\cite{Ishikawa:2017faj,Ji:2017oey,Green:2017xeu} and is  currently known to three-loops (two-loops) in the quark-antiquark (gluon), respectively~\cite{Braun:2020ymy}. The same
renormalization therefore applies in forward and nonforward kinematics. At finite lattice spacing,
however, reduced symmetries may induce mixing for certain choices of
the Dirac structure~\cite{Green:2017xeu,Constantinou:2017sej}.
For a choice of \(\Gamma\) that avoids such mixing, the renormalized
matrix element can be written as
\begin{equation}
 \widetilde h_{\Gamma}^q(z,P',P,\mu)
 =
 Z_R(z,a,\mu)\,
 e^{\delta m(a)|z|}
 \widetilde h_{\Gamma,\mathrm{bare}}^q(z,P',P,a),
\label{eq:qgpd_renormalization}
\end{equation}
where \(\delta m(a)\) removes the Wilson-line self-energy divergence
and \(Z_R\) accounts for the remaining logarithmic renormalization.

Several renormalization prescriptions have been proposed, including mass subtraction~\cite{Chen:2016fxx}, RI/MOM scheme~\cite{Stewart:2017tvs,Alexandrou:2017huk}, ratio scheme~\cite{Radyushkin:2017cyf}, and the hybrid renormalization scheme~\cite{Ji:2020brr}. Mass subtraction remains sensitive to discretization effects at short distances, while RI/MOM and ratio scheme can introduce unwanted nonperturbative effects at large Wilson-line separations. The hybrid renormalization scheme treats the short distance and long distance regions separately and has been widely adopted in recent high precision LaMET calculations. Its typical implementation combines ratio scheme at short distances with self renormalization at large distances~\cite{LatticePartonLPC:2021gpi}.
Gradient-flow renormalization and matching have been studied for quasi-PDFs and quasi-DAs
~\cite{Monahan:2016bvm,Brambilla:2023vwm,Zhang:2025mer,Zhang:2025npd}.
Since the matching is performed at the operator level, the same
framework can in principle be applied to quasi-GPD matrix elements,
although this has not yet been studied.

The quasi-GPD is obtained
by Fourier transforming the renormalized matrix element.  For the symmetric frame, the quasi-GPD matrix
element can be parametrized as
\begin{equation}
\label{eq:def_qgpd}
\begin{aligned}
 \widetilde F_q
 (x,\widetilde\xi,&t,\bar P^z,\mu)\\
 ={}&
 \frac{1}{2\bar{P}^0}\int_{-\infty}^{\infty}
 \frac{d\lambda}{2\pi}\,
 e^{ix\lambda}\,
 \widetilde h_{\gamma^0}^q
 \left(
 \frac{\lambda}{\bar P^z},
 P',P,\mu
 \right)\\
 ={}&
 \frac{1}{2\bar{P}^0}\bar{U}(P')
 \left[
 \gamma^0\widetilde H_q
 +
 \frac{i\sigma^{0\mu}\Delta_\mu}{2M}\widetilde E_q
 \right]U(P).
\end{aligned}
\end{equation}
Unlike physical light-cone GPDs, the quasi momentum fraction \(x\),
conjugate to \(\lambda\), has support on the full real axis at finite
\(\bar P^z\). 

In lattice calculations, however, the matrix element is available only
at a discrete set of spatial separations and over a finite range, as
the signal-to-noise ratio deteriorates rapidly with increasing
separation.  A direct
truncated Fourier transform can therefore generate unphysical
oscillations in \(x\) and has limited resolution, particularly in the
small-\(x\) region. Various reconstruction methods have been employed to address the problem, including the derivative method~\cite{Lin:2017ani}, the Backus--Gilbert method~\cite{Backus:1968svk}, and the Bayes--Gauss--Fourier transform (BGFT), based on GPR~\cite{Alexandrou:2020tqq}.
However, the asymptotic large-distance behavior of Euclidean
partonic correlators, including quasi-GPD matrix elements, has been
derived, providing theoretically constrained information on the
long-distance tail and a systematically controlled route to the Fourier transformation~\cite{Ji:2020brr,Ji:2026vir}.

For the unpolarized isovector GPDs, for example, the large-momentum
expansion can be written as
\begin{align}
F_{u-d}(&y,\xi,t,\mu)\\
={}&
\int_{-\infty}^{\infty}
\frac{dx}{|x|}\,
C\left(
\frac{y}{x},
\frac{\xi}{x},
\frac{\mu}{x\bar P^z}
\right)
\widetilde F_{u-d}
(x,\xi,t,\bar P^z,\mu)
+\cdots .
\label{eq:qgpd_matching}
\end{align}
Although the quasi-GPD extends outside the physical interval
\(-1\leq x\leq1\), its one-loop collinear singularities coincide with
those of the light-cone GPD in the PDF and DA regions and cancel
in the matching coefficient. At leading power, the coefficient is independent of \(t\)~\cite{Yao:2022vtp}
and reduces to the quasi-PDF matching coefficient at zero skewness, as expected.
The \(t\) dependence then remains entirely in
the nonperturbative GPD, up to finite-momentum corrections.

\subsubsection{Matching, resummation, and power corrections}
\label{sec4B2}

The one-loop matching for nonsinglet unpolarized and helicity
quasi-GPDs was first derived in~\textcite{Ji:2015qla} and subsequently
extended to the transversity case~\cite{Xiong:2015nua}. RI/MOM
matching for the three leading-twist quark GPDs was studied in~\textcite{Liu:2019urm}. The complete NLO matching coefficients were later given in~\textcite{Yao:2022vtp} for all channels, including all regions in the momentum space and uncovered a kinematic region
missed in earlier results due to incomplete analytic continuation. Although the quasi-GPD
extends outside the physical interval \(-1\leq x\leq1\), its one-loop
collinear singularities coincide with those of the light-cone GPD in
the PDF and DA regions and cancel in the matching coefficient.

The nonsinglet coordinate-space matching coefficients for the
unpolarized and helicity quasi-GPDs are identical to the corresponding
quasi-DA coefficient and are known to two-loop order~\cite{Ji:2025mvk}. 
Beyond the nonsinglet quark case, one-loop matching has also been
worked out for leading-twist singlet quark and gluon quasi-GPDs,
including quark--gluon mixing
~\cite{Ma:2022ggj,Ma:2022gty,Yao:2022vtp}.  
A recent prescription
further makes the Fourier transform of the off-diagonal coefficient
\(C_{gq}\) well defined in momentum space~\cite{Ji:2025cbb}.
At zero skewness, since the matching is reduced to the quasi-PDF case, the
available two-loop quasi-PDF matching can be directly used in GPD
analyses~\cite{Yao:2022vtp,Holligan:2023jqh,Ding:2024saz}.

Numerical extractions have nevertheless
remained limited to quark GPDs. Gluon nonlocal operators generally have
poorer signal-to-noise ratios and require the simultaneous treatment
of gluon--quark mixing, and as a result, no \(x\)-dependent gluon GPD has yet been
obtained from lattice QCD.

Renormalization-group resummation (RGR) of the large-momentum expansion has
been implemented in LaMET for both PDFs and DAs~\cite{Su:2022fiu,Holligan:2023rex}. At zero skewness, where the
quasi-GPD matching reduces to the quasi-PDF case, the same treatment
can be applied directly and has been used in lattice GPD calculations
~\cite{Holligan:2023jqh}. At nonzero skewness, however, the matching
contains several partonic scales associated with the incoming and
outgoing partons and with soft radiation. A dedicated threshold
factorization has therefore been developed to resum the corresponding
large logarithms~\cite{Holligan:2025baj}. 

Another issue is the power accuracy of the LaMET expansion, whose
structure can be studied through renormalon analysis; see \textcite{Beneke:1998ui} for a comprehensive review. The subtraction
of the Wilson-line linear divergence carries a leading UV-renormalon
ambiguity of \(\mathcal O(z\Lambda_{\mathrm{QCD}})\), corresponding to
\(\mathcal O(\Lambda_{\mathrm{QCD}}/(x\bar P^z))\) in momentum space.
To remove this ambiguity from the LaMET expansion, leading-renormalon
resummation (LRR) was developed by imposing a common renormalon
prescription in the self-renormalization of the Wilson line and
perturbative matching~\cite{Holligan:2023rex,Zhang:2023bxs}. This removes the
leading UV-renormalon ambiguity from the LaMET expansion.

Independently of the Wilson-line renormalon cancellation, quasi-GPDs are subject to genuine infrared power corrections. Renormalon analysis indicates that these
corrections are generically of order
\(\Lambda_{\mathrm{QCD}}^2/(\bar P^z)^2\), with kinematic
enhancements near the large-\(x\) endpoint and possibly, the crossover points
\(x=\pm\xi\).  In the crossover region,
\(|x\mp\xi|\lesssim\mathcal O(\Lambda_{\mathrm{QCD}}/\bar P^z)\),
it has been suggested that the leading nonperturbative correction changes from
\(\mathcal O(\Lambda_{\mathrm{QCD}}^2/(\bar P^z)^2)\) to
\(\mathcal O(\Lambda_{\mathrm{QCD}}/\bar P^z)\)
~\cite{Braun:2024snf}. However, a recent study shows that the infrared
contribution near \(x=\pm\xi\) is suppressed by a linear factor
\(|x\pm\xi|\), analogous to the power suppression of soft contributions
in DVCS factorization~\cite{Su:2026GPD}.
Kinematic power corrections of order
\(m_N^2/(\bar P^z)^2\) and \((-t)/(\bar P^z)^2\) have also been
studied~\cite{Braun:2026rgj}. 

\begin{table*}[t]
\caption{\label{tab:gpd_lattice_results}
Representative lattice calculations of \(x\)-dependent quark GPDs
within LaMET. All nucleon results refer to the isovector \(u-d\)
combination. Blank entries within an ensemble block share the pion
mass and lattice spacing specified in its first row.}
\centering
\scriptsize
\setlength{\tabcolsep}{3.2pt}
\renewcommand{\arraystretch}{1.15}
\begin{ruledtabular}
\begin{tabular}{
  p{3.25cm}
  p{0.90cm}
  p{2.35cm}
  p{1.75cm}
  p{1.65cm}
  p{4.15cm}
}
GPDs & Hadron & \(\xi\) & \(m_\pi\) & \(a\) & Source \\
\hline

\(H_v^\pi\)
&
\(\pi\)
&
\(0\)
&
\(310~\mathrm{MeV}\)
&
\(0.12~\mathrm{fm}\)
&
\cite{Chen:2019lcm}
\\

\(H_v^\pi\)
&
\(\pi\)
&
\(0\)
&
\(135~\mathrm{MeV}\)
&
\(0.09~\mathrm{fm}\)
&
\cite{Lin:2023gxz}
\\

\(H_v^\pi\)
&
\(\pi\)
&
\(0\)
&
\(300~\mathrm{MeV}\)
&
\(0.04~\mathrm{fm}\)
&
\cite{Ding:2024saz}
\\[2pt]

\(H,E,\widetilde H,\widetilde E\)
&
\(N\)
&
\(0,\ 1/3\)
&
\(260~\mathrm{MeV}\)
&
\(0.093~\mathrm{fm}\)
&
\cite{Alexandrou:2020zbe}
\\

\(H_T,E_T,\widetilde H_T,\widetilde E_T\)
&
\(N\)
&
\(0,\ 1/3\)
&
\(260~\mathrm{MeV}\)
&
\(0.093~\mathrm{fm}\)
&
\cite{Alexandrou:2021bbo}
\\

\(H,E\)
&
\(N\)
&
\(0\)
&
\(260~\mathrm{MeV}\)
&
\(0.093~\mathrm{fm}\)
&
\cite{Bhattacharya:2022aob}
\\

\(\widetilde H\)
&
\(N\)
&
\(0\)
&
\(260~\mathrm{MeV}\)
&
\(0.093~\mathrm{fm}\)
&
\cite{Bhattacharya:2023jsc}
\\

\(H_T,E_T,\widetilde H_T\)
&
\(N\)
&
\(0\)
&
\(260~\mathrm{MeV}\)
&
\(0.093~\mathrm{fm}\)
&
\cite{Bhattacharya:2025yba}
\\

\(H,E\)
&
\(N\)
&
\(1/7,\ \pm1/5,\ \pm1/2\)
&
\(260~\mathrm{MeV}\)
&
\(0.093~\mathrm{fm}\)
&
\cite{Chu:2025kew}
\\

Twist-three GPDs
&
\(N\)
&
\(0\)
&
\(260~\mathrm{MeV}\)
&
\(0.093~\mathrm{fm}\)
&
\cite{Bhattacharya:2023nmv}
\\[2pt]

\(H,E\)
&
\(N\)
&
\(0\)
&
\(135~\mathrm{MeV}\)
&
\(0.09~\mathrm{fm}\)
&
\cite{Lin:2020rxa}
\\

\(\widetilde H\)
&
\(N\)
&
\(0\)
&
\(135~\mathrm{MeV}\)
&
\(0.09~\mathrm{fm}\)
&
\cite{Lin:2021brq}
\\

\(H,E\)
&
\(N\)
&
\(0,\ 0.1\)
&
\(135~\mathrm{MeV}\)
&
\(0.09~\mathrm{fm}\)
&
\cite{Holligan:2023jqh}
\\
\end{tabular}
\end{ruledtabular}
\end{table*}

\subsection{Lattice results for \(x\)-dependent GPDs}

With the LaMET framework established, studies of \(x\)-dependent GPDs
have progressed from theoretical development to lattice calculations.  LaMET calculations of \(x\)-dependent quark GPDs have been performed
for both the pion and the nucleon, with most studies focusing on
isovector combinations. These studies include calculations at the
physical pion mass or with fine lattice spacings. For the nucleon,
unpolarized, helicity, and transversity GPDs have been studied at
leading twist, and exploratory calculations have also calculated
axial twist-three GPDs. 

Despite this progress, current calculations remain exploratory,
with systematic uncertainties not yet fully controlled.
The available results are summarized in
Table~\ref{tab:gpd_lattice_results}.

\subsubsection{Pion}

The pioneering lattice application of LaMET to GPDs determined the
pion valence GPD \(H_v^\pi(x,0,t)\) at zero skewness, with results
obtained at two nonzero momentum transfers, \(-t\simeq0.36\) and
\(0.91~\mathrm{GeV}^2\)~\cite{Chen:2019lcm}. As a consistency check, the \(t=0\) result agreed with the earlier
LaMET pion PDF~\cite{Zhang:2018nsy}, while the integral over \(x\)
reproduced the pion form factor within uncertainties. The sparse \(t\) coverage, however, provided only limited information
on the transverse spatial structure.

\begin{figure}[!b]
    \centering
\includegraphics[width=\columnwidth]{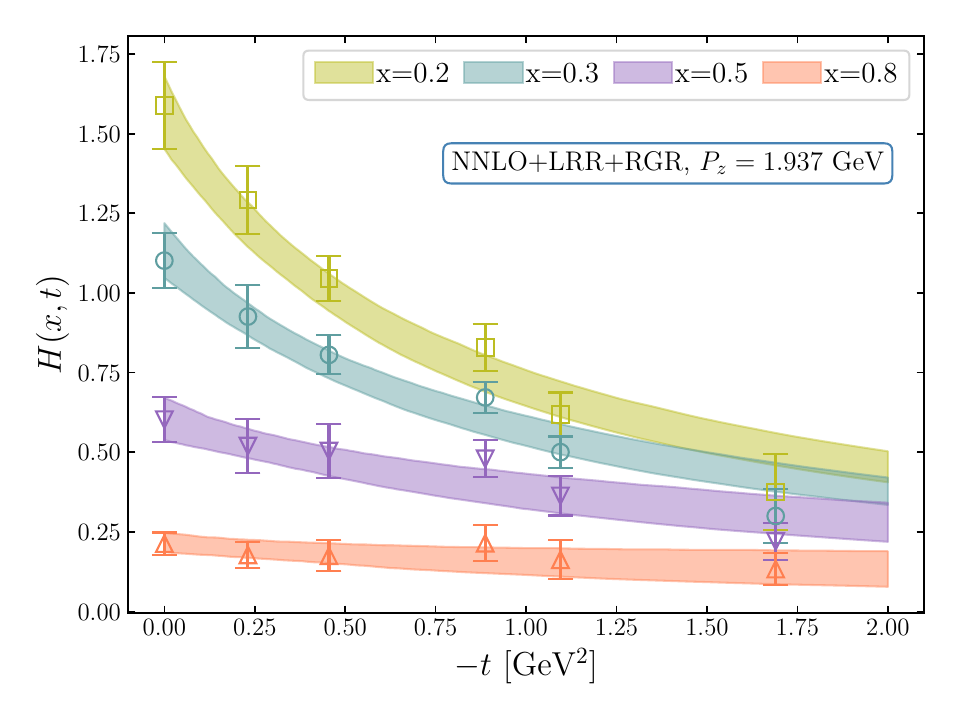}
\caption{\label{fig:Ding_GPD}
Zero-skewness pion valence GPD as a function of
$-t$ for several values of $x$. The shaded bands show monopole fits to
the $t$ dependence. From~\textcite{Ding:2024saz}.}
\end{figure}

Two subsequent studies extended the pion GPD calculation in different
directions. Lin determined \(H_v^\pi\) directly at the physical pion mass
on a \(0.09~\mathrm{fm}\) lattice, with four nonzero momentum transfers
up to \(-t=0.97~\mathrm{GeV}^2\), using hybrid renormalization and
NNLO matching~\cite{Lin:2023gxz}.  Ding et al.\ studied \(H_v^\pi\)  on a much finer lattice,
\(a=0.04~\mathrm{fm}\)~\cite{Ding:2024saz}.
 They employed the asymmetric frame, which enabled access to six
momentum transfers up to \(-t=1.69~\mathrm{GeV}^2\), and used RGR and
LRR on top of hybrid renormalization and NNLO matching. The resulting
\(t\) dependence at  is shown in Fig.~\ref{fig:Ding_GPD}. Both studies further reconstructed impact-
parameter-space distribution, providing early lattice-QCD tomographic images of the pion.
The resulting distributions directly show that quarks carrying larger
longitudinal momentum fractions are more localized in the transverse
plane.
Despite their different systematic uncertainties, the qualitative
agreement between the two calculations is encouraging for the use of
LaMET in lattice GPD studies.

\subsubsection{Nucleon}

The first nucleon calculations considered the isovector unpolarized
and helicity GPDs, followed by the transversity GPDs in the same
lattice setup
~\cite{Alexandrou:2020zbe,Alexandrou:2021bbo}. Both studies used an
ensemble with \(m_\pi\simeq260~\mathrm{MeV}\) and proton boosts up to
\(\bar P^z=1.67~\mathrm{GeV}\). At zero skewness,
\(H\), \(E\), and \(\widetilde H\) were obtained at
\(-t=0.69~\mathrm{GeV}^2\), with \(E\) showing noticeably larger
uncertainties than \(H\). The analysis was also extended to
\(|\xi|=1/3\), providing the first lattice results in the DA region.
Using the same ensemble, the subsequent transversity calculation
determined all four chiral-odd GPDs. The clearest signal was found for
\(H_T\), which showed good agreement between the two largest boosts,
while \(E_T\) and \(\widetilde H_T\) were substantially noisier and
\(\widetilde E_T\) remained consistent with zero.

Around the same time, Lin carried out an independent calculation of
the isovector unpolarized nucleon GPDs \(H\) and \(E\) directly at the
physical pion mass~\cite{Lin:2020rxa}. The calculation was performed
at zero skewness with a proton boost of about \(2.2~\mathrm{GeV}\) and
four nonzero momentum transfers up to
\(-t=0.97~\mathrm{GeV}^2\). The helicity GPD \(\widetilde H\) was subsequently calculated in the same lattice setup
~\cite{Lin:2021brq}. These two studies provided the first lattice-QCD tomography of the nucleon from unpolarized and helicity GPDs.
\begin{figure}[!b]
    \centering
\includegraphics[width=\columnwidth]{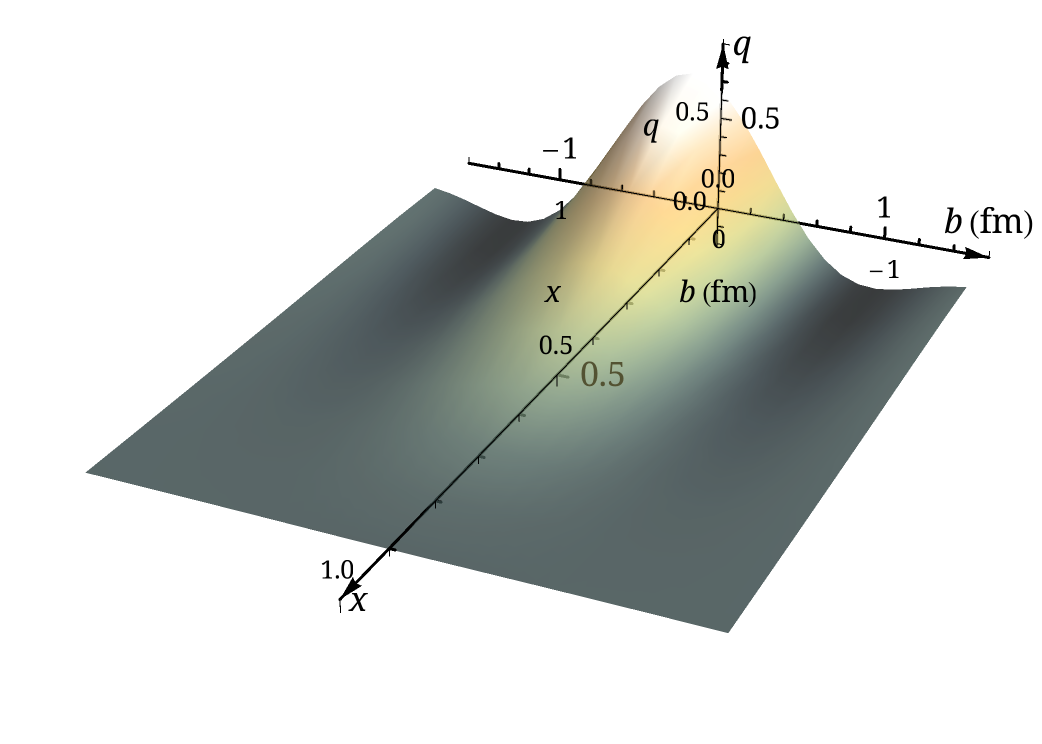}
    \caption{The first lattice determination of the nucleon
    impact-parameter-space distribution as a function of
    \(x\) and \(b_\perp\), obtained from \(H\) at the physical pion
    mass. From~\textcite{Lin:2020rxa}.}
    \label{fig:lin_nucleon_ipd}
\end{figure}

Holligan and Lin later reanalyzed the data in~\textcite{Lin:2020rxa}
using hybrid renormalization and NNLO matching supplemented by RGR and
LRR~\cite{Holligan:2023jqh}. At zero skewness, the resummation
substantially reduced the perturbative uncertainty without producing a
comparable shift in the central values of \(H\) and \(E\). They also
reported results at \(\xi=0.1\) and
\(-t=0.23~\mathrm{GeV}^2\), where the analysis was restricted to
one-loop matching with LRR because RGR was not yet available at
nonzero skewness.

Bhattacharya et al.\ introduced an asymmetric frame approach for lattice GPD calculations~\cite{Bhattacharya:2022aob}, in which the final state momentum is kept fixed and the momentum transfer is assigned entirely to the initial state.
With the final-state momentum fixed, the same sequential propagator can
be reused for different momentum transfers, allowing several values of
\(t\) to be accessed at reduced computational cost.

The asymmetric frame also makes explicit the frame dependence of the
conventional quasi-GPD definition. At finite \(P^z\),
\(\widetilde{h}_{\gamma^0}^q(z,P',P,\mu)\) in the symmetric frame is
related by a transverse Lorentz boost to a linear combination of Dirac
components in the asymmetric frame. The authors therefore parametrized
the nonlocal vector matrix element in terms of eight Lorentz-invariant
amplitudes \(A_i\), which can be extracted in either frame and used to
construct the quasi-GPDs. This formulation also motivates a
Lorentz-invariant quasi-GPD definition, which is consistent with the
standard quasi-GPD up to finite-momentum power corrections. The
invariant amplitudes extracted independently in the symmetric and
asymmetric frames were found to agree within uncertainties. The
standard and Lorentz-invariant quasi-GPD definitions were then compared
directly. They gave compatible results for \(H\), whereas a more
visible difference was observed for \(E\), indicating a stronger
sensitivity of the latter to finite-momentum power corrections.

The same framework was later extended to the helicity and transversity
GPDs~\cite{Bhattacharya:2023jsc,Bhattacharya:2025yba}, which are
parametrized by eight \(\widetilde A_i\) and twelve \(A_{T i}\)
Lorentz-invariant amplitudes, respectively. Unlike
\(\widetilde h_{\gamma^0}^q(z,P',P,\mu)\) in the unpolarized case, the
standard helicity matrix element
\(\widetilde h_{\gamma^3\gamma_5}^q(z,P',P,\mu)\) is already frame
independent, since the transverse boost leaves the 3-direction
unchanged. There is nevertheless an ambiguity in defining a
Lorentz-invariant quasi-GPD. Such nonuniqueness is natural for a
finite-momentum LaMET construction, since the quasi-observables
themselves are frame dependent. The two definitions considered in the
helicity study gave consistent results for \(\widetilde H\) within
uncertainties. The asymmetric frame also substantially enlarged the
momentum-transfer coverage, with results presented at eight values of
\(-t\) spanning \(0.17\)--\(2.29~\mathrm{GeV}^2\) for both the
helicity and transversity GPDs.

Chu et al.\ subsequently extended the asymmetric-frame
calculation of unpolarized nucleon GPDs to nonzero skewness, with
\(\xi=1/7,\,\pm1/5,\) and \(\pm1/2\)~\cite{Chu:2025kew}.
They also considered purely longitudinal momentum transfer, for
which \(H\) and \(E\) cannot be separated. Instead, this setup gives
access to a combination \(H_L\), whose light-cone limit is
proportional to
\(H-\frac{\xi^2}{1-\xi^2}E\).
The authors found \(H_L\simeq H\) for the kinematics studied.
These configurations extend the calculation toward the minimum
\(-t\) allowed at fixed skewness, with the full calculation covering
approximately \(0.06\lesssim -t\lesssim2.25~\mathrm{GeV}^2\).

More recently, Chu et al.\ explored a combined
LaMET and SDF analysis of the same bare lattice
matrix elements using a neural-network fit~\cite{Chu:2025jsi}. At zero skewness, they
reconstructed the joint \(x\)- and \(t\)-dependence of the valence
components of \(H\) and \(E\), and derived the corresponding
impact-parameter-space distributions.

The LaMET approach has also been explored beyond leading twist.
The first lattice calculation of axial twist-three GPDs was carried out
by Bhattacharya et al.~\cite{Bhattacharya:2023nmv} at zero skewness,
with proton momenta up to \(\bar P^z=1.67~\mathrm{GeV}\) and momentum
transfers up to \(-t=2.76~\mathrm{GeV}^2\). Nonvanishing twist-three
contributions were observed, and several consistency checks based on
GPD sum rules were found to be satisfied. The matching, however, involves two major unresolved issues.
First, the calculation employed only the two-parton one-loop kernel
derived for the forward twist-three distribution \(g_T\), while the
mixing with three-parton quark--gluon correlators, which is intrinsic
to twist-three factorization, was not included.  Second, the matching
kernel contains a perturbative zero-mode term proportional to
\(\delta(y/x)\).  In the large-momentum expansion,
the delta function probes the quasi-GPD at
\(\lvert x\rvert\to\infty\), which cannot be determined from
the lattice calculation, and this contribution was therefore omitted.
These two limitations prevent a complete implementation of
twist-three GPD extractions and remain important targets for future
developments.

Although the introduction of asymmetric frame has improved
the statistical precision and momentum-transfer coverage of GPD
calculations, present LaMET studies remain exploratory. Most are still
based on single ensembles, without a renormalization scheme with controlled continuum and
large-momentum extrapolations. Although several studies have examined
the dependence on the hadron momentum, such comparisons alone cannot
establish convergence toward the large-momentum limit, since increasing
\(\bar P^z\) suppresses power corrections while simultaneously enhancing
discretization effects through \(a\bar P^z\). 

\begin{figure}[!t]
    \centering
    \includegraphics[width=0.493\columnwidth]{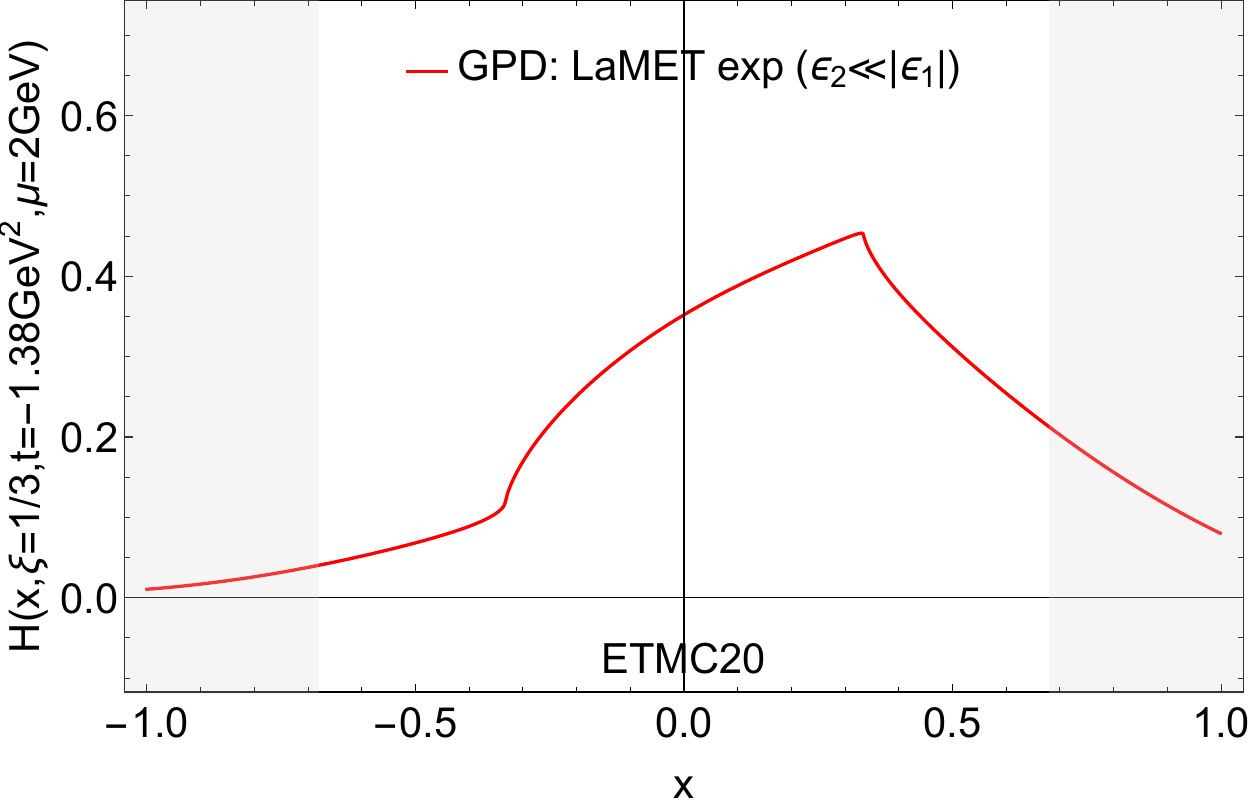}
    \includegraphics[width=0.493\columnwidth]{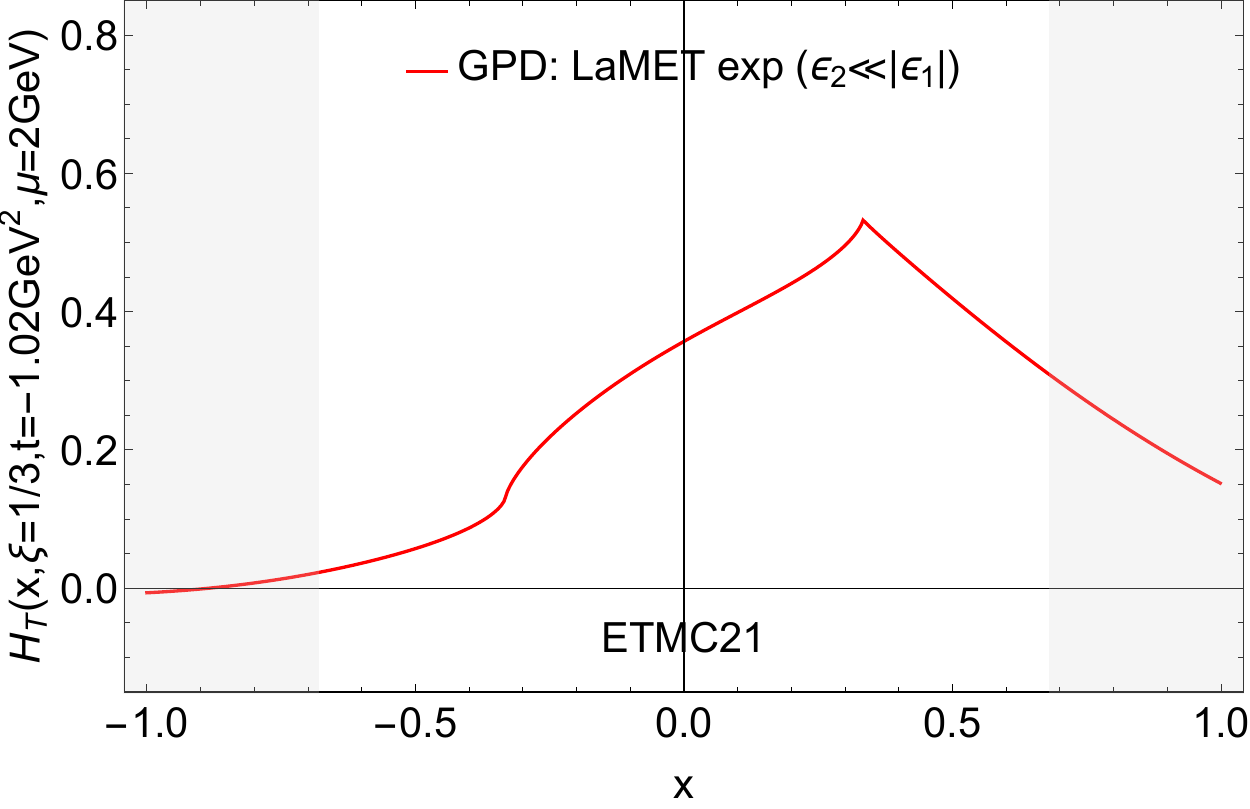}
    \includegraphics[width=0.493\columnwidth]{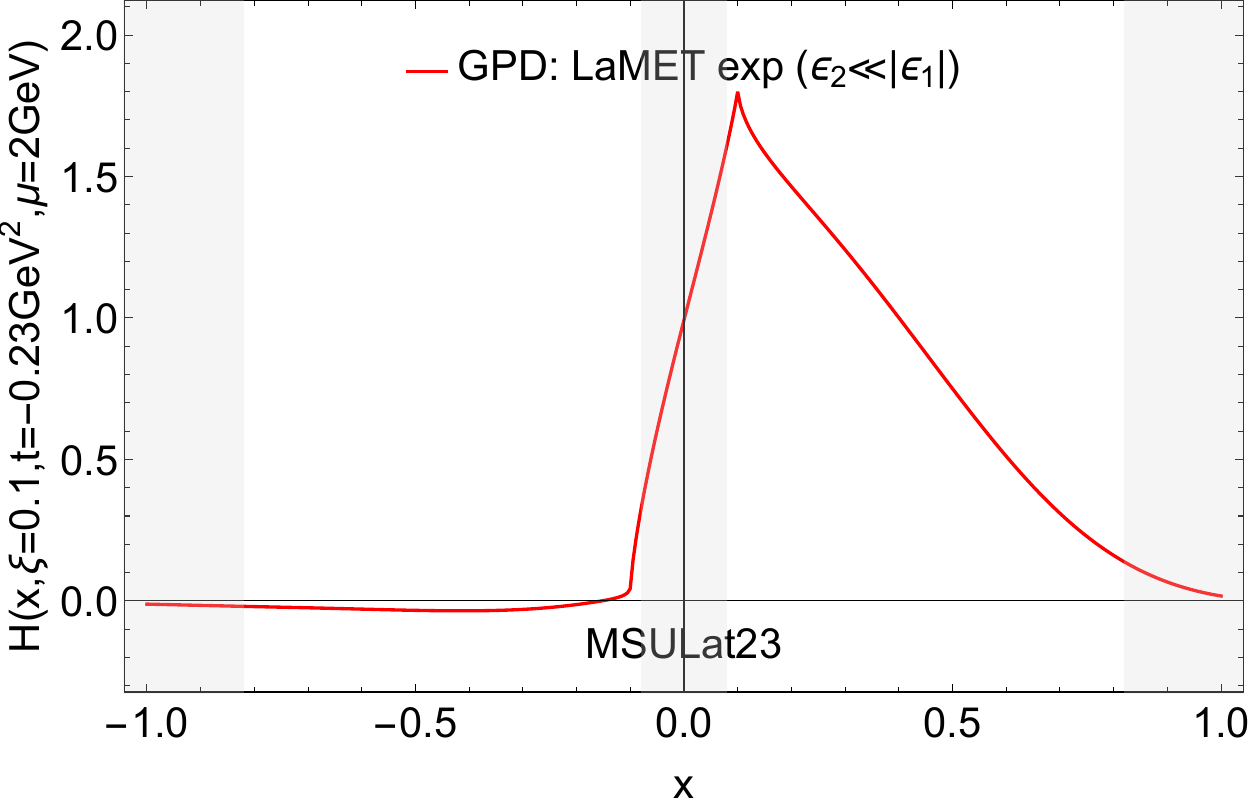}
    \includegraphics[width=0.493\columnwidth]{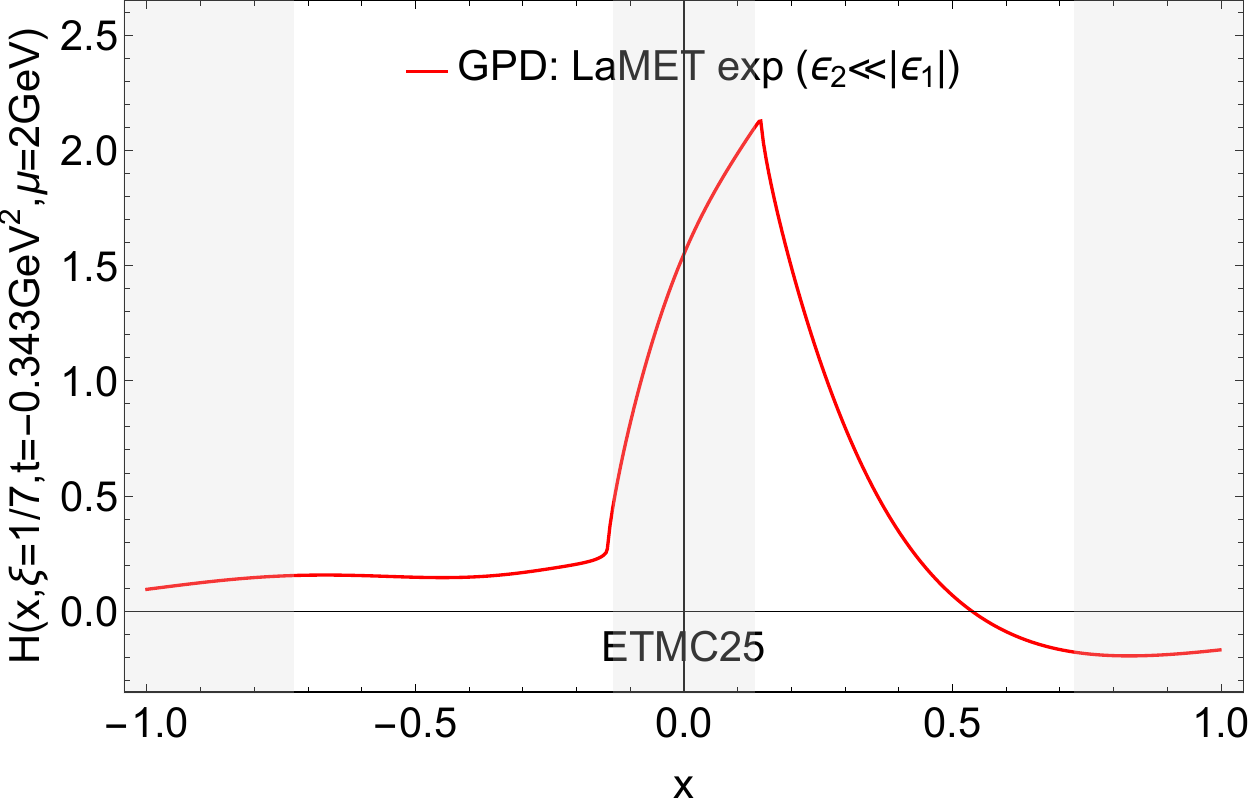}
    \caption{
    The reanalyzed non-zero skewness unpolarized $H$ or transversity $H_T$ GPDs from \textcite{Alexandrou:2020zbe} (ETMC20), \textcite{Alexandrou:2021bbo} (ETMC21), \textcite{Holligan:2023jqh} (MSULat23), and \textcite{Chu:2025kew} (ETMC25), which are calculated by performing one-loop matching on the quasi-GPDs under proper regulators $\epsilon_2 \ll |\epsilon_1| \rightarrow 0$. The gray bands indicate regions where either $2|1\pm x|P^z < 0.8 \, {\rm GeV}$ or $2(|x|+|\xi|)P^z < 0.8 \, {\rm GeV}$, for which the large-momentum expansion is not expected to be reliable. From~\textcite{Su:2026GPD}.
    }
    \label{fig:su_gpd}
\end{figure}

Beyond these general lattice uncertainties,
nonzero-skewness calculations have exhibited pronounced unphysical
behavior near the PDF--DA boundaries, \(x=\pm\xi\).
Recent work has shown that the crossover points
do not introduce additional infrared sensitivity at leading twist, and
that the unphysical behavior can instead arise from the treatment of the
regulators in the perturbative matching~\cite{Su:2026GPD}.
In particular, when regulating the plus distributions appearing in the
matching kernel, the order of limits is important. The plus-distribution
regulator should be removed before taking \(x\to\pm\xi\). Otherwise,
ill-defined or spurious behavior may arise at the PDF--DA boundaries.
With the matching implemented using the proper order of limits,
Su et al.\ reanalyzed existing lattice data and obtained GPDs that are
continuous across \(x=\pm\xi\), while their derivatives remain
discontinuous, as shown in Fig.~\ref{fig:su_gpd}.

\section{Advances in Experiment and Phenomenology}
\label{sec:5_exp_and_pheno}

Substantial experimental efforts have been devoted to unraveling the multidimensional structure of the nucleon through GPDs over the past decades. HERA~\cite{Abramowicz:1998ii, Klein:2008di} was built in 1992 and operated until 2007 as the world’s first lepton--hadron collider at DESY in Hamburg. Beyond the vast inclusive and semi-inclusive measurements there that facilitated precise determinations of PDFs~\cite{H1:2009pze, H1:2015ubc}, the high-energy lepton and hadron beams also enabled early measurements of DVCS~\cite{HERMES:2001bob, H1:2001nez, ZEUS:2003pwh} and DVMP~\cite{ZEUS:1998xpo,H1:1999pji,H1:2000kis,ZEUS:2002wfj,H1:2005dtp,H1:2009cml}. Building upon this, the planned EIC with notably higher luminosity and versatile polarization capabilities, as well as the proposed EicC~\cite{Anderle:2021wcy, Xiao:2026tbs}, will offer new opportunities to deepen our understanding of strong interactions and multidimensional nucleon structures~\cite{Accardi:2012qut, AbdulKhalek:2021gbh}.

In parallel, the Continuous Electron Beam Accelerator Facility (CEBAF) and facilities in multiple experimental halls at JLab have enabled high-precision measurements of lepton--hadron scattering~\cite{Leemann:2001dg}, especially following its 12~GeV upgrade in the early 2010s~\cite{Dudek:2012vr}, benefiting from its fixed-target configuration. This also enables the measurements of near-threshold $J/\psi$ production~\cite{GlueX:2019mkq,Duran:2022xag,GlueX:2023pev,007:2026dow,Chatagnon:2026qsv,Tyson:2026gnd} as a probe of large-skewness gluon GPDs and EMT form factors of the nucleon. The proposed 22~GeV upgrade~\cite{Accardi:2023chb} could further complement the future EIC program, together providing a comprehensive picture of the nucleon. The COMPASS experiment at CERN~\cite{COMPASS:2007rjf} further bridges them with high-energy muon-induced DVCS~\cite{Joerg:2016hhs} and DVMP measurements~\cite{COMPASS:2019fea}, while the RHIC spin program had offered a unique opportunity to investigate the hadron spin structures through proton--proton collisions, particularly in the gluon-dominated high-energy region~\cite{Bunce:2000uv, Aschenauer:2015eha}.

Correspondingly, phenomenological investigations have made significant progress over the past decades as well. One of the earliest approaches to parameterizing GPDs while preserving their physical constraints is the double-distribution framework, which expresses GPDs as an integral transform together with a profile function and a possible $D$-term that models the skewness dependence and ensures the polynomiality condition~\cite{Radyushkin:1997ki,Radyushkin:1998es,Musatov:1999xp,Polyakov:1999gs}. The Vanderhaeghen--Guichon--Guidal (VGG)~\cite{Vanderhaeghen:1999xj} and Goloskokov--Kroll (GK)~\cite{Goloskokov:2005sd,Goloskokov:2007nt} models were developed upon this in early studies of GPDs and remain widely used in contemporary analyses. Meanwhile, considerable effort has been devoted to developing more general descriptions of GPDs, since commonly used factorized profile ansatzes have limited flexibility and are generally not preserved in the same form under QCD evolution~\cite{Radyushkin:1998es,Musatov:1999xp}. Approaches based on the Shuvaev transform~\cite{Shuvaev:1999ce,Noritzsch:2000pr} and dual parameterization~\cite{Polyakov:2002wz} were among the first to construct more general GPDs while maintaining connections to the double-distribution framework~\cite{Muller:2014wxa}. Still, their phenomenological application remains limited due to their integral-transform structure. The Kumeri\v{c}ki--M\"uller (KM) model~\cite{Kumericki:2007sa,Kumericki:2009uq} instead parameterizes GPDs and observables directly through moments, particularly the conformal moments~\cite{Mueller:2005ed}, and pioneered global phenomenological analyses of GPDs. With subsequent refinements, it remains an important baseline for modern analyses~\cite{Kumericki:2016ehc,Cuic:2020iwt,Cuic:2023mki}. 

More recently, progress has been driven by two complementary developments: lattice-QCD calculations of parton distributions~\cite{Ji:2013dva,Ji:2020ect,Constantinou:2020pek}, discussed in the previous section, and modern computational methods such as neural networks. Following their success in PDF extraction and uncertainty quantification~\cite{NNPDF:2021njg}, neural-network techniques have been applied to GPD phenomenology, including neural-network extractions of CFFs~\cite{Kumericki:2011rz,Moutarde:2018kwr,Moutarde:2019tqa,Almaeen:2024guo, Mezrag:2026wcf}, theory-constrained representations of GPDs~\cite{Dutrieux:2021wll,Xu:2026lko}, and modular numerical frameworks such as PARTONS~\cite{Berthou:2015oaw}. Finally, a powerful and versatile global analysis framework based on conformable moment expansion, GPDs through Universal Moment Parameterization (GUMP), has carried out the first global extractions of GPDs in combination with lattice QCD simulations, demonstrating that the inverse problem in the phenomenological determination of GPDs can be effectively addressed by incorporating complementary lattice QCD inputs~\cite{Guo:2022upw,Guo:2023ahv,Guo:2025muf}.

In this section, we review major experimental and phenomenological advances in GPD studies, together with complementary constraints from nonperturbative approaches. We focus on leading-twist, chiral-even proton GPDs, which have been the primary target of theoretical and experimental efforts, and highlight the remaining challenges toward a precise, model-independent mapping of multidimensional nucleon structure. The forthcoming EIC, with its broad kinematic reach and high luminosity, is expected to usher in a new era of precision three-dimensional nucleon imaging.

\subsection{Collinear parton distribution function constraints}

\begin{figure}[t]
    \centering
    \includegraphics[width=0.483\textwidth]{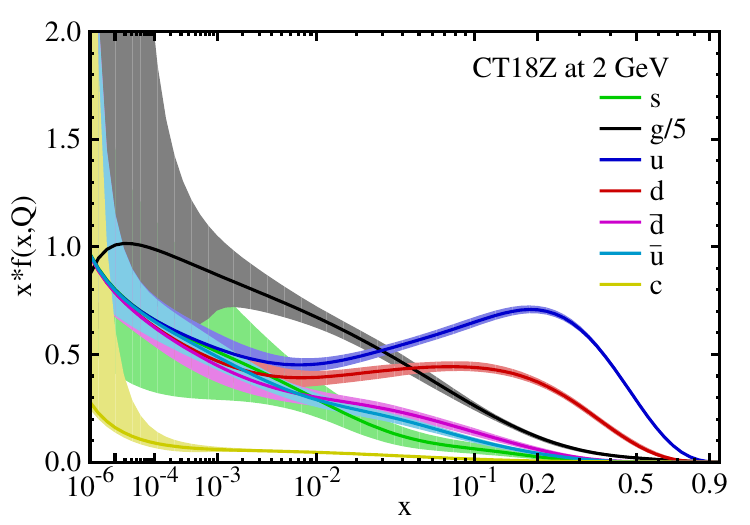}
    \caption
    {\raggedright An example of the extracted unpolarized proton PDFs at scale $Q=2~\mathrm{GeV}$ and NNLO accuracy by the CTEQ-TEA collaboration. From~\textcite{Hou:2019efy}.}
    \label{fig:sec5:CTEQPDF}
\end{figure}

PDFs correspond to the forward limit of GPDs and have long been the most explored and phenomenologically well-determined parton distributions in high-energy particle and nuclear physics. Substantial progress has been achieved in their precise determinations and uncertainty quantification lately by several collaborations, including ABMP~\cite{Alekhin:2017kpj}, CJ~\cite{Accardi:2016qay}, CTEQ-TEA~\cite{Hou:2019efy}, HERA~\cite{H1:2015ubc}, JAM~\cite{Moffat:2021dji}, MSHT~\cite{Bailey:2020ooq,McGowan:2022nag}, and NNPDF~\cite{NNPDF:2021njg}. These analyses reach theoretical accuracies at NNLO in perturbative QCD and, in some cases, beyond. They impose the most stringent constraints on the forward limits of GPDs because of their high precision and extensive phase-space coverage. Figure~\ref{fig:sec5:CTEQPDF} shows an example of the unpolarized proton PDFs obtained by the CTEQ-TEA collaboration~\cite{Hou:2019efy}. %

\begin{figure}[t]
    \centering
    \includegraphics[width=0.483\textwidth]{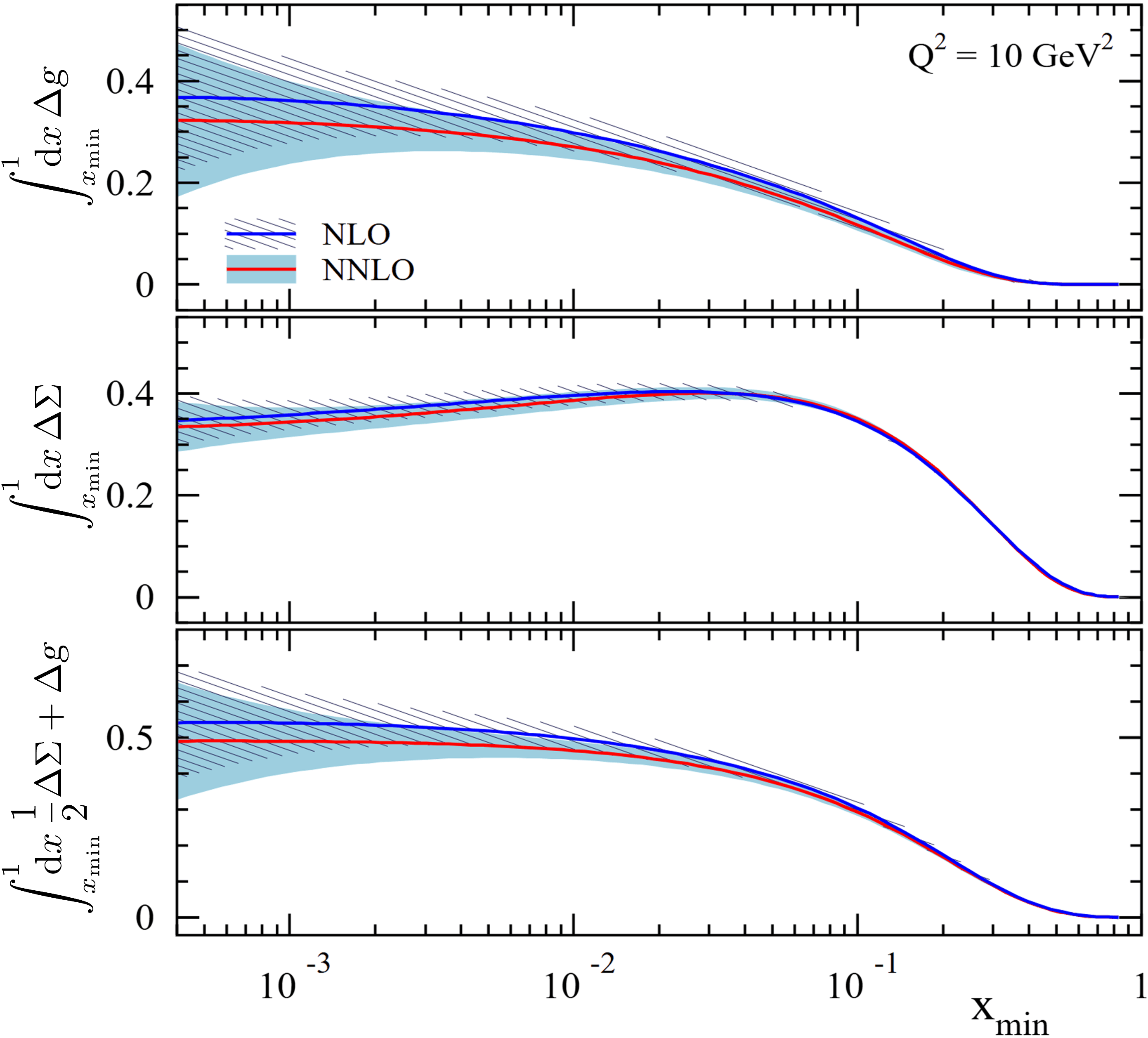}
    \caption
    {\raggedright An example of the constraints on the helicity contributions of quarks, gluons and their combinations based on the extracted polarized proton PDFs by the BDSSV collaboration at scale $Q^2=10~\mathrm{GeV}^2$ and NNLO accuracy. Adapted from~\textcite{Borsa:2024mss}}
    \label{fig:sec5:BDSSVspin}
\end{figure}

Beyond unpolarized PDFs, the polarized (helicity) PDFs, which encode the spin-dependent parton structure of the nucleon, are equally important. Nevertheless, helicity PDFs are considerably less constrained due to the limited availability of polarized experimental data, rendering their precise extraction particularly challenging. Substantial progress has been made lately by several groups, including DSSV/BDSSV~\cite{deFlorian:2008mr,deFlorian:2009vb,Borsa:2024mss}, JAM~\cite{Ethier:2017zbq,Cocuzza:2022jye}, NNPDF~\cite{Nocera:2014gqa,Cruz-Martinez:2025ahf}, and MAP~\cite{Bertone:2024taw}. The first moments of the helicity PDFs, obtained by integration over $x$, provide the total contributions of quark and gluon helicities to the proton spin, which are essential components of the proton spin sum rule~\cite{Jaffe:1989jz}. An illustrative example of the quark and gluon helicity contributions in the proton is shown in Fig.~\ref{fig:sec5:BDSSVspin}, based on the polarized proton PDFs extracted in the BDSSV24 analysis~\cite{Borsa:2024mss} at $Q^2=10~\mathrm{GeV}^2$ with NNLO accuracy, suggesting potentially small contributions from OAM. Recent efforts have also focused on incorporating small-$x$ evolution in global analyses of helicity PDFs, leading to improved constraints in the small-$x$ region~\cite{Adamiak:2021ppq,Adamiak:2023yhz,JAMCollaborationSmall-xAnalysisGroup:2025tfa}.

The unpolarized and helicity PDFs correspond to the forward limits of the $H$ and $\widetilde{H}$ GPDs, respectively; hence, their extraction constitutes a fundamental component of any GPD analysis. Unlike many (semi-)inclusive observables, exclusive processes typically involve both vector GPDs ($H$, $E$) and axial-vector GPDs ($\widetilde{H}$, $\widetilde{E}$) as demonstrated in e.g., Sec.~\ref{sec:3_DVCS}. This motivates the consistent, simultaneous determination of both unpolarized and helicity PDFs for precision studies of GPDs, further highlighting the importance of precise helicity PDF extractions. 

\subsection{Measurements of hard exclusive processes}\label{sec:5_hard_ex_mea}

Besides (semi-)inclusive measurements, GPDs are primarily constrained through high-energy (hard) exclusive (diffractive) production processes off the nucleon, where the initial nucleon remains intact in the final states, and all final-state particles are fully identified. Two quintessential processes, DVCS~\cite{Ji:1996nm} and DVMP~\cite{Radyushkin:1996ru, Collins:1996fb}, constitute the key experimental inputs for GPD extractions as discussed in Sec.~\ref{sec:3_hard_proc}. Here we review the major advances in the experimental measurements of hard exclusive processes and discuss their implications in phenomenological studies.

\subsubsection{Deeply virtual Compton scattering}

Deeply virtual exclusive processes, initiated by a virtual photon with momentum $q$ and virtuality $Q^2 \equiv -q^2$ in the deeply virtual limit ($Q^2 \to \infty$), were among the earliest proposed observables for elucidating the multidimensional nucleon structures with GPDs. Two prototypical channels, DVCS~\cite{Ji:1996nm} and DVMP~\cite{Radyushkin:1996ru,Collins:1996fb}, rapidly emerged as primary experimental probes, with extensive measurements performed at HERA, JLab, and other facilities.

Early DVCS measurements were made by HERMES, H1, and ZEUS at DESY~\cite{HERMES:2001bob,H1:2001nez,H1:2005gdw,H1:2007vrx,H1:2009wnw,ZEUS:2003pwh,ZEUS:2008hcd} and by CLAS and Hall A at JLab~\cite{CLAS:2001wjj,CLAS:2006krx,JeffersonLabHallA:2006prd,CLAS:2008ahu}. These measurements were analyzed with GPD models such as the VGG model, among other theoretical models available at the time~\cite{Kivel:2000fg,Vanderhaeghen:1999xj,Frankfurt:1997at,Donnachie:2000px}. Subsequently, the KM framework was developed to incorporate them into systematic global analyses based on parametrized GPDs~\cite{Kumericki:2007sa,Kumericki:2009uq}.

Despite these developments, the global determinations of GPDs have been significantly hindered by their dependence on three kinematic variables $(x,\xi,t)$ and by the limited multidimensional and polarization information in the early measurements. At HERA, the azimuthal dependence was commonly integrated over, while $x_B$ and $Q^2$ were strongly correlated. In particular, it is now well established that azimuthal angular modulations in $\phi$, together with measurements using polarized beams and targets, are essential for a robust determination of the CFFs, especially because four leading-twist chiral-even CFFs ($\mathcal{H}$, $\mathcal{E}$, $\widetilde{\mathcal{H}}$, and $\widetilde{\mathcal{E}}$), each generally complex, can contribute simultaneously~\cite{Belitsky:2001ns,Belitsky:2010jw,Shiells:2021xqo} as discussed in Sec.~\ref{sec:3_DVCS}. Experimentally, these modulations have been studied through beam-charge, beam-spin, longitudinal- and transverse-target-spin, and double-spin observables at HERMES~\cite{HERMES:2001bob,HERMES:2006pre,HERMES:2008abz,HERMES:2009cqe,HERMES:2010dsx,HERMES:2011bou,HERMES:2012gbh,HERMES:2012idp} and JLab~\cite{CLAS:2006krx,CLAS:2008ahu,CLAS:2015bqi,CLAS:2015uuo,CLAS:2018bgk,JeffersonLabHallA:2006prd,JeffersonLabHallA:2015dwe,Defurne:2017paw}.

In the valence region, these limitations have been largely reduced by the high-precision measurements performed in different experimental halls at JLab. The CLAS collaboration has provided beam-spin asymmetries, polarized-target observables, and polarized and unpolarized cross sections over an extensive kinematic domain~\cite{CLAS:2001wjj,CLAS:2006krx,CLAS:2008ahu,CLAS:2015bqi,CLAS:2015uuo,CLAS:2018bgk}. Complementary Hall A measurements have provided high-precision helicity-dependent and helicity-independent cross sections and their azimuthal dependence~\cite{JeffersonLabHallA:2006prd,JeffersonLabHallA:2015dwe,Defurne:2017paw}. Measurements on neutron and nuclear targets have further broadened the flavor and nuclear sensitivity of the DVCS program, including the Hall A neutron measurements, the first CLAS12 measurement with detection of the active neutron, and coherent and incoherent measurements on ${}^{4}\mathrm{He}$ targets~\cite{JeffersonLabHallA:2007jdm,Benali:2020vma,CLAS:2024qhy, CLAS:2017udk,CLAS:2018ddh}. This capability to systematically explore the kinematic phase space of hard exclusive processes, including their detailed azimuthal $\phi$ dependence, together with polarization measurements using both polarized beams and targets, makes JLab an essential complement to HERA and a crucial precursor to the future EIC program, particularly in view of a potential 22~GeV upgrade~\cite{Accardi:2023chb}.

\begin{figure}[t]
    \centering
    \includegraphics[width=0.483\textwidth]{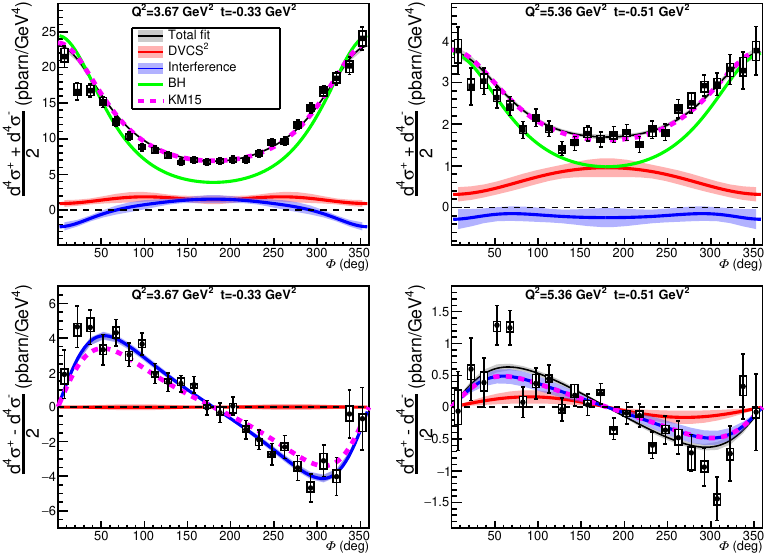}
    \caption{\raggedright High-precision measurements of beam-polarized and unpolarized $\phi$-dependent DVCS cross sections at high Bjorken $x_B$ at JLab Hall A. Adapted from~\textcite{JeffersonLabHallA:2022pnx}.}
    \label{fig:sec5:JLabDVCS}
\end{figure}

Even higher-precision data at higher virtuality have been obtained following the 12~GeV beam upgrade~\cite{CLAS:2022syx,JeffersonLabHallA:2022pnx,CLAS:2024qhy}. For a comprehensive collection of DVCS measurements and references, see the recently developed GPD databases~\cite{Burkert:2025gzu,qgtdatabase}. Figure~\ref{fig:sec5:JLabDVCS} shows examples of high-precision measurements of beam-polarized and unpolarized $\phi$-dependent DVCS cross sections at high Bjorken $x_B$ at JLab Hall A~\cite{JeffersonLabHallA:2022pnx}. These comprehensive, high-precision datasets, including measurements with polarized beams and targets, have substantially improved constraints on the CFFs and advanced the global extraction of GPDs~\cite{Kumericki:2016ehc,Cuic:2020iwt,Guo:2023ahv,Guo:2025muf}. Additionally, the COMPASS experiment at CERN bridges the low-$x_B$ regime explored at HERA and the valence region studied at JLab~\cite{COMPASS:2018pup}. 

Beyond DVCS, many other processes have been proposed in the literature to further explore GPDs. In particular, TCS, the time-reversed counterpart of DVCS, probes the complex conjugate of the DVCS amplitudes at leading order in $\alpha_s$~\cite{Berger:2001xd}. In light of its close connection to DVCS, TCS not only provides complementary experimental input for GPD extractions but also offers a stringent test of the universality of GPD collinear factorization. The first experimental measurement of TCS was recently reported by the CLAS Collaboration at JLab~\cite{CLAS:2021lky}. Incorporating them into future analyses, as additional data become available, will represent an important step forward in the phenomenological extraction of GPDs. Moreover, other hard exclusive processes such as DDVCS~\cite{Muller:1994ses, Belitsky:2002tf,Guidal:2002kt}, provide additional measurable kinematic variables that can enhance sensitivity to the $x$ dependence of GPDs and help mitigate the inverse problem inherent in DVCS and DVMP observables. Experimentally, however, their larger phase space and additional kinematic suppression reduce event rates and require higher luminosities. Recent theoretical developments and a proposed SoLID positron-beam measurement are discussed in \textcite{Deja:2023ahc,Zhao:2021zsm}.

\subsubsection{Deeply virtual meson production}

At leading power, longitudinal DVMP amplitudes factorize in terms of GPDs and a meson DA~\cite{Radyushkin:1996ru,Collins:1996fb} as discussed in Sec.~\ref{sec:3_DVMP}. Extensive exclusive meson measurements were performed at HERA. Light-vector-meson measurements included exclusive $\rho^{0}$ production~\cite{H1:1996gwv,H1:1997gev,H1:1999pji,ZEUS:1998xpo,ZEUS:2007iet,H1:2009cml} and $\phi$ production~\cite{H1:1997gev,ZEUS:2005bhf,H1:2009cml}. Heavy-quarkonium measurements included $J/\psi$ electroproduction and photoproduction~\cite{H1:1996gwv,H1:2000kis,ZEUS:2002wfj,ZEUS:2004yeh,H1:2005dtp,H1:2013okq,ZEUS:1998xpo}, together with $\Upsilon$ photoproduction~\cite{ZEUS:1998cdr,H1:2000kis,ZEUS:2009asc}. Exclusive meson production was also studied at HERMES, including $\rho^0$ production~\cite{HERMES:2000jnb} and measurements of $\pi^+$ production~\cite{HERMES:2007hrc,HERMES:2009gtv}. However, the phenomenological application of these measurements to GPD extractions is more involved and faces several extra challenges, see~\cite{Favart:2015umi} for an earlier experimental and phenomenological review.

One important subtlety arises from the formation of the meson in the final state, which is described by its light-cone DA~\cite{Lepage:1980fj,Brodsky:1994kf} within the framework of collinear factorization~\cite{Radyushkin:1996ru,Collins:1996fb}. Beyond the limited knowledge of meson DAs, quantitative analyses are complicated by the fact that amplitudes for transversely polarized mesons are suppressed by $1/Q$ relative to longitudinal ones. Because collinear factorization has been rigorously established only at leading power for longitudinal meson production, applying the GPD framework requires an experimental separation of the longitudinal and transverse cross sections and verification of the expected leading-power scaling. These requirements are experimentally demanding and limit the precision of phenomenological analyses~\cite{Goloskokov:2007nt,Cuic:2023mki}.

\begin{figure}[t]
    \centering
    \includegraphics[width=0.483\textwidth]{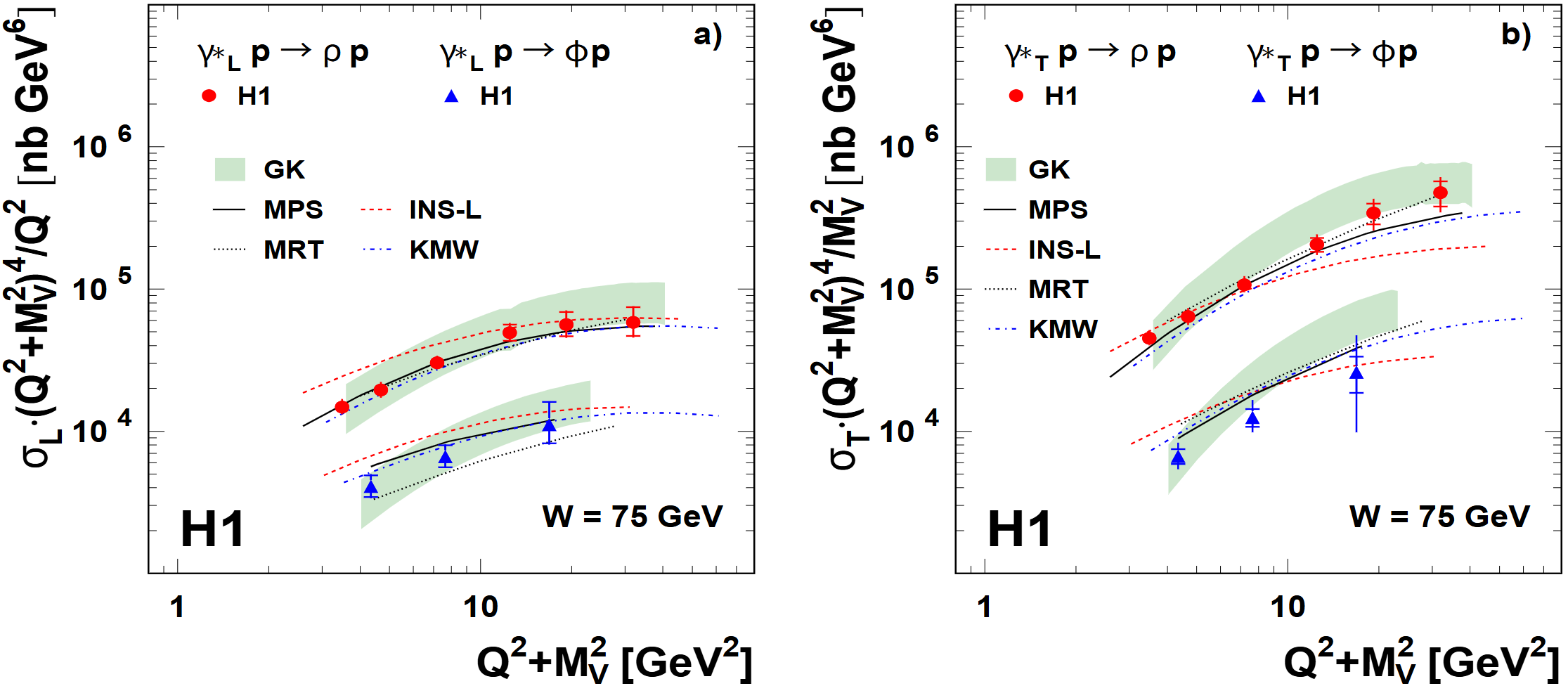}
    \caption{\raggedright Measurements of the $(Q^2+M_V^2)$ dependence of rescaled longitudinal and transverse cross sections for exclusive $\rho$ and $\phi$ electroproduction. The observation departs from the LO scaling expectation, particularly at moderate to low $Q^2$. From~\textcite{H1:2009cml}.}
    \label{fig:sec5:H1DVMP}
\end{figure}

Specifically, Figure~\ref{fig:sec5:H1DVMP} shows the H1 Collaboration measurements of the $\left(Q^2+M_V^2\right)$ dependence of the rescaled DVMP total cross sections at HERA for $\rho$ and $\phi$ meson production, for longitudinally and transversely polarized virtual photons, respectively~\cite{H1:2009cml}. At LO in perturbative QCD, these rescaled cross sections are expected to be approximately independent of $\left(Q^2+M_V^2\right)$ in the large-$Q^2$ limit. In contrast, the experimental data exhibit a clear deviation from this scaling behavior, particularly in the medium- to low-$Q^2$ region, invoking the potential importance of higher-order perturbative contributions and power-suppressed effects~\cite{Goloskokov:2007nt,Cuic:2023mki}.

\begin{table*}[t]
\centering
\caption{Representative DVCS measurements, organized by kinematic regime,
experiment, target, and principal observables. Qualitatively, HERA probes predominantly small $x_B$, HERMES and COMPASS
intermediate $x_B$, and JLab the valence region.}
\label{tab:sec5:DVCSexp}
\scriptsize
\setlength{\tabcolsep}{3.3pt}
\renewcommand{\arraystretch}{1.15}
\begin{ruledtabular}
\begin{tabular}{@{}p{2.45cm}p{1.95cm}p{0.9cm}p{6.45cm}p{5.25cm}@{}}
\textbf{Kinematic regime}
& \textbf{Experiment}
& \textbf{Target}
& \textbf{Principal measurements}
& \textbf{References}
\\
\hline
Small $x_B$ (HERA)
& H1
& $p$
& Unpolarized differential cross sections with $Q^2$, $W$, and $t$ dependences
& \cite{H1:2001nez,H1:2005gdw,H1:2007vrx,H1:2009wnw}
\\

Small $x_B$ (HERA)
& ZEUS
& $p$
& Unpolarized differential cross sections with $Q^2$, $W$, and $t$ dependences
& \cite{ZEUS:2003pwh,ZEUS:2008hcd}
\\

Intermediate $x_B$
& HERMES
& $p,d$
& Beam-charge, beam-spin, and double-spin asymmetries; longitudinal- and
transverse-target-spin asymmetries;
& \cite{HERMES:2001bob,HERMES:2006pre,HERMES:2008abz,
HERMES:2009cqe,HERMES:2010dsx,HERMES:2011bou,
HERMES:2012gbh,HERMES:2012idp}
\\

Intermediate $x_B$
& COMPASS
& $p$
& Differential cross sections and their $t$ dependence
& \cite{COMPASS:2018pup}
\\

Valence
& CLAS
& $p$
& Beam- and target-spin and double-spin asymmetries;
polarized and unpolarized $\phi$-dependent cross sections
& \cite{CLAS:2001wjj,CLAS:2006krx,CLAS:2008ahu,
CLAS:2015bqi,CLAS:2015uuo,CLAS:2018bgk}
\\

Valence
& Hall A
& $p$
& Unpolarized and beam-polarized
$\phi$-dependent differential cross sections
& \cite{JeffersonLabHallA:2006prd,JeffersonLabHallA:2015dwe,
Defurne:2017paw}
\\

Valence
& CLAS\&Hall A
& $p$
& Higher-$Q^2$ proton differential cross sections and asymmetries with extended
multidimensional coverage
& \cite{CLAS:2022syx,JeffersonLabHallA:2022pnx,Lee:2026osn}
\\

Valence, neutron
& CLAS\&Hall A
& $n$
& DVCS differential cross-section off the neutron; Beam-spin asymmetry with exclusive detection of the active neutron
& \cite{JeffersonLabHallA:2007jdm,Benali:2020vma, CLAS:2024qhy}
\\

Valence, nuclear
& CLAS
& ${}^{4}\mathrm{He}$
& Coherent DVCS on ${}^{4}\mathrm{He}$ and incoherent DVCS from
bound protons
& \cite{CLAS:2017udk,CLAS:2018ddh}
\\

\end{tabular}
\end{ruledtabular}
\end{table*}

\begin{table*}[t]
\centering
\caption{Representative exclusive meson-production measurements, organized
by final-state meson and kinematic regime.}
\label{tab:sec5:DVMPexp}
\scriptsize
\setlength{\tabcolsep}{3.2pt}
\renewcommand{\arraystretch}{1.15}
\begin{ruledtabular}
\begin{tabular}{@{}p{2.45cm}p{1.95cm}p{0.9cm}p{6.45cm}p{5.25cm}@{}}
\textbf{Kinematic regime}
& \textbf{Experiment}
& \textbf{Meson}
& \textbf{Principal measurements}
& \textbf{References}
\\
\hline

Small $x$ (HERA)
& H1, ZEUS
& $\rho^{0}$
& Exclusive electroproduction differential cross sections with $Q^2$, $W$, and $t$
dependences; polarization observables and spin-density matrix elements
& \cite{H1:1996gwv,H1:1997gev,H1:1999pji,ZEUS:1998xpo,
ZEUS:2007iet,H1:2009cml}
\\

Small $x$ (HERA)
& H1, ZEUS
& $\phi$
& Exclusive electroproduction differential cross sections; longitudinal--transverse
separation and spin-density matrix elements
& \cite{H1:1997gev,ZEUS:2005bhf,H1:2009cml}
\\

Small $x$ (HERA)
& H1\&ZEUS
& $J/\psi$
& Exclusive electroproduction and photoproduction differential cross sections and
their $Q^2$, $W$, and $t$ dependences
& \cite{H1:1996gwv,H1:2000kis,ZEUS:1998xpo,ZEUS:2002wfj,
ZEUS:2004yeh,H1:2005dtp,H1:2013okq}
\\

Small $x$ (HERA)
& H1, ZEUS
& $\Upsilon$
& Exclusive photoproduction differential cross sections
& \cite{ZEUS:1998cdr,H1:2000kis,ZEUS:2009asc}
\\

Intermediate $x_B$
& HERMES
& $\rho^{0}$
& Exclusive electroproduction differential cross sections at intermediate
virtual-photon energies
& \cite{HERMES:2000jnb}
\\

Intermediate $x_B$
& HERMES
& $\pi^{+}$
& Differential cross sections and transverse-target-spin asymmetries
& \cite{HERMES:2007hrc,HERMES:2009gtv}
\\

Intermediate $x_B$
& COMPASS
& $\pi^{0}$
& Differential cross sections
& \cite{COMPASS:2019fea,COMPASS:2024hvm}
\\

Intermediate $x_B$
& COMPASS
& $\omega$
& Transverse-target-spin asymmetries and spin-density matrix elements
& \cite{COMPASS:2016ium,COMPASS:2020zre}
\\

Intermediate $x_B$
& COMPASS
& $\rho^{0}$
& Transverse-target-spin asymmetries and spin-density matrix elements
& \cite{COMPASS:2013fsk,COMPASS:2022xig}
\\

Valence
& CLAS
& $\rho^{0}$
& Differential cross sections and polarization observables
& \cite{CLAS:2008rpm}
\\

Valence
& CLAS
& $\phi$
& Differential cross sections
& \cite{CLAS:2001zwd}
\\

Valence
& CLAS\&Hall A
& $\pi^{0}$
& Unseparated and Rosenbluth-separated differential cross sections, structure
functions, and beam-spin asymmetries
& \cite{CLAS:2014jpc,JeffersonLabHallA:2016wye,
JeffersonLabHallA:2017hky,JeffersonLabHallA:2020dhq,
CLAS:2023wda}
\\

Valence
& CLAS\&Hall C
& $\pi^{+}$
& Differential cross sections,
longitudinal--transverse separations, and $Q^2$-scaling studies
& \cite{CLAS:2022iqy,Volmer:2000ek,Horn:2006tm,
Tadevosyan:2007yd,Horn:2007ug,Blok:2008jy,Huber:2008id,
Huber:2014ius,Huber:2014kar}
\\

Valence
& CLAS
& $\eta$
& Beam-spin asymmetries in exclusive electroproduction
& \cite{CLAS:2019uzc}
\\
\end{tabular}
\end{ruledtabular}
\end{table*}

The measurements also indicate transverse-polarization dominance at low $Q^2$, complicating the application of conventional collinear factorization. This issue is even more pronounced in DVMP measurements at JLab, including exclusive $\rho^0$, $\phi$, $\pi^0$, $\eta$, and charged-pion electroproduction~\cite{CLAS:2001zwd,CLAS:2008rpm,CLAS:2014jpc,JeffersonLabHallA:2016wye,JeffersonLabHallA:2017hky,CLAS:2019uzc,CLAS:2022iqy,CLAS:2023wda} and at COMPASS including the exclusive $\pi^0$ production~\cite{COMPASS:2019fea,COMPASS:2024hvm}, spin-dependent $\omega$ production~\cite{COMPASS:2016ium}, and spin-density matrix elements in exclusive $\omega$ and $\rho^0$ muoproduction~\cite{COMPASS:2013fsk, COMPASS:2020zre,COMPASS:2022xig}. Complementary Hall C measurements have provided longitudinal--transverse-separated charged-pion cross sections and tests of their $Q^2$ dependence~\cite{Volmer:2000ek,Horn:2006tm,Tadevosyan:2007yd,Horn:2007ug,Blok:2008jy,Huber:2008id,Huber:2014ius,Huber:2014kar}. This has remained an active topic of investigation following the 12~GeV upgrade, which extends the kinematic reach to higher $Q^2$~\cite{JeffersonLabHallA:2020dhq}. Consequently, achieving a quantitative understanding of the production mechanism and constructing a comprehensive theoretical framework for DVMP remain significant theoretical challenges, with important implications for the interpretation of current and future measurements.

DVCS and DVMP have emerged as precision probes of GPDs, driven by pioneering measurements from H1, ZEUS, and HERMES at DESY, high-precision programs in JLab, and complementary measurements at COMPASS. Representative measurements of hard exclusive processes are summarized in Tables~\ref{tab:sec5:DVCSexp} and~\ref{tab:sec5:DVMPexp}. While challenges remain---including the disentanglement of multiple CFFs and, for DVMP, control of longitudinal--transverse separation and higher-order and power-suppressed effects---these measurements have significantly advanced the global extraction of GPDs in the past few decades~\cite{Cuic:2023mki,Guo:2025muf}. The future EIC is expected to provide the kinematic reach, polarization capability, and luminosity required for more in-depth investigations of GPDs with these processes, among others.

\subsubsection{Exclusive heavy-quarkonium production}

\label{sec:5:subsec_HeavyQuark}
Another important channel for GPD studies is the exclusive production of heavy quarkonium; see Sec.~\ref{sec:3_DVMP_heavy} for detailed discussions. Here, the large heavy-quark mass provides a natural hard scale for a perturbative treatment, while the formation of the quarkonium can be organized in NRQCD~\cite{Bodwin:1994jh,Ivanov:2004vd}. Unlike light-meson DVMP, this treatment does not require a large photon virtuality, and the meson light-cone distribution amplitude is replaced by NRQCD long-distance matrix elements. It can therefore be applied to both photo- and leptoproduction, for longitudinally and transversely polarized photons~\cite{Flett:2021ghh,Koempel:2011rc,Chen:2019uit}. Heavy-quarkonium production is particularly sensitive to gluon GPDs and thus provides an important probe of the nucleon's gluon structure, although perturbative and heavy-quark corrections must be controlled.

\begin{figure}[t]
    \centering
    \includegraphics[width=0.483\textwidth]{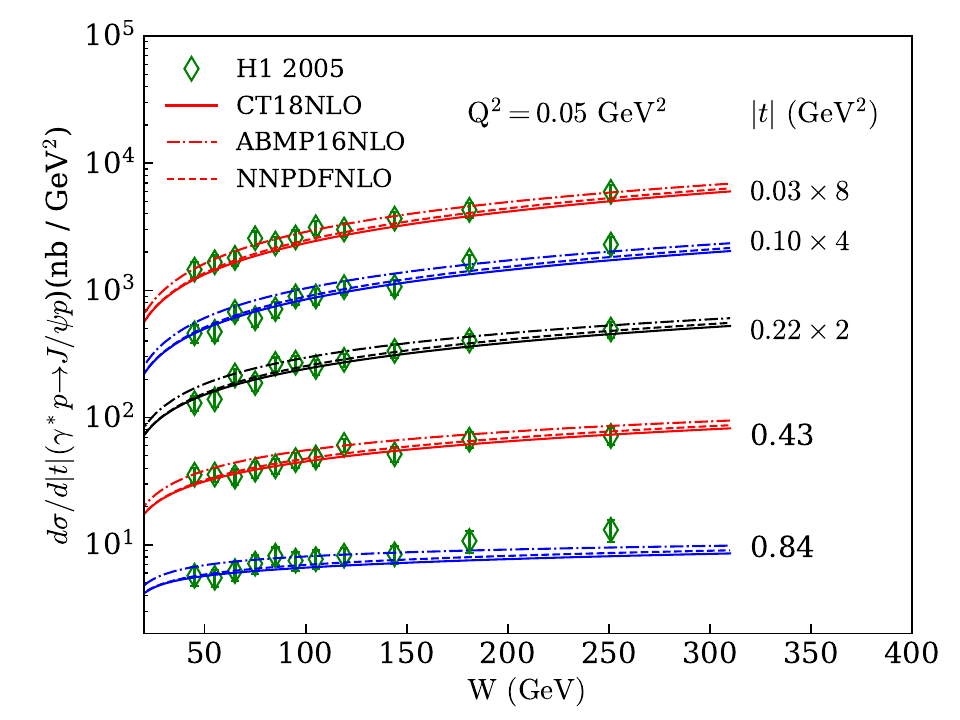}
    \caption{\raggedright An example of GPD-based analysis of exclusive $J/\psi$ electroproduction cross-section measurements by the H1 Collaboration at HERA~\cite{H1:2005dtp}, for quasi-real ($Q^2=0.05~\mathrm{GeV}^2$) photons at different center-of-momenta energies $W$ across different PDF inputs. From~\textcite{Goloskokov:2024egn}.}
    \label{fig:sec5:HERAJpsi}
\end{figure}

Two complementary kinematic regimes of exclusive heavy-quarkonium production, namely the high-energy and near-threshold regions, have been extensively studied over the past decades for the extraction of GPDs. The high-energy regime is characterized by its sensitivity to small-$x$ gluon distributions. Early fixed-target measurements of exclusive $J/\psi$ photoproduction were performed at SLAC, Cornell, and Fermilab~\cite{Camerini:1975cy,Gittelman:1975ix,Binkley:1981kv,Denby:1983wv}. These were followed by comprehensive photo- and electroproduction measurements at HERA by the H1 and ZEUS collaborations~\cite{H1:1996kyo,H1:2000kis,ZEUS:2002wfj,ZEUS:2004yeh,H1:2005dtp,H1:2013okq}. Exclusive $\Upsilon$ photoproduction, which provides a larger hard scale but a substantially lower production rate, was also measured at HERA~\cite{ZEUS:1998cdr,H1:2000kis,ZEUS:2009asc,ZEUS:2011spj}. Figure~\ref{fig:sec5:HERAJpsi} shows a recent GPD-based analysis of exclusive $J/\psi$ electroproduction cross section measurements by the H1 Collaboration~\cite{H1:2005dtp}, demonstrating an excellent description of the data for both quasi-real ($Q^2 = 0.05~\mathrm{GeV}^2$) photons at different center-of-momenta energies $W$ across different PDF inputs~\cite{Goloskokov:2024egn}. Nevertheless, sizable perturbative corrections and strong factorization-scale dependence have been reported at NLO~\cite{Ivanov:2004vd,Jones:2015nna,Jones:2016icr,Flett:2021ghh}, motivating optimized scale choices, the resummation of logarithmically enhanced small-$x$ contributions, or related high-energy approaches~\cite{Flett:2019pux,Flett:2020duk,Flett:2024htj}.

\begin{figure}[t]
    \centering
    \includegraphics[width=0.36\textwidth]{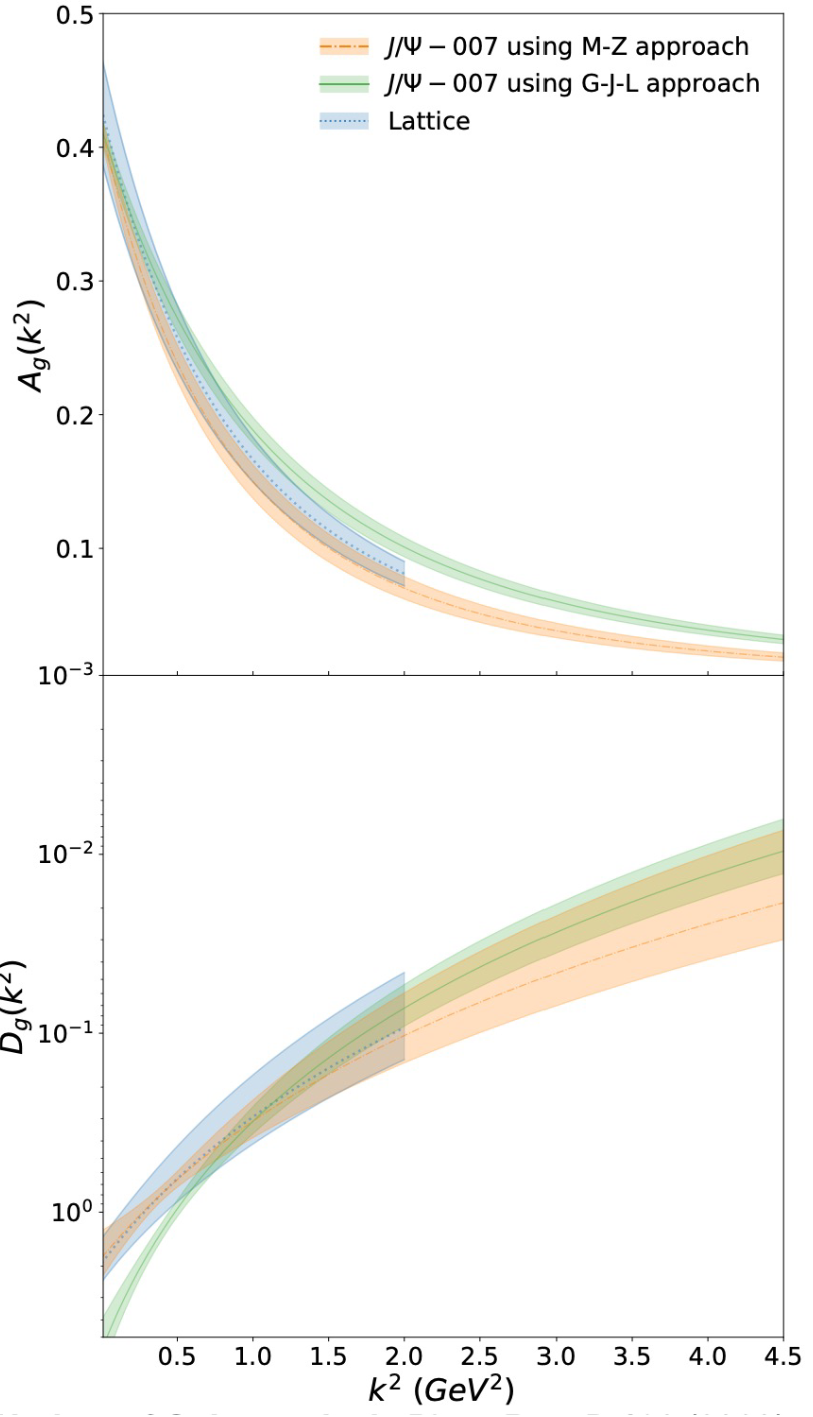}
    \caption{\raggedright Phenomenological extractions of the proton gluon EMT form factors using near-threshold $J/\psi$ production at JLab~\cite{Duran:2022xag}, based on GPD frameworks~\cite{Guo:2021ibg, Guo:2023pqw} and holographic QCD approaches~\cite{Mamo:2019mka, Mamo:2021krl}, compared with lattice QCD results~\cite{Hackett:2023rif}. Adapted from~\textcite{Duran:2022xag} with revisions from~\textcite{Guo:2023pqw} by S.~Joosten \textit{et al.}.}
    \label{fig:sec5:JLabJpsi}
\end{figure}

Near-threshold heavy-quarkonium productions have attracted substantial attention from both experimental~\cite{GlueX:2019mkq,Duran:2022xag,GlueX:2023pev,007:2026dow,Chatagnon:2026qsv,Tyson:2026gnd} and theoretical studies~\cite{Gryniuk:2016mpk,Hatta:2018ina,Hatta:2019lxo,Mamo:2019mka,Mamo:2021krl,Guo:2021ibg,Guo:2023pqw,Sun:2021pyw,Sun:2021gmi,Kharzeev:2021qkd} with similar developments on near-threshold exclusive $\phi$ electroproduction~\cite{Hatta:2025vhs}. With a large-skewness expansion, they probe nucleon EMT form factors, i.e., moments of GPDs; see Sec.~\ref{sec:3_DVMP_heavy}. The first near-threshold measurement was reported by the GlueX Collaboration in Hall D~\cite{GlueX:2019mkq}, followed by high-statistics measurements from the $J/\psi$-007 experiment in Hall C~\cite{Duran:2022xag,007:2026dow} and an extended GlueX analysis covering the full near-threshold kinematic region~\cite{GlueX:2023pev}. More recently, CLAS12 reported complementary total and differential cross-section measurements using electroproduced quasi-real photons~\cite{Chatagnon:2026qsv} and extended the measurements to the neutron target~\cite{Tyson:2026gnd}. These experiments provide supplemental kinematic coverage and permit important cross-checks of the production mechanism. Figure~\ref{fig:sec5:JLabJpsi} illustrates representative extractions of the gluon EMT form factors in the proton from near-threshold $J/\psi$ production data at JLab, obtained using both GPD-based and holographic QCD approaches, and benchmarked against state-of-the-art lattice QCD simulations, showing good overall agreement~\cite{Duran:2022xag}, noting that relativistic corrections and competing mechanisms could also be important~\cite{Du:2020bqj,Strakovsky:2019bev,JointPhysicsAnalysisCenter:2023qgg,Blask:2025jua}.

Except for some exploratory investigations~\cite{Guo:2024wxy,Goloskokov:2024egn}, exclusive heavy-quarkonium production has nevertheless not been incorporated into most GPD global analyses~\cite{Kumericki:2007sa,Kumericki:2009uq,Cuic:2023mki,Guo:2024wxy,Guo:2025muf}. In addition to subtleties associated with perturbative and heavy-quark corrections, the NRQCD formalism is also more complicated to implement. Incorporating exclusive heavy-quarkonium production into global fitting frameworks is therefore expected to constitute a central element of future GPD extractions, particularly for constraining gluon GPDs, and will be essential for fully exploiting the precision program of the EIC.

\subsection{Phenomenological and global extractions of GPDs}
\label{subsec:5c}

GPD phenomenology has long faced two major challenges: the construction of sufficiently flexible parameterizations that respect all theoretical constraints, and the inverse problem that limits GPD determination from conventional hard exclusive $2\to2$ processes. A particularly important property of GPDs is the polynomiality constraint on their Mellin moments as discussed in Sec. \ref{subsec:2_polynomiality}. These polynomiality constraints constitute an infinite tower of integral relations among GPDs, making it highly nontrivial to construct realistic parameterizations or other descriptions that satisfy them exactly. Meanwhile, it is well-known that GPD extractions from hard exclusive $2\to2$ processes such as DVCS and DVMP suffer from a severe inverse problem, as observables provide only limited sensitivity to the full $x$ dependence. Consequently, distinct GPDs can reproduce the same PDFs and exclusive observables, as illustrated by the so-called ``shadow'' GPDs~\cite{Bertone:2021yyz}, preventing a unique phenomenological determination. On the other hand, this ambiguity/indetermination can be substantially reduced by incorporating complementary lattice-QCD constraints, as discussed in Sec.~\ref{sec:4_lattice_qcd}. In the following, we review progress made in the past decades in addressing these issues in GPD determinations from both the perspective of global analysis and other nonperturbative approaches that provide complementary insights into GPDs.

\subsubsection{Double-distribution representation}

One of the earliest nontrivial constructions of nucleon GPDs is based on the double-distribution representation~\cite{Radyushkin:1997ki, Radyushkin:1998es}, where GPDs, taking the quark ones for illustration, are written as,
\begin{equation}
H_q(x,\xi,t)=H_q^{\mathrm{DD}}(x,\xi,t)+\theta(|\xi|-|x|)\,D_q(x,\xi,t)\,.
\end{equation}
The double-distribution (DD) contribution can be further expressed as an integral transform,
\begin{align} 
\begin{split} H_{q}^{\mathrm{DD}}(x,\xi,t)=\int_{-1}^1 \text{d}\beta &\int_{-1+|\beta|}^{1-|\beta|}\text{d}\alpha\delta(x-\beta-\alpha\xi)\\ &\times \pi_q(\alpha,|\beta|) f_q(|\beta|,t)\ , 
\end{split} 
\end{align}
and the $D$-term~\cite{Polyakov:1999gs} reads
\begin{equation}
D_q(x,\xi,t)=\left[1-\left(\frac{x}{\xi}\right)^2\right]
\sum_{n=0} d_q^{(2n+1)}(t)\, C_{2n+1}^{3/2}\left(\frac{x}{\xi}\right)\,.
\end{equation}
These $d_q^{(2n+1)}(t)$ will generate the $C^{(2n+2)}_q(t)$ form factors in the polynomiality constraints shown in eqs.~\eqref{eq:quark_H_polynomiality} and~\eqref{eq:quark_E_polynomiality}. Here, the Gegenbauer polynomials $C_{n}^{3/2}\left(x\right)$ are introduced as discussed previously in Sec.~\ref{subsec:2_evolution_conf}.

The profile function $\pi_q(\alpha,|\beta|)$ encodes the skewness dependence. Polynomiality follows from the DD support and the separate $D$-term, while a normalized profile preserves the forward limit. A commonly used normalized ansatz is~\cite{Musatov:1999xp}
\begin{equation}
\pi^{(b)}(\alpha,|\beta|)=
\frac{\Gamma(2b+2)}{2^{2b+1}\Gamma^2(b+1)}
\frac{\big[(1-|\beta|)^2-\alpha^2\big]^b}{(1-|\beta|)^{2b+1}}\,,
\end{equation}
with $b=1$ and $b=2$ conventionally adopted for quark and gluon GPDs, respectively. The function $f_q(|\beta|,t)$ denotes the zero-skewness GPD or $t$-dependent PDF, often modeled with global fits of PDFs and phenomenological input for the $t$ dependence. 

Within this framework, GPDs are constructed from PDF inputs, selected profile functions, and a modeled $t$ dependence, supplemented by an essentially unconstrained $D$-term. Prominent phenomenological implementations include the VGG~\cite{Vanderhaeghen:1999xj} and GK~\cite{Goloskokov:2005sd, Goloskokov:2007nt} models, which remain widely used in contemporary studies. In the GK model, the $t$ dependence is introduced via a Regge-inspired parametrization of the PDFs,
\begin{equation}
    f_q(x,t) = e^{b t}\, x^{-\alpha' t}\, f_q(x)\,,
\end{equation}
where $f_q(x)$ is the forward gluon PDF, and the exponential and Regge terms capture the observed phenomenological $t$ dependence in the high-energy, small-$t$ region. The VGG model, on the other hand, implements a slightly more complicated $t$ dependence suitable across both the small- and large-$t$ regions~\cite{Guidal:2004nd}, whereas the $D$-terms can always be added separately.

One key limitation of the double-distribution representation concerns the choice of profile functions, which limits the flexibility in the $\xi$ dependence of GPDs. In particular, the restricted functional form can hinder a sufficient description of experimental data at small Bjorken-$x$~\cite{Freund:2002qf}. This has motivated the development of alternative approaches that provide a more general and flexible framework for constructing GPDs.

\subsubsection{Conformal-moment framework and Kumeri\v{c}ki--M\"uller analysis}

Early attempts to generalize GPD parameterizations include the dual parameterization~\cite{Polyakov:2002wz} and the conformal moment representation~\cite{Mueller:2005ed}. Both approaches exploit the fact that direct parameterization of GPD moments naturally implements polynomiality constraints and facilitates QCD evolution. In particular, Gegenbauer moments diagonalize the LO QCD evolution equations~\cite{Belitsky:1997pc}, providing an ideal basis for GPD parameterizations; see Sec. \ref{sec:3_DVCS} for detailed discussion.

In the conformal-moment framework, observables such as DVCS and DVMP amplitudes, besides GPDs themselves, can be expressed as contour integrals, enabling global analyses of GPDs without explicit $x$-space expressions. Building on this framework, the Kumeri\v{c}ki--M\"uller (KM) analyses~\cite{Kumericki:2007sa,Kumericki:2009uq} were among the first systematic GPD fits based on conformal moments, with NLO Wilson coefficients and QCD evolution for DVCS. The conformal-moment framework was subsequently developed for NLO DVMP in \textcite{Muller:2013jur}, and more recently employed in a simultaneous NLO analysis of GPDs incorporating DIS, DVCS, and DVMP data in \textcite{Cuic:2023mki}.

In conformal moment space, the polynomiality constraints on GPDs immediately imply that the conformal moments of GPDs defined in Eq.~\eqref{eq:GPD_conformal_expansion} also acquire a polynomial form, 
\begin{equation}
\mathcal{F}_{j}(\xi,t) = \sum_{\substack{k=0, \,\mathrm{even}}}^{j+1} \xi^k \mathcal{F}_{j,k}(t)\,,
\end{equation}
which truncates at $j+1$. Here, the polynomiality in terms of GPD's Mellin moments is directly reflected in the conformal moments. These generalized conformal form factors $\mathcal{F}_{j,k}(t)$, to be distinguished from the Mellin ones, can be classified into three distinct categories: the forward $\mathcal{F}_{j,0}(t)$, the regular off-forward $\mathcal{F}_{j,k}(t)$ with $j \ge k \ge 2$, and the $\mathcal{F}_{j,j+1}(t)$ for odd $j$. 

The forward moments are simply the Mellin moments of the $t$-dependent PDFs. Motivated by the frequently used $x$-space GPD ansatz that incorporates Regge-inspired small-$x$ behavior through an effective trajectory $\alpha_i(t)$,
\begin{equation}
    F(x,t) = \sum_i N_i\, x^{-\alpha_i(t)} (1-x)^{\beta_i} R_i(t)\ ,
\end{equation}
where multiple components, each labeled by $i$, can be used for a given GPDs for improved flexibility. The equivalent parameterization in moment space can be constructed using the Euler beta function $B(x,y)$,
\begin{equation}
    \mathcal{F}_{j,0}(t) = \sum_i N_i \,
    \frac{B\!\left(j+1-\alpha_i(t),\,1+\beta_i\right)}
         {B\!\left(2-\alpha_i(0),\,1+\beta_i\right)}\,
    R_i(t)\ .
\end{equation}
Variants of this GPD forward moments ansatz are commonly used in GPD analyses~\cite{Kumericki:2007sa, Kumericki:2009uq, Muller:2014wxa, Guo:2022upw, Guo:2023pqw}. Commonly adopted parameterizations for the residual $t$-dependence factor $R_i(t)$ include an exponential form, $R_i(t)=\exp(b_i t)$, and dipole or tripole forms, $R_i(t) = (1 - t/m_i^2)^{-p}$ with $p=2,3$, motivated by phenomenological observations.

The regular off-forward moments $\mathcal{F}_{j,k}(t)$ with $j\ge k\ge2$ are much less constrained and can therefore be parameterized flexibly. Cross-channel partial-wave expansions provide one systematic approach~\cite{Kumericki:2009uq,Muller:2014wxa}. Additional guidance on their skewness dependence can come from theoretical approaches such as holographic QCD~\cite{Mamo:2024jwp}. The moments $\mathcal{F}_{j,j+1}(t)$ associated with the $D$ term are typically treated separately from the regular off-forward moments. They include the EMT form factor $C(t)$, whose extraction, including through dispersion-relation methods, provides an important experimental probe of nucleon mechanical properties~\cite{Burkert:2018bqq,Kumericki:2019ddg,Dutrieux:2021nlz,Burkert:2023wzr}.

\begin{figure}[t]
    \centering
    \includegraphics[width=0.483\textwidth]{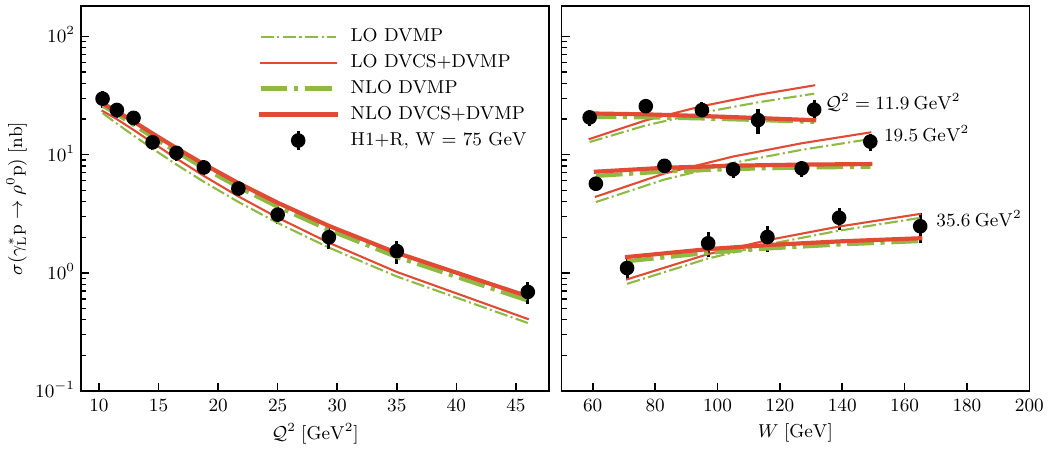}
    \includegraphics[width=0.483\textwidth]{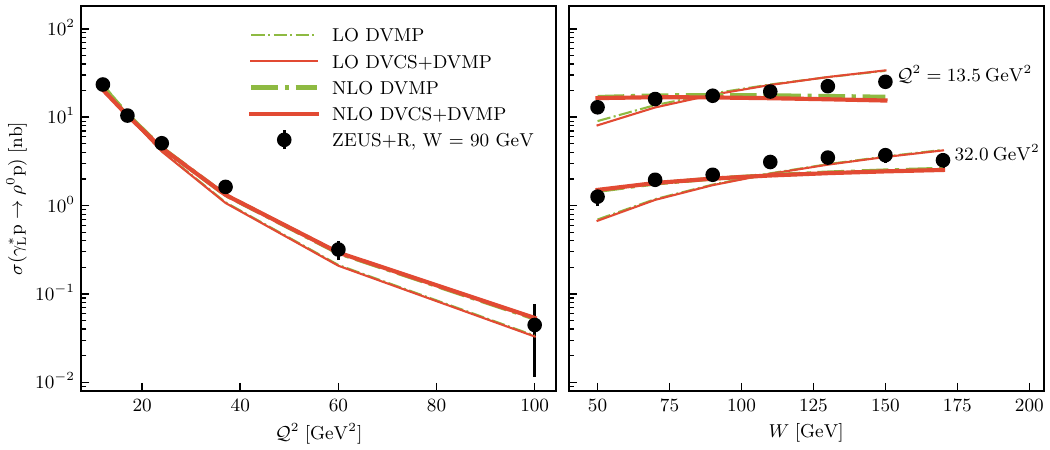}
    \caption{\raggedright Examples of recent conformal-moment based fits to small-$\xi$ DVMP measurements by H1~\cite{H1:2009cml} and ZEUS~\cite{ZEUS:2007iet}. From~\textcite{Cuic:2023mki}.}
    \label{fig:sec5:KMDVMPFit}
\end{figure}

The KM model was formulated within the framework described above, incorporating two major simplifications. First, the skewness-dependent terms are truncated and modeled through ratios to the forward ones, reflecting the fact that the available experimental data predominantly probe kinematic regions with small $\xi$. Second, the valence contributions are modeled solely through the imaginary parts of their corresponding CFFs and corresponding dispersion relations, consistent with the model's focus on the high-energy region where valence contributions are expected to be suppressed. Early KM analyses used HERA DVCS measurements~\cite{Kumericki:2007sa,Kumericki:2009uq}; later studies incorporated HERMES and JLab DVCS observables, flavor separation, and HERA DVMP data~\cite{Kumericki:2016ehc,Cuic:2020iwt}. More recently, the conformal-moment framework has been applied to a simultaneous NLO analysis of GPDs incorporating DVMP measurements from H1~\cite{H1:2009cml} and ZEUS~\cite{ZEUS:2007iet}, DVCS data from HERA~\cite{H1:2009wnw,ZEUS:2008hcd,ZEUS:2003pwh,H1:2005gdw}, and additional inputs including DIS measurements~\cite{Cuic:2023mki}, as illustrated in Fig.~\ref{fig:sec5:KMDVMPFit}. A particularly interesting observation therein is that both the fit quality and the universality of the extracted GPDs, as quantified by the skewness ratio, are significantly improved when the analysis is performed at NLO. This highlights the importance of perturbative corrections in the relevant kinematic regions and indicates potential perturbative convergence.

Given their notable phenomenological success, the KM model and other conformal-moment frameworks continue to provide an essential theoretical baseline for contemporary studies of GPDs. On the other hand, owing to the limited information on valence-quark distributions incorporated in the model and the weak sensitivity of the measurements to the $\tilde E$ GPD in the small-$x_B$ region, they do not permit a complete determination of the full set of GPDs, particularly in the valence domain. Achieving such a comprehensive description would require an expanded set of phenomenological inputs together with a more flexible and general parameterization of the GPDs.

\subsubsection{GPDs through universal moment parameterization}
\label{sec:GUMP}

Building on the same conformal moment representation of GPDs, the GPDs through Universal Moment Parameterization (GUMP) program was proposed to perform a global analysis of all four chiral-even leading-twist GPDs~\cite{Guo:2022upw, Guo:2023ahv}. A key advance of the GUMP program is the complete parameterization of all four leading-twist GPDs for valence quarks, sea quarks, and gluons, including separate $u$- and $d$-quark contributions. This provides a unified description of the $x$ dependence and Mellin moments of GPDs within a common parameterization whose flexibility can be systematically increased as increasingly stringent phenomenological constraints become available. In particular, it enables lattice-QCD inputs and other complementary observables to be incorporated consistently into global GPD analyses. Such information is essential for stabilizing GPD extractions, which otherwise constitute a severely ill-posed inverse problem when constrained by experimental observables alone as discussed above.

The GUMP framework also maintains explicit parameterizations of the $E$ and $\widetilde{E}$ GPDs, which play important roles in certain polarization observables and meson-production channels, especially in the lower-energy valence region. Their more limited phenomenological constraints, however, necessitate additional empirical assumptions, owing, for example, to the absence of forward-PDF constraints analogous to those available for $H$ and $\widetilde{H}$. In the small-$x$ region, where $E$ and $\widetilde{E}$ remain largely unconstrained, one therefore adopts relations such as $E_{\bar q}=R_{\bar q}^E H_{\bar q}$ for $q=u,d$, with analogous assumptions for $\widetilde{E}$. Despite these limitations, a unified treatment of all four leading-twist GPDs is essential, since experimental observables generally receive contributions from multiple GPDs simultaneously, and their consistent extraction requires that they be analyzed on an equal footing.

For the off-forward $\xi$ and $t$ dependence, GUMP adopts a simple ratio ansatz for the $\xi$-dependent moments, $\mathcal{F}_{j,k}(t)\equiv R_k\mathcal{F}_{j-k,0}(t)$, with the expansion truncated at $\xi^4$ in the absence of direct experimental or theoretical constraints. The $t$ dependence is modeled through factorized $R_i(t)$ factors, using exponential forms for sea distributions and dipole forms for valence distributions, supplemented by a Regge trajectory $\alpha(t)$ in a manner similar to the KM model. These choices are intended as pragmatic placeholders rather than fundamental restrictions of the framework. As increasingly precise experimental data, lattice-QCD calculations, and theoretical constraints become available, the assumed relations among moments, the truncation in skewness, and the functional forms governing the $t$ dependence can be systematically relaxed or replaced by more flexible parameterizations. In this way, GUMP provides a natural path toward global analyses with progressively reduced model dependence and a more complete quantification of the uncertainties associated with the presently weakly constrained off-forward structure.

\begin{figure}[t]
    \centering
    \includegraphics[width=0.483\textwidth]{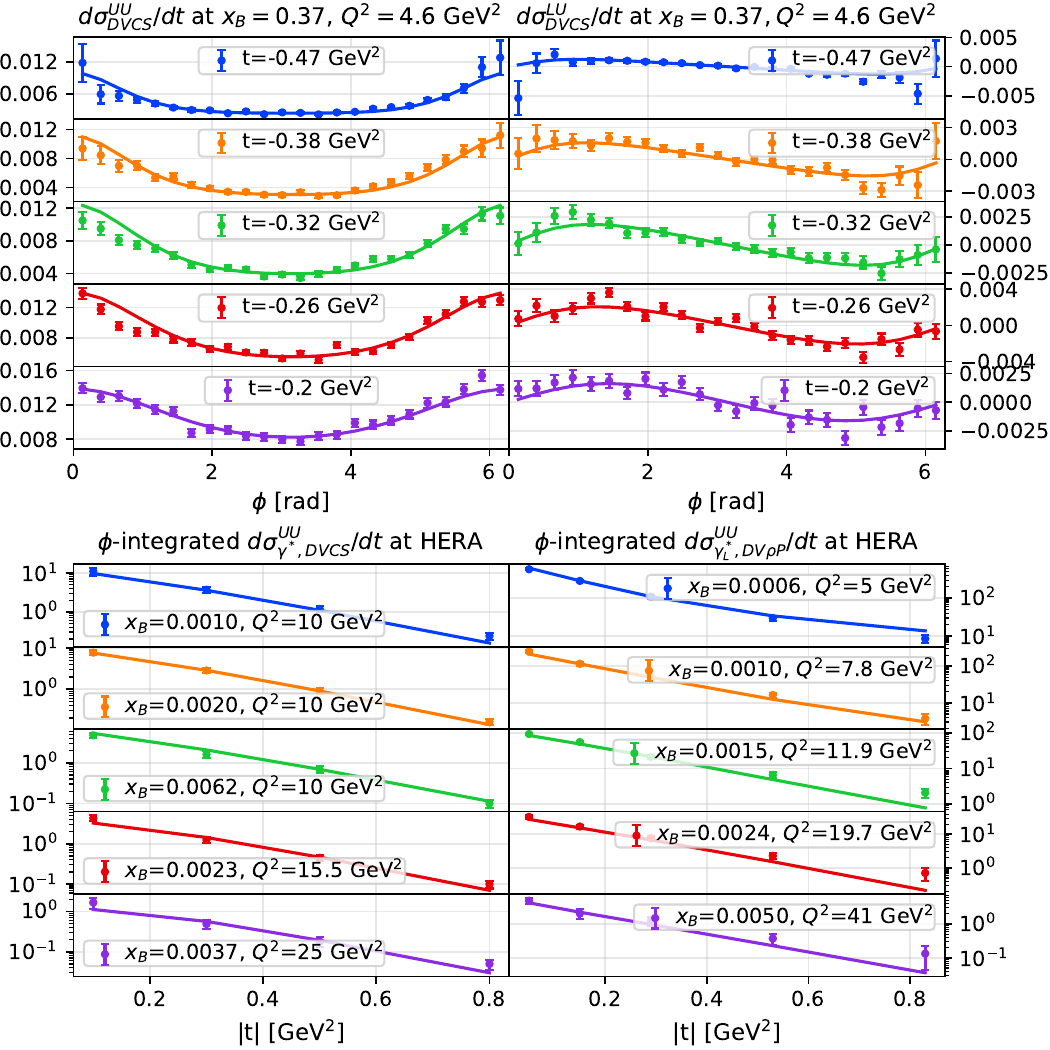}
    \caption{Representative GUMP1.0 fits to selected hard exclusive measurements: unpolarized DVCS at JLab (upper left), beam-polarized DVCS at JLab (upper right), unpolarized DVCS at HERA (lower left), and unpolarized DV$\rho$P at HERA (lower right). The JLab panels show $\phi$-differential cross sections~\cite{CLAS:2018bgk,Georges:2017xjy,JeffersonLabHallA:2022pnx}, while the HERA panels show $\phi$-integrated cross sections~\cite{ZEUS:2007iet,H1:2009cml,H1:2009wnw}. From~\textcite{Guo:2025muf}.}
    \label{fig:sec5:GUMPExpfit}
\end{figure}

Within this framework, the latest development, GUMP1.0~\cite{Guo:2025muf}, systematically implements lattice-QCD constraints on the $x$ dependence of GPDs at both zero and nonzero skewness~\cite{Bhattacharya:2022aob,Bhattacharya:2023jsc,Chu:2025kew}, with NLO QCD evolution consistently applied to both the experimental observables and the GPDs themselves. Additionally, it incorporates constraints from globally extracted polarized and unpolarized PDFs by JAM~\cite{Cocuzza:2022jye}, global fits of nucleon electromagnetic form factor~\cite{Ye:2017gyb}, lattice calculations of generalized form factors up to the fifth Mellin moment~\cite{Bhattacharya:2023ays}, along with other experimental measurements of exclusive production cross sections. Figure~\ref{fig:sec5:GUMPExpfit} shows representative fits to experimental observables in GUMP1.0 including unpolarized and beam-polarized DVCS differential cross sections~\cite{CLAS:2018bgk,Georges:2017xjy,JeffersonLabHallA:2022pnx} and beam-spin asymmetries~\cite{CLAS:2022syx} measured at JLab, as well as $\phi$-integrated differential cross sections for DVCS and deeply virtual $\rho$-meson production (DV$\rho$P) measured by H1 and ZEUS at HERA~\cite{ZEUS:2007iet,H1:2009cml,H1:2009wnw}. Overall, the extracted GPDs provide an excellent description of the data; in particular, they reproduce DVCS and DV$\rho$P measurements simultaneously at HERA, consistent with earlier findings in the KM framework~\cite{Cuic:2023mki}. On the other hand, potential higher-twist contributions not included in the analysis may still be relevant and warrant further investigation.

\begin{figure}[t]
    \centering
     \includegraphics[width=0.483\textwidth]{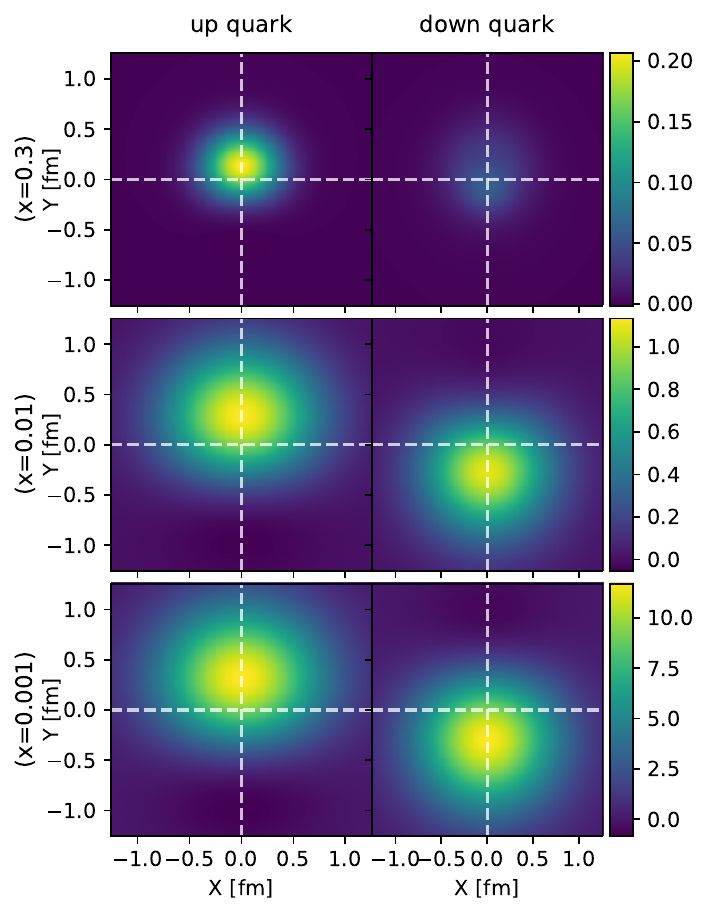}
    \caption{\raggedright The intrinsic transverse-space distributions $q_{\rm{Intrin}}^{\perp}(x,\boldsymbol b_\perp)$ of up and down quarks in a proton polarized along the $X$ direction for $x=0.3$, $0.01$, and $0.001$ with GUMP1.0 GPDs at $\mu=2$ GeV. From~\textcite{Guo:2025muf}.}
    \label{fig:sec5:GUMPTransverse}
\end{figure}

The comprehensive set of inputs employed in the analysis, not shown here in full, enables a quantitative extraction of the full GPDs. Figure~\ref{fig:sec5:GUMPTransverse} shows the inferred proton structure in transverse coordinate space from the GUMP1.0 extraction, illustrating a key motivation for GPD studies. The intrinsic transverse-space quark distributions $q_{\rm{Intrin}}^{\perp}(x,\boldsymbol b_\perp)$ in a nucleon transversely polarized along the $\hat{e}_x$ direction have been defined in Sec. \ref{sec:2_IPS},
where the contribution from the center-of-mass motion has been removed~\cite{Ji:2020hii}. As shown in Fig.~\ref{fig:sec5:GUMPTransverse}, the up-quark distribution exhibits a sizable transverse shift in a transversely polarized nucleon and dominates the transverse spin contribution. The down-quark contribution is smaller and negative, while the remaining contribution originates from gluons. These results provide a spatial representation of the Ji sum rule for transverse spin~\cite{Ji:1996ek} and illustrate the concept of nucleon tomography in the GPD framework.

With all the advances, GUMP1.0 still retains pragmatic modeling assumptions for GPD species and flavor components that remain weakly constrained. In addition, lattice-QCD systematic uncertainties are approximated empirically as 30\% of the mean values, with correlations neglected. A more complete treatment of these uncertainties and their correlations will be important for a statistically consistent combination of lattice and experimental constraints and will benefit from continued efforts across the lattice QCD and phenomenology communities.

\subsubsection{Other frameworks and computational advances}

We also highlight other recent frameworks as well as computational advances in GPD phenomenology. An important development has been the construction of common numerical infrastructures connecting GPD models, perturbative coefficient functions, CFFs, and experimental observables. Early numerical fitting tools were developed for the analysis of DVCS data~\cite{Guidal:2008ie}, while the open-source C++ framework PARTONS provides a modular environment for implementing GPD models and analyzing deeply virtual exclusive processes~\cite{Berthou:2015oaw}. It has supported phenomenological studies ranging from conventional GPD fits to neural-network extractions of CFFs~\cite{Moutarde:2018kwr,Moutarde:2019tqa, Mezrag:2026wcf}. Dedicated event generators have also been developed for exclusive processes sensitive to GPDs~\cite{Aschenauer:2022aeb}.  Additional efforts are devoted to parameterizing~\cite{Hashamipour:2021kes, Kriesten:2021sqc, Panjsheeri:2025vpa, Irani:2026ogc} and perturbatively evolving GPDs~\cite{Vinnikov:2006xw,Bertone:2022frx, Freese:2024ypk} directly in $x$-space, providing independent cross-checks and complementary insights to the moment-space techniques discussed above. More recently, the development of an open database for GPD analyses has aimed to promote reproducibility and consistent comparisons among experimental data, phenomenological models, and lattice-QCD inputs~\cite{Burkert:2025gzu}.

The inverse problem of GPD extraction naturally separates into two levels: the extraction of CFFs from experimental observables and the subsequent reconstruction of the underlying $x$-dependent GPDs. The first problem is already nontrivial because DVCS observables depend both linearly and quadratically on four complex CFFs in the leading-twist chiral-even sector alone. Although a sufficiently complete set of polarization observables can, in principle, determine all leading-twist CFF components~\cite{Shiells:2021xqo}, currently available measurements generally leave weakly constrained directions and may admit continuous families of solutions. Neural-network parametrizations, pioneered in Ref.~\cite{Kumericki:2011rz}, reduce the functional bias associated with fixed analytic ansatze and permit global descriptions across different kinematic regions~\cite{Moutarde:2019tqa}. Related developments include neural-network emulators of multidimensional cross sections~\cite{Grigsby:2020auv}, flavor-separated CFF analyses~\cite{Cuic:2020iwt}, physics-informed benchmarking and uncertainty studies~\cite{Almaeen:2022imx}, generative and Bayesian treatments of nonunique solutions~\cite{Almaeen:2024guo}, and global neural-network fits constrained by local likelihood or $\chi^2$ information~\cite{CaleroDiaz:2025luc}.

Reconstructing GPDs from CFFs is more severely ill posed, which leads to the shadow GPDs as discussed at the beginning of this subsection~\cite{Bertone:2021yyz,Bertone:2021wib,Moffat:2023svr}. Theory-constrained neural networks have therefore been developed to incorporate support, polynomiality, positivity, and known reduction limits while retaining substantial functional flexibility~\cite{Dutrieux:2021wll}. Bayesian studies have investigated the complementary constraints provided by lattice-QCD information~\cite{Riberdy:2023awf}, whereas symbolic regression offers compact and interpretable representations of selected GPD combinations~\cite{Dotson:2025omi}. Recent physics-informed approaches embed the principal-value convolution directly into the learning architecture~\cite{Watkins:2025apc}, constrain GPD networks simultaneously through CFFs and lattice Mellin moments~\cite{Xu:2026lko}, or combine neural-network representations with numerical QCD information~\cite{Chu:2025jsi}.

That said, systematic uncertainties from model architecture, training, theoretical constraints, and perturbative accuracy remain to be fully quantified. Moreover, flexible parametrizations cannot compensate for limited experimental constraints, underscoring the need for uncertainty quantification, benchmarking, and comprehensive analyses with combined inputs. Nevertheless, these advances provide an increasingly mature foundation for multidimensional nucleon imaging at JLab and the future EIC.

\subsection{Nucleon models for GPDs}

Furthermore, we highlight model-based approaches to the multidimensional structure of the nucleon. Owing to their dependence on the three kinematic variables $(x,\xi,t)$ and the limited existing constraints, GPDs cannot yet be determined in a fully model-independent manner. Nonperturbative models therefore remain valuable complements to phenomenological extractions and lattice QCD. Besides providing qualitative insight into the underlying quark and gluon dynamics, they supply information in regions where experimental or lattice inputs are sparse.

Light-front quantization~\cite{Dirac:1949cp} provides a natural framework for partonic structure. At fixed light-front time $x^+=0$, hadron states admit a Fock-space expansion in multiparton light-front wave functions (LFWFs)~\cite{Brodsky:1997de}. GPDs can then be represented as overlaps of LFWFs with different parton momenta and, at nonzero skewness, possibly different Fock-particle numbers~\cite{Brodsky:2000xy,Diehl:2000xz}. Constituent-quark, quark--diquark, spectator, bag, and phenomenological light-front models have been widely used to study chiral-even and chiral-odd GPDs, transverse densities, and EMT form factors~\cite{Ji:1997gm,Boffi:2002yy,Pasquini:2005dk,Goldstein:2010gu,Lorce:2011dv,Gutsche:2013zia,Mondal:2015uha,Chakrabarti:2015ama,Neubelt:2019sou,Kriesten:2021sqc,Tezgin:2024tfh}. Meson-cloud extensions additionally incorporate higher Fock sectors and provide insight into sea-quark and peripheral nucleon structure~\cite{Pasquini:2006dv,Pasquini:2006ib,Strikman:2009bd}.  A more dynamical light-front approach is basis light-front quantization (BLFQ), in which a model Hamiltonian is diagonalized in a systematically enlarged basis. The BLFQ approach has been widely applied to investigate nucleon GPDs and associated angular-momentum and mechanical distributions~\cite{Mondal:2019jdg,Xu:2021wwj,Liu:2022fvl,Kaur:2023lun,Liu:2024umn}; see~\textcite{Vary:2025yqo} for a recent review. Despite truncation and effective-interaction uncertainties, the BLFQ approach offers valuable inputs toward more complete partonic descriptions.

Effective chiral models provide complementary low-scale descriptions of the nucleon. The chiral quark-soliton model treats the nucleon as a large-$N_c$ pion-field soliton containing relativistic valence quarks and a polarized Dirac sea, and has been applied to GPDs and their flavor, spin, and mechanical structure~\cite{Petrov:1998kf,Goeke:2001tz,Schweitzer:2002nm,Ossmann:2004bp,Wakamatsu:2006dy,Goeke:2007fp,Schweitzer:2016jmd,Won:2023ial,Kim:2024ibz}. Furthermore, the Nambu–Jona–Lasinio model incorporates dynamical chiral symmetry breaking and, for baryons, scalar and axial-vector diquark correlations, providing predictions for finite-skewness GPDs, form factors, and quark angular momentum~\cite{Mineo:2005qr,Freese:2019bhb,Freese:2020mcx,Chandra:2026smf}.

Another important direction is holographic QCD. Motivated by gauge--gravity duality~\cite{Maldacena:1997re}, holographic models describe aspects of strongly coupled QCD through bulk gravitational fields in a higher-dimensional spacetime. Top-down and bottom-up constructions reproduce important features of low-energy hadron physics, including confinement spectra and chiral symmetry breaking~\cite{Sakai:2004cn,Erlich:2005qh,Karch:2006pv}; see \textcite{Kim:2012ey} for a review. In light-front holography, the holographic coordinate is identified with a boost-invariant transverse partonic separation, leading to an effective light-front Schr\"odinger equation and analytic approximations to hadronic LFWFs~\cite{Brodsky:2006uqa,Brodsky:2014yha}. Early holographic studies inferred zero-skewness nucleon GPDs by matching electromagnetic form factors to GPD sum rules~\cite{Vega:2010ns,Chakrabarti:2013gra}. Later developments related their $x$ and $t$ dependence to Regge behavior, constituent-counting rules, and generalized Veneziano amplitudes~\cite{deTeramond:2018ecg}. Recent string-inspired constructions extend the description to arbitrary skewness while implementing polynomiality and permit comparisons of Mellin moments and form factors with lattice QCD~\cite{Mamo:2024jwp,Mamo:2024vjh,Hechenberger:2025wnz}. Holographic amplitudes have also been applied to near-threshold heavy-quarkonium production and the extraction of gluonic EMT form factors~\cite{Mamo:2019mka,Mamo:2021krl}. Their quantitative predictions, however, depend on the holographic background and the matching of bulk fields to QCD operators.

Instanton-based descriptions of the QCD vacuum provide a different
perspective on nonperturbative nucleon structure. In a distinct semiclassical picture, the instanton liquid model treats the vacuum as an ensemble of instantons and anti-instantons whose fermionic zero modes generate effective quark interactions. It describes spontaneous chiral symmetry breaking and anomalous $U_A(1)$ dynamics and provides low-resolution inputs for hadronic wave functions, distributions, and form factors~\cite{Shuryak:2021fsu,Shuryak:2022thi}. Instanton-based studies have addressed nucleon EMT form factors and quark--gluon forces~\cite{Polyakov:2018exb}, as well as light-front wave functions, quark and gluon distributions, GPDs, and hadronic form factors~\cite{Shuryak:2023siq,Liu:2024jno,Liu:2024vkj,Kim:2023pll}; see~\textcite{Shuryak:2026pqt} for a recent review.

Continuum functional approaches based on Dyson--Schwinger and Bethe--Salpeter equations provide a direct connection between hadron structure and the underlying QCD correlation functions~\cite{Roberts:1994dr,Maris:2003vk}; see \textcite{Eichmann:2016yit} for a review. They have been developed extensively for pion and kaon GPDs, yielding predictions for their spatial and mechanical structure~\cite{Mezrag:2014tva,Mezrag:2016hnp,Shi:2020pqe,Zhang:2021mtn,Raya:2021zrz}.The extension to nucleon GPDs remains considerably less developed because of the additional three-body dynamics associated with the baryon bound state.

These approaches capture complementary aspects of nonperturbative QCD, from partonic structure and chiral dynamics to confinement, vacuum topology, and dressed QCD interactions. Although consistent comparisons require further theoretical development, their combination with experimental and lattice-QCD constraints helps identify robust features of nucleon structure and guide future measurements at JLab and the Electron--Ion Collider.

\subsection{Nucleon spin, mass, and forces}

\begin{figure}[t]
    \centering
     \includegraphics[width=0.9\columnwidth]{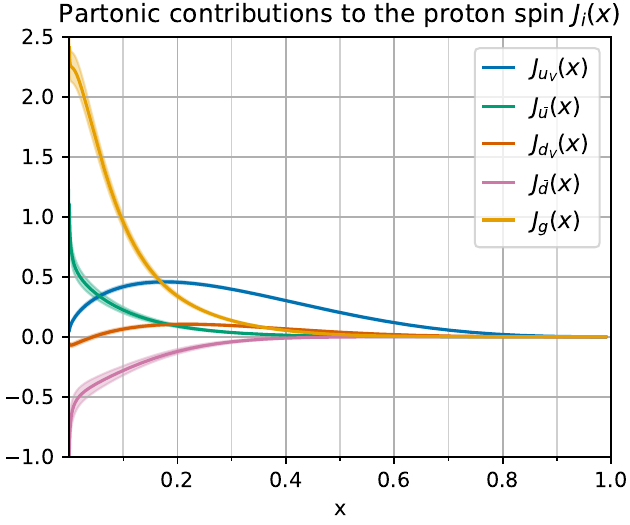}
    \caption{\raggedright Proton transverse spin decomposition in terms of quark and gluon parton angular momenta, obtained from the GUMP1.0 GPDs at $\mu=2~\mathrm{GeV}$, $J(x)=\sum_i J_i(x)$, and $\int^1_0 J(x)dx=1/2$. The bands represent $90\%$ confidence intervals obtained via Monte Carlo propagation of the quoted Hessian uncertainties. Parameter correlations, theoretical systematics, and parameterization uncertainties are not included. Adapted from~\textcite{Guo:2025muf}.}
    \label{fig:sec5:spinsumrule}
\end{figure}

We conclude this section with the implications of GPD phenomenologies for the proton's spin, mass, and internal-force structure. These properties are encoded in GPDs and the associated EMT form factors,  thereby connecting measurements of exclusive processes with the most fundamental and interesting aspects of the nucleon structure physics. 

As discussed in Sec.~\ref{subsec:2_spin_t3GPD}, the combinations of twist-two GPDs $J_{q,g}(x)=x[H_{q,g}(x)+E_{q,g}(x)]/2$ admit an interpretation as the partonic angular-momentum densities of quarks and gluons in a transversely polarized proton. Figure~\ref{fig:sec5:spinsumrule} presents the proton spin decomposition obtained from the GUMP1.0 GPDs at $\mu=2$ GeV~\cite{Guo:2025muf}. This global analysis combines hard exclusive measurements with constraints from collinear PDFs, electromagnetic form factors, and lattice-QCD calculations of GPD moments and related observables. It thus provides one of the most comprehensive phenomenological determinations of the complete quark and gluon angular-momentum distributions currently available. 

The uncertainty bands in Fig.~\ref{fig:sec5:spinsumrule} should nevertheless be interpreted with caution. They reflect only the propagation of the quoted Hessian uncertainties of the fitted parameters and do not incorporate their full correlations, theoretical uncertainties, or the dependence on the chosen GPD parameterization. These limitations are particularly relevant at small $x$, where direct experimental constraints remain sparse and the extrapolation toward $x=0$ can significantly affect the integrated quark and gluon angular momenta. Extending the kinematic reach toward small $x$ and improving the flavor and gluon sensitivity of exclusive measurements will therefore be essential for a quantitatively robust spin decomposition.

\begin{figure}[t]
    \centering
     \includegraphics[width=\columnwidth]{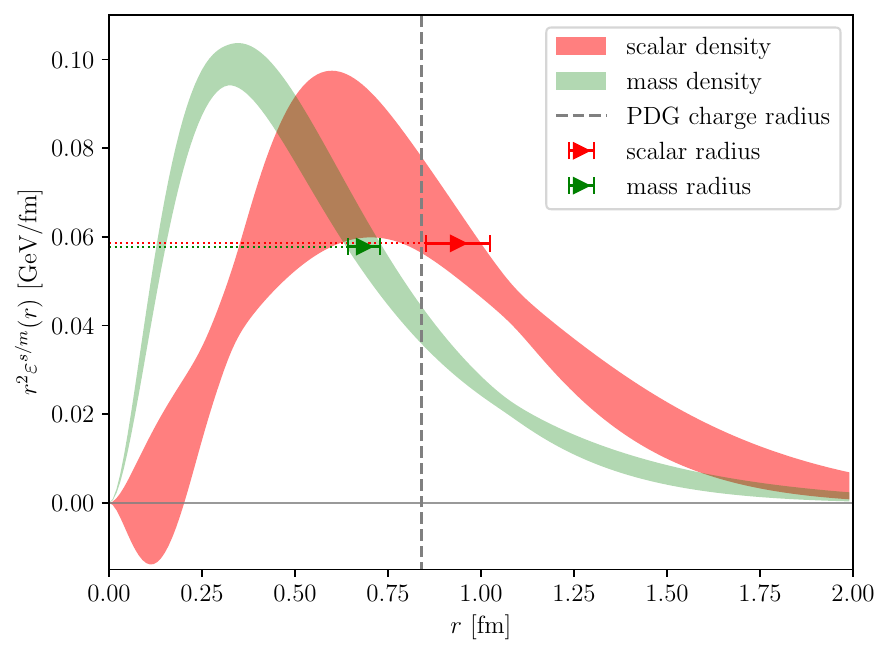}
    \caption{\raggedright A comparison of different measures of the proton size; the scalar and mass radii, with their corresponding scalar ($\varepsilon^s(r)$) and mass ($\varepsilon^m(r)$) density profiles (See Sec.~\ref{subsec:2_mass_scalar_force}), are obtained by combining gluon contributions extracted from a global analyses near-threshold $J/\psi$ photoproduction and with quark contributions from~\textcite{Guo:2025jiz}. These are compared against the canonical proton size measure, its electric charge radius ($0.8409 \pm 0.0004\,\mathrm{fm}$), using the PDG average~\cite{ParticleDataGroup:2024cfk}. The scalar strong-interaction radius ($0.94 \pm 0.09\,\mathrm{fm}$) is the largest of the three, reflecting the realistic spatial extent of the nucleon. From~\textcite{Ji:2026vjl}.}
    \label{fig:sec5:radii}
\end{figure}

\begin{figure}[t]
    \centering
     \includegraphics[width=0.9\columnwidth]{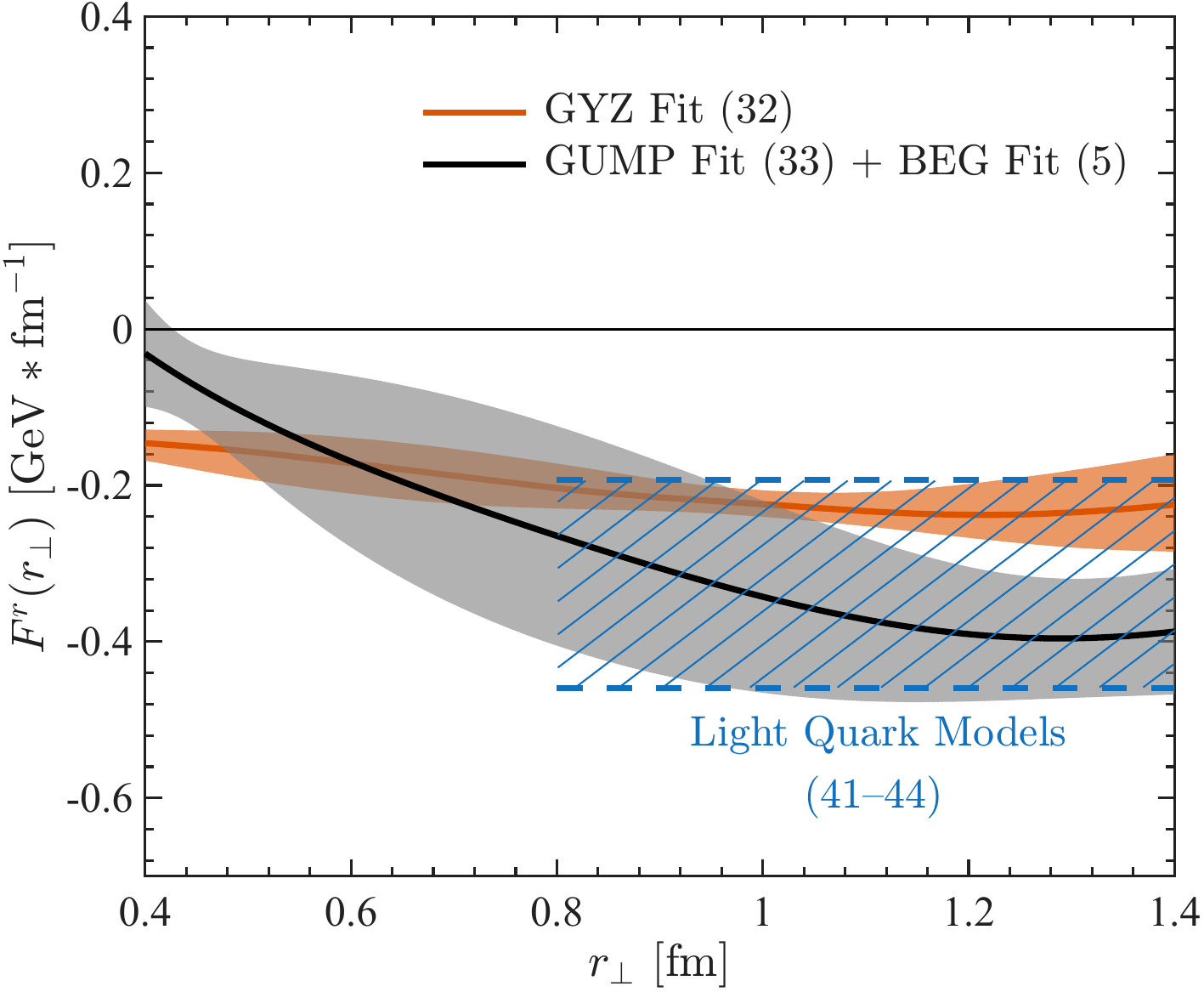}
    \caption{\raggedright Strong force on quarks in the proton in the transverse plane of the infinite momentum frame, obtained from the latest global analyses of the quark GPDs and EMT form factors. The figure shows the total force from combining the \textcite*{Burkert:2018bqq} fit and the GUMP1.0 fit~\cite{Guo:2025muf} and from the fit dominated by LQCD calculations (orange curve and band)~\cite{Hackett:2023rif}. The blue hatched band shows the string tensions from several relativistic quark models~\cite{LealFerreira:1979xq,Eich:1983kg,Dziembowski:1996cv,Glozman:1997ag}. From~\textcite{Ji:2026lyj}.}
    \label{fig:sec5:force}
\end{figure}

Within the GPD and EMT framework, the size of the proton can be characterized not only through its electromagnetic charge distribution, but also through the spatial distributions of energy and scalar density as introduced in Sec.~\ref{subsec:2_mass_scalar_force}. The corresponding mass and scalar radii probe, respectively, the extent of the energy distribution and the strong-interaction scale~\cite{Ji:2021mtz}. These measures encode distinct aspects of nucleon structure and therefore need not coincide. A recent study combines gluon EMT form factors extracted from a global analysis of near-threshold $J/\psi$ photoproduction within a holographic QCD framework with quark contributions from a Bayesian GFF analysis to reconstruct the total mass and scalar density profiles~\cite{Ji:2026vjl}. The extracted scalar radius, $0.94\pm0.09\,\mathrm{fm}$, exceeds in central value both the mass radius and the electromagnetic charge radius, $0.8409\pm0.0004\,\mathrm{fm}$, suggesting a broader spatial distribution of scalar density than of energy or charge. This comparison illustrates how EMT form factors complement electromagnetic measurements by probing the spatial structure associated with nucleon mass generation and strong interactions. A quantitative interpretation nevertheless requires control over the theoretical relation between near-threshold quarkonium production and gluon EMT form factors, including its dependence on the assumed production mechanism.

Beyond angular momentum and mass, the EMT form factors also characterize the spatial distributions of the stress tensor or more accurately, the momentum-current distributions~\cite{Polyakov:2002yz,Polyakov:2018zvc,Lorce:2018egm,Freese:2021czn}. Experimentally, the quark $C/D$ form factor has been constrained through dispersive analyses of DVCS measurements~\cite{Burkert:2018bqq}. Near-threshold $J/\psi$ photoproduction measurements from GlueX, Hall~C, and CLAS12 provide complementary sensitivity to gluonic structure~\cite{GlueX:2019mkq,Duran:2022xag,GlueX:2023pev,007:2026dow,Chatagnon:2026qsv}. Lattice QCD, in turn, provides first-principles calculations of the quark and gluon EMT form factors and their flavor decomposition~\cite{Shanahan:2018nnv,Shanahan:2018pib,Pefkou:2021fni,Hackett:2023rif}.

As discussed in Sec.~\ref{subsec:2_mass_scalar_force}, the QCD equations of motion relate the divergence of the quark EMT to the color-Lorentz force density between the quarks and gluons. Figure~\ref{fig:sec5:force} compares two determinations of the effective transverse force on quarks in the infinite-momentum frame~\cite{Ji:2026lyj}: one based on the experiment-constrained GUMP and dispersive-DVCS inputs~\cite{Burkert:2018bqq,Guo:2025muf}, and another based on the lattice-dominated GYZ analysis~\cite{Guo:2025jiz,Hackett:2023rif}. The former gives an average force of $-0.382(92)~\mathrm{GeV/fm}$ over $r_\perp=1.0$--$1.4~\mathrm{fm}$, whereas the latter yields $-0.217^{+0.016}_{-0.015}~\mathrm{GeV/fm}$ over $r_\perp=0.7$--$1.2~\mathrm{fm}$. The negative sign denotes an inward, attractive force. Within uncertainties, both determinations are compatible with an approximately constant force over an intermediate distance range and with the string tensions of the relativistic quark model,
providing force-based evidence for confinement. It should not, however, be identified directly with the static heavy-quark potential, nor does it constitute a formal proof of confinement~\cite{Ji:2026lyj}. Its quantitative determination remains limited by uncertainties in the EMT form factors, particularly at large momentum transfer, spatial reconstruction, force normalization, and lattice systematics, motivating broader exclusive measurements and improved lattice calculations.

Taken together, the spin decomposition, force distribution, and comparison of radii illustrate how GPDs and their EMT moments provide a unified description of several fundamental properties of the proton. A more complete picture will require consistent treatments of three-dimensional Breit-frame and two-dimensional light-front densities, improved control of experimental and lattice-QCD uncertainties, and joint analyses capable of resolving the quark and gluon contributions over a broad range of momentum fractions and momentum transfers.

\section{Conclusion and Future Prospects}
\label{sec:6_conclude}

To conclude, over the three decades since their introduction, GPDs have evolved from a largely qualitative concept into a quantitative framework for probing the quark and gluon structure of the nucleon within QCD. By correlating partonic momentum with spatial information, they provide a unified foundation for nucleon tomography and for addressing fundamental questions concerning the spin, mass, and internal forces of the nucleon. This progress has been driven by the cumulative constraints from a new class of hard exclusive measurements, progressively precise theoretical calculations, steadily maturing first-principles lattice-QCD results, and increasingly sophisticated phenomenological analyses.

Considerable experimental progress has established DVCS as the leading channel for constraining GPDs~\cite{Aschenauer:2025cdq}, while TCS, DVMP, and heavy-quarkonium production provide complementary sensitivity to flavor, polarization, and gluon structures. Additionally, DDVCS and certain $2\to3$ processes may offer additional sensitivity to the parton momentum fraction, though they remain experimentally challenging. The broad kinematic reach and precision anticipated from the JLab program, the future EIC, and the proposed EicC have enabled and will further promote the transition from qualitative imaging toward more complete quark and gluon tomography. This requires coordinated measurements across processes, targets, and polarization observables, together with careful control of experimental systematics and theoretical uncertainties.

Perturbative QCD studies have also advanced significantly in parallel, bringing hard exclusive processes into the domain of precision phenomenology. Leading-power DVCS coefficient functions are now known through NNLO, complemented by increasingly accurate GPD evolution and threshold resummation. Two-loop results are also available for nonsinglet DDVCS and pion DVMP, while QCD factorization has been established for a broad class of $2\to3$ processes, subject to possible Glauber-mode obstructions. Kinematic power corrections, including finite-$t$ and target-mass effects, have been derived through twist-six accuracy at tree level and could play an important role at typical fixed-target kinematics. Matching the precision anticipated from future data will require higher-order calculations of hard coefficient functions and GPD evolution, together with improved control over twist-three and other power corrections. Integrating these developments consistently across processes, clarifying the limits of factorization, and providing public, validated tools for evolution and observables evaluation will be crucial for further phenomenological analyses.

Lattice QCD has become an equally important component of the GPD program. Recent advances in calculations of Mellin moments and generalized form factors have provided increasingly comprehensive information on the quark and gluon contributions to the nucleon mass, angular momentum, internal force structures, and other properties. Meanwhile, LaMET has opened a direct path toward $x$-dependent GPDs and first-principles nucleon tomography. Although current direct lattice calculations of GPDs remain exploratory and have not yet achieved comprehensive control over systematic uncertainties, particularly those associated with finite lattice spacing and hadron momentum, they are becoming increasingly mature. Further refinements in the perturbative matching of LaMET, especially near the crossover point $x=\xi$, together with direct calculations of $x$-dependent gluon GPDs, will further extend their reach. As these developments continue, lattice-QCD constraints are expected to play an increasingly important role alongside experimental measurements in extracting GPDs and elucidating the associated nucleon structure.

GPD phenomenology has developed into a powerful framework for integrating diverse experimental and theoretical information on the multidimensional structure of the nucleon. Its central challenge is to balance the intrinsically high-dimensional nature of GPDs against the lower-dimensional information provided by experimental observables, lattice-QCD calculations, and other theoretical inputs. Flexible yet systematically improvable representations that incorporate these constraints while preserving fundamental properties---such as polynomiality, positivity, and the correct forward limits---are therefore essential. Rigorous uncertainty quantification remains an important priority, particularly when combining inputs with distinct systematic uncertainties and correlations, and will be essential for reliable precision extractions of nucleon properties from GPDs.

Taken together, these advances bring the field closer than ever to answering the questions that originally motivated GPD studies: how the nucleon spin is assembled, how its mass and spatial structure emerge, and how quarks and gluons generate the internal forces that bind it. Realizing this potential will require a coherent program uniting precision measurements, systematically improved theory, first-principles lattice calculations, and global analyses with controlled uncertainties. Such a synthesis can establish GPDs as a quantitatively validated framework for nucleon structure and, ultimately, reveal how the complexity of hadronic matter emerges from QCD.

\appendix

\begin{acknowledgments}
We thank many of our colleagues for discussions and correspondences related to the main topic of this review, including V. Braun, L. Chen, K. Cichy, G. Duplan\v{c}i\'{c}, Y. Jia, K. Kumeri\v{c}ki, A. Manashov, K. Passek-Kumeri\v{c}ki, A. Schaefer, J. Schoenleber, P. Sznajder, L. Szymanowski, W. Wang, X. Xiong, F. Yuan, and Y. Zhao. We also thank the QGT collaboration for the initial motivation of this work.  
Y. G. is supported by the Office of Science of the U.S. Department of Energy under Contract No. DE-AC02-05CH11231. and No. DE-SC0012704, and by LDRD funds from Brookhaven Science Associate. X. J. is supported in part by NSFC Grant No. 12635004, the State Key Laboratory of Dark Matter Physics, and Thomas and Linda Lau Family Foundation. Y.J. is grateful for the support in part by the National Natural Science Foundation of China with Grant No.12535006, by the Guangdong Basic and Applied Basic Research Foundation under grant No. 2026A1515011306, %
and by the University Development Fund of the Chinese University of Hong Kong, Shenzhen, under the grant No. UDF0100386. This work is partially supported by the National Natural Science Foundation of China under Grant Nos. 12125503 and 12305103.
\end{acknowledgments}

\bibliography{refs}

\end{document}